\documentclass[fleqn,usenatbib]{mnras}

\usepackage{newtxtext,newtxmath}

\usepackage[T1]{fontenc}

\DeclareRobustCommand{\VAN}[3]{#2}
\let\VANthebibliography\thebibliography
\def\thebibliography{\DeclareRobustCommand{\VAN}[3]{##3}\VANthebibliography}

\usepackage{graphicx}	
\usepackage{amsmath}	
\usepackage{bm}
\usepackage{booktabs}
\usepackage{threeparttable}
\usepackage[dvipsnames]{xcolor}
\usepackage[normalem]{ulem}
\usepackage[export]{adjustbox}

\newcommand{\email}[1]{\mbox{\href{mailto:#1}{#1}}}

\makeatletter
\newlength{\abovecaptionskip}%
\makeatother

\defcitealias{mandel20}{M20}

\DeclareMathOperator{\exponential}{Exponential}

\DeclareMathOperator{\vectorise}{vec}

\title[Foundation DR1 with \textsc{BayeSN}]{Testing the Consistency of Dust Laws in SN Ia Host Galaxies:\\ A \textsc{BayeSN} Examination of Foundation DR1}

\author[S. Thorp et al.]{
Stephen Thorp,$^{1}$\thanks{E-mail: \email{sjt202@cam.ac.uk}}
Kaisey S. Mandel,$^{1,2}$ David O. Jones,$^{3\thanks{NASA Einstein Fellow}}$ Sam M. Ward,$^{1}$
and Gautham Narayan$^{4,5}$
\\
$^{1}$Institute of Astronomy and Kavli Insititute for Cosmology, Madingley Road, Cambridge, CB3 0HA, UK\\
$^{2}$Statistical Laboratory, DPMMS, University of Cambridge, Wilberforce Road, Cambridge, CB3 0WB, UK\\
$^{3}$Department of Astronomy and Astrophysics, University of California, Santa Cruz, CA 95064, USA\\
$^{4}$University of Illinois at Urbana-Champaign, 1003 W. Green St., IL 61801, USA\\
$^{5}$Center for Astrophysical Surveys, National Center for Supercomputing Applications, Urbana, IL 61801, USA
}

\date{Accepted XXX. Received YYY; in original form ZZZ}

\pubyear{2021}

\begin{document}
\label{firstpage}
\pagerange{\pageref{firstpage}--\pageref{lastpage}}
\maketitle

\begin{abstract}
We apply \textsc{BayeSN}, our new hierarchical Bayesian model for the SEDs of Type Ia supernovae (SNe Ia), to analyse the $griz$ light curves of 157 nearby SNe Ia ($0.015<z<0.08$) from the public Foundation DR1 dataset. We train a new version of \textsc{BayeSN}, continuous from 0.35--0.95~$\upmu$m, which we use to model the properties of SNe Ia in the rest-frame $z$-band, study the properties of dust in their host galaxies, and construct a Hubble diagram of SN Ia distances determined from full $griz$ light curves. Our $griz$ Hubble diagram has a low total RMS of 0.13~mag using \textsc{BayeSN}, compared to 0.16~mag using \textsc{SALT2}. Additionally, we test the consistency of the dust law $R_V$ between low- and high-mass host galaxies by  using our model to fit the full time- and wavelength-dependent SEDs of SNe Ia up to moderate reddening (peak apparent $B-V \lesssim 0.3$). Splitting the population at the median host mass, we find $R_V=2.84\pm0.31$ in low-mass hosts, and $R_V=2.58\pm0.23$ in high-mass hosts, both consistent with the global value of $R_V=2.61\pm0.21$ that we estimate for the full sample. For all choices of mass split we consider, $R_V$ is consistent across the step within $\lesssim1.2\sigma$. Modelling population distributions of dust laws in low- and high-mass hosts, we find that both subsamples are highly consistent with the full sample's population mean $\mu(R_V) = 2.70\pm0.25$ with a 95\% upper bound on the population $\sigma(R_V) < 0.61$. The $R_V$ population means are consistent within $\lesssim1.2\sigma$. We find that simultaneous fitting of host-mass-dependent dust properties within our hierarchical model does not account for the conventional mass step.
\end{abstract}




\section{Introduction}
From the landmark discovery of the Universe's accelerating expansion \citep{riess98, perlmutter99}, to state-of-the-art efforts measuring the Hubble constant \citep{dhawan18, burns18, riess19} and dark energy equation-of-state parameter \citep{scolnic18, abbott19}, Type Ia supernovae (SNe Ia) have been a key pillar in our understanding of cosmology. The ongoing development of a homogeneous low redshift SN Ia sample \citep{foley18,jones19,jones20}, as well as future high redshift data from the \textit{Nancy Grace Roman Space Telescope} \citep{spergel15, hounsell18} and the Vera C.\ Rubin Observatory's Legacy Survey of Space and Time \citep{ivezic19}, will enable a new era of precision cosmology. To make the most of the incoming data, however, SN Ia models must be improved, and subtleties such as SN--host correlations \citep[e.g.][]{kelly10,sullivan10} must be better understood. In this paper, we apply the state-of-the-art hierarchical model, \textsc{BayeSN} \citep{mandel20}, to analyse current low-redshift data from the Foundation Supernova Survey \citep{foley18, jones19}, and to explore the relation between the dust properties and stellar masses of SN Ia host galaxies.

The host galaxy `mass step' -- a tendency for post-standardisation SN Ia magnitudes to be brighter in more massive galaxies \citep{kelly10, sullivan10, smith20} -- is well established, and routinely corrected for, but its cause remains uncertain. Another question is the proper interpretation of the SN Ia apparent colour--luminosity relation, and the value(s) of the $R_V$ parametrizing the extinction and reddening law of host galaxy dust affecting the light of SNe Ia. For some highly reddened SNe Ia with peak apparent $(B-V) \gg 0.3$ (e.g. SN 2003hg, SN 2002cv, \citealp{eliasrosa06, eliasrosa08}; SN 2006X, \citealp{wang08}; SN 2014J, \citealp{amanullah14}; \citealp{hoang17}), unusually low individual $R_V \approx 1.5-2$ values have been inferred -- far from the mean Galactic value of $\approx3.1$. However, such reddened SNe are typically excluded from cosmological analyses, due to a standard colour cut $(B-V) \lesssim 0.3$. Still, there is ongoing debate over whether `normal' samples of SNe Ia with $(B-V)\lesssim0.3$ are consistent with a low $R_V\approx\text{1--2}$ \citep[e.g.][]{nobili08, stanishev18}, a higher $R_V$ of $\approx\text{2.5--3}$ \citep[e.g.][]{folatelli10, chotard11, foley11, mandel11, mandel17, mandel20, burns14, sasdelli16, leget20, arima21}, or a wider range of values \citep[e.g.][]{amanullah15}. These studies are made difficult by the limited range of reddening within the $(B-V)\lesssim0.3$ colour cut, and limited wavelength range of conventional optical analyses. The confounding of intrinsic variation with extrinsic dust effects in the apparent colours of SNe Ia also complicates the interpretation of the data \citep[e.g.][]{freedman09}. For example, \citet{mandel17} showed that the convolution of an intrinsic SN colour-luminosity trend with dust reddening-extinction effects generically results in a nonlinear effective mean apparent colour-luminosity curve, and the conventional linear fit approximates an average of the physically-distinct intrinsic and dust slopes.

Recently, it was proposed by \citet{brout20} that a difference in the dust properties of low- and high-mass host galaxies is the root cause of the mass step. They use survey simulations with an input $B-V$ colour distribution consisting of an intrinsic and extrinsic component, the latter obeying a colour--luminosity relation parametrized by $R_V+1$. A population distribution of these $R_V$ values is included, and a version of their model where this is permitted to differ between low and high host galaxy stellar masses is also considered. They iterate over survey simulations, attempting to find the values of the input population parameters for which the simulation outputs best agree with observed supernova data. In particular, for each simulation they compare its colour distribution, its Hubble residual bias and scatter in different colour bins, and its overall fitted colour--luminosity trend, to the equivalent quantities obtained from \textsc{SALT2} \citep{guy07, guy10, betoule14} fits to a compilation of supernova light curves. At low redshift, they used data from the CfA Supernova Program \citep[CfA;][]{hicken09, hicken12}, Carnegie Supernova Project \citep[CSP;][]{contreras10, folatelli10, stritzinger11, krisciunas17} and Foundation Supernova Survey \citep{foley18, jones19}, with high redshift data coming from the Pan-STARRS-1 Medium Deep Survey \citep[MDS;][]{rest14, scolnic18}, Supernova Legacy Survey \citep[SNLS;][]{astier06,betoule14}, SDSS-II Supernova Survey \citep{frieman08,sako11,sako18} and the first three years of Dark Energy Survey data \citep[DESYR3,][]{des16, brout19}, for a total of 1445~SNe. Comparing their simulations with SALT2 parameter fits to these data, they infer an $R_V$ population distribution peaking at $2.75\pm0.35$ in low-mass hosts ($M_*<10^{10}\mathrm{M}_\odot$), and $1.50\pm0.25$ in high-mass hosts, with wide standard deviations of $1.3\pm0.2$ in both mass bins. They conclude that including these different $R_V$ distributions eliminates the need for a 0.06~mag mass step which they otherwise observed in the data \citep[see also][]{popovic21}. 

However, the observation of a mass step in near-infrared (NIR) data, where sensitivity to dust should be reduced, disfavours a difference in host dust properties as an explanation of the step. Recently, \citet{ponder20} used separate \textsc{SNooPy} \citep{burns11} fits to optical and $H$-band light curves of SNe Ia to explore how correlations with host mass change between the optical and NIR. They analysed the $H$-band (optical) light curves of 99 (71) supernovae from the CfA/CfAIR2 \citep{woodvasey08,friedman15}, CSP, SweetSpot \citep{weyant18}, and elsewhere in the literature \citep{jha99, hernandez00, krisciunas00, krisciunas03, krisciunas04a, krisciunas04b, krisciunas07, valentini03, phillips06, pastorello07a, pastorello07b, stanishev07, pignata08}, finding an $H$-band mass step of $0.18\pm0.05$~mag ($0.10\pm0.04$~mag after an outlier cut) at a best fit location of $10^{10.44}\mathrm{M}_\odot$, with a comparably sized $0.17\pm0.05$~mag step in the optical. Contemporary work by \citet{uddin20} recovered a similar result in an analysis of the CSP-I sample. Using independent \textsc{SNooPy} fits to each of the $uBgVriYJH$ light curves of every supernova, they measure comparable steps for all passbands at the median host mass of $10^{10.48}\mathrm{M}_\odot$, with step sizes of between $0.07\pm0.03$ ($V$-band) and $0.15\pm0.04$~mag ($u$-band). They also computed \citep[see][fig.\ 13]{uddin20} the step size vs.\ wavelength behaviour that would be expected from the $R_V$ and $E(B-V)$ population distributions implied by the \citet{brout20} results, finding this to be in poor agreement with the behaviour observed in the CSP data. 

Recently, however, \citet{johansson21} analysed a sample from the literature (including CfA/CfAIR, CSP-I, and others from \citealp{baronenugent12,stanishev18, amanullah15}), along with 42 SNe Ia from their own intermediate Palomar Transient Factory \citep[iPTF;][]{rau09} survey whose NIR data were obtained with the Reionization and Transients InfraRed camera \citep[RATIR;][]{butler12}. They estimated NIR mass steps consistent with zero (although not entirely inconsistent with previous literature estimates), and claimed that fitting for $R_V$ on a supernova-by-supernova basis eliminated the mass step in the optical.

Other recent work by \citet{gonzalezgaitan20} studied the mass dependence of the apparent SN Ia colour--luminosity relation. They fit extensions of the \citet{tripp98} standardisation formula to the \textsc{SALT2} parameters of 740 supernovae from the Joint Lightcurve Analysis \citep[JLA;][]{betoule14}. The \citet{tripp98} formula,
\begin{equation}\label{eqn:tripp}
    \mu_s = m_B^s - M_B + \alpha x_1^s - \beta c^s,
\end{equation}
expresses the distance modulus, $\mu_s$, of a supernova, $s$, as a linear combination of its $B$-band apparent magnitude $m_B^s$, light curve stretch $x_1^s$, and apparent $B-V$ colour $c^s$, with stretch--luminosity and colour--luminosity coefficients $\alpha$ and $\beta$, and an absolute magnitude constant $M_B$. The mean magnitude offset, or mass step, between SNe Ia in low- and high-mass host galaxies is typically accounted for with two different values of $M_B$ for each subsample. In their analysis, \citet{gonzalezgaitan20} allow for different values of the apparent colour--luminosity coefficient, $\beta$, in different mass and colour bins. They find a significant relation between $\beta$ and host galaxy mass. Since $\beta$ averages the intrinsic colour--luminosity relation and extrinsic dust law \citep{mandel17}, this could be explained by either a difference in $R_V$, or a difference in intrinsic properties. They also see a significant relation between $\beta$ and apparent colour -- likely driven by a tendency for the reddest supernova to follow an $R_V$ driven colour--luminosity trend, with the bluest following an intrinsic trend \citep[expected from][]{mandel17}. Unlike \citet{brout20}, \citet{gonzalezgaitan20} find that find that the mass step is preserved even when allowing for different apparent colour--luminosity relations in different mass bins.

In the near future, the Vera C.\ Rubin Observatory's Legacy Survey of Space and Time \citep[LSST;][]{ivezic19}, and supernova programs on the \textit{Nancy Grace Roman Space Telescope} \citep[\textit{Roman Space Telescope};][]{spergel15, hounsell18}, will massively expand the high-redshift SN Ia dataset which informs dark energy analyses. However, these projects are likely to yield small or sub-optimal low redshift samples \citep{foley18, hounsell18, jones20}, meaning these data must come from elsewhere. The low-redshift sample used in many previous analyses \citep[e.g.][]{betoule14, rest14, scolnic14, scolnic18} has come from a variety of sources -- particularly the CfA \citep{riess99, jha06, woodvasey08, hicken09, hicken12, friedman15}, CSP \citep{contreras10, folatelli10, stritzinger11, krisciunas17}, and Cal\'an/Tololo Survey \citep{hamuy96a}. This heterogeneity gives rise to calibration and cross-calibration systematics which contribute greatly to the error budget \citep{conley11, scolnic14, scolnic18, brout19}, in spite of efforts to remedy this \citep[e.g.][]{scolnic15,currie20}. Since these data are often critical to training SN Ia models (e.g. \textsc{BayeSN}, \citealp{mandel11, mandel20}; \textsc{SALT2}, \citealp{guy07,guy10,betoule14}), the uncertainty associated with their calibration can bleed into cosmological analyses in non-trivial ways. An additional complication is that the CfA \citep[see][]{hicken09, hicken12} and CSP \citep[see][]{krisciunas17} samples, which form a large fraction of the current low-$z$ dataset, are dominated by supernovae discovered through galaxy-targeted monitoring programs such as the Lick Observatory Supernova Search \citep[LOSS;][]{li00, filippenko01, leaman11, li11}. This creates a complex selection function, and probes a distribution of host galaxies unrepresentative of the true population.

To address these problems, considerable ongoing effort is being applied to replacing the existing low redshift dataset with a more homogeneous sample observed entirely on Pan-STARRS-1 \citep[PS1;][]{kaiser10, chambers16}. The first data release of the Foundation Supernova Survey \citep[Foundation DR1;][]{foley18,jones19} represents the current result of these efforts, compiling cosmology-ready photometry of 180 SNe Ia at $z\lesssim0.1$. By following up SN Ia discoveries that mainly come from `untargeted' surveys -- primarily the All-Sky Automated Search for Supernovae \citep[ASAS-SN;][]{shappee14} and Pan-STARRS Survey for Transients \citep[PSST;][]{huber15} -- this sample should be considerably less biased in terms of host galaxy properties than existing low redshift data. Future data releases of the Foundation program and the Young Supernova Experiment \citep[YSE;][]{jones20} will enlarge this further. This will provide a homogeneous, well-calibrated low-$z$ SN Ia sample to anchor the LSST and \textit{Roman Space Telescope} datasets of the future, and accompany the considerable high-$z$ sample already observed in the PS1 MDS \citep{rest14, scolnic18, villar20}.

The excellent calibration properties, and internal consistency, of the Foundation data make them an ideal training set for SN Ia light curve and spectral energy distribution (SED) models. These models are fundamental to any cosmological analysis, as the route by which distances are estimated from photometric light curves. The conventional model used in most recent analyses is \textsc{SALT2} \citep{guy07, guy10, betoule14}, whose parameter estimates (peak apparent $B$-band magnitude $m_B$; stretch $x_1$; and apparent colour $c$) are converted to a distance via a linear \citet{tripp98} standardisation. This paradigm has a number of known weaknesses, including its limited wavelength coverage, confounding of intrinsic colour and extrinsic dust effects \citep[see][]{mandel17}, and uncertain characterisation of residual scatter \citep[a key systematic, see][]{scolnic14a}. Moreover, the most recent \textsc{SALT2} training \citep{betoule14} was carried out using a heterogeneous sample of SNe Ia with known calibration and cross-calibration issues, baking these systematics into the model. In their recent Pantheon analysis, \citet{scolnic18} found that these inherited systematics in the \textsc{SALT2} model (which cannot easily be corrected) likely dominate over the calibration systematics in the data they were fitting \citep[which can be at least partly alleviated,][]{scolnic15} .

Some of the limitations of the \textsc{SALT2} SED model have been addressed by alternatives. In particular, the \textsc{BayeSN} model \citep{mandel20} is a coherent framework for optical and near-infrared (NIR) SEDs, continuous in wavelength from 0.35--1.8~$\upmu$m, with distinct treatments of dust and intrinsic variability. In \citet{mandel20}, this model was trained on a compilation \citep{avelino19} of 79 low-$z$ optical and NIR SNe Ia, mainly from the CSP \citep{contreras10, stritzinger11, krisciunas17} and CfA  \citep{jha99, woodvasey08, hicken09, hicken12, friedman15} surveys, as well as earlier data from the Las Campanas Observatory \citep[LCO;][]{krisciunas04a, krisciunas04b} and elsewhere in the literature \citep{krisciunas03, krisciunas07, stanishev07, pignata08, leloudas09}. Whilst this dataset was over two times larger than those used in training earlier optical and NIR light curve models \citep{mandel09, mandel11, burns11}, the heterogeneity of data sources is subject to cross-calibration issues. Additionally, host galaxy mass estimates \citep[from e.g.][]{ponder20, uddin20}, indicate a considerable bias towards high masses (around 80\% have host galaxy stellar masses $>10^{10}\mathrm{M}_\odot$).

The \textsc{BayeSN} framework is well suited to the incorporation and analysis of correlations of SN Ia properties with host galaxies. Additionally, the supernovae in Foundation DR1 benefit from calibration homogeneity, consistently determined host galaxy masses \citep{jones19}, and a less biased host distribution than other low-$z$ SN Ia samples. This makes them an ideal dataset for studying SN--host correlations amongst the low--moderate reddening ($B-V\lesssim0.3$) SNe Ia typical of cosmological samples. Therefore, in this work, we present a first \textsc{BayeSN} analysis of Foundation DR1. As part of this study, we repeat the mass-agnostic \citet{mandel20} analysis on the Foundation data, verifying \textsc{BayeSN}'s performance, and modelling for the first time the statistical properties of SNe Ia in the rest-frame $z$-band. Additionally, we extend the \textsc{BayeSN} model to allow population dust properties to be split by host galaxy mass. By including this alongside a conventional step-like brightness offset when training our SED model, we are able to test if dust properties differ significantly in low and high mass SN Ia host galaxies, and if there is any interplay between dust properties and the mass step. In previous \textsc{SALT2}-based analyses of this problem, all inferences about $R_V$ have been made from the apparent colour--luminosity relation between extinguished absolute magnitude in $B$ and apparent $B-V$ colour at the time of maximum light. In contrast, the hierarchical framework of \textsc{BayeSN} embeds a distinct dust law within the time- and wavelength-dependent SED model, inferring $R_V$ by drawing on the colour--luminosity information contained in the full $griz$ light curves. This properly leverages the fact that the dust law impacts the SED at all times and wavelengths, enabling us to better discern intrinsic SN variations from the effects of dust.  Performing this inference within a Bayesian framework enables us to marginalise over other effects in the problem in order to draw robust conclusions about the dust and probabilistically quantify our uncertainties about its properties.

In Section \ref{sec:model}, we recap the details of the \textsc{BayeSN} model \citep[fully described in][]{mandel20}, and discuss the modifications we employ in our treatment of host galaxy mass. In Section \ref{sec:data} we describe the Foundation dataset, and the data cuts we applied. This section also discusses several choices of host galaxy mass split that we consider throughout. Section \ref{sec:results} presents and discusses our results, covering SN Ia $z$-band behaviour, the mass step and residual scatter, population dust properties, the distribution of light curve shape parameters, intrinsic colour and extrinsic dust, and Hubble diagram scatter in turn throughout Sections \ref{sec:w1}--\ref{sec:hubblediagrams}. Finally, in Section \ref{sec:conclusions}, we summarise our results, and discuss the outlook for future work.

\section{The \textsc{BayeSN} Model}
\label{sec:model}

Hierarchical Bayes provides a natural probabilistic framework for modelling and inference of populations and their constituent individuals \citep[e.g.][]{gelman_bda}, and was first applied to SN Ia modelling by \citet{mandel09}. Throughout this work, we carry out our analysis in several different configurations of the \textsc{BayeSN} hierarchical model \citep[hereafter \citetalias{mandel20}]{mandel20}, detailed fully in Sections \ref{sec:nosplit}--\ref{sec:fullsplit}. Our \textit{No-Split} model (Section \ref{sec:nosplit}) treats all SNe in the sample as a single population with parameters agnostic of host galaxy mass. Our \textit{Partial-Split} model (Section \ref{sec:partialsplit}) allows key parameters to differ between the low- and high-mass subsamples, whilst keeping others common to all supernovae irrespective of host mass. Finally, our \textit{Full-Split} model (Section \ref{sec:fullsplit}) treats the subsamples in the two mass bins as entirely separate populations. Within the \textit{No-Split} configuration, we carry out parallel analyses where different assumptions are made about the $R_V$ value(s) parametrizing host galaxy dust laws. The first of these (Section \ref{sec:globalRV}) assumes a single value of $R_V$ for all SNe in the sample, whilst the second (Section \ref{sec:popRV}) assumes a population distribution of $R_V$, with individual values permitted for each SN. Within the \textit{Partial-Split} configuration, we run equivalent analyses where either a separate single $R_V$, or a separate population distribution of $R_V$, is allowed in each mass bin.

As in \citetalias{mandel20}, the models are all implemented in the Stan probabilistic programming language \citep{carpenter17, stan20}, with the joint posterior over all global and individual SN parameters (eqs. 27-28 in \citetalias{mandel20}, with modifications as stated below) sampled using Stan's advanced Hamiltonian Monte-Carlo algorithm \citep{hoffman14,betancourt16}. During training, we adopt external distance constraints from the distance-redshift relation based on a fiducial $\Lambda$CDM cosmology \citep[$H_0=73.24$, $\Omega_M=0.28$, $\Omega_\Lambda=0.72$;][]{riess16}, with external distance uncertainties derived from peculiar velocity and spectroscopic redshift uncertainties. We assume a peculiar velocity uncertainty of $\sigma_{\text{pec}}=150~\text{km\,s}^{-1}$ \citep{carrick15}.

\subsection{\textit{No-Split} Model}
\label{sec:nosplit}
\subsubsection{Global Dust Law}
\label{sec:globalRV}
As a starting point, we consider the configuration of the \textsc{BayeSN} model described in \citetalias{mandel20} that is agnostic with respect to host galaxy mass. For full details of the model, we refer the reader to the \citetalias{mandel20} paper. We briefly summarise the essentials here. Throughout the paper, we will refer to this as the \textit{No-Split} model, since it ignores host galaxy mass and treats all SNe in the sample jointly as a single population.

The \textsc{BayeSN} forward model specifies a time- and wavelength-varying surface describing the intrinsic SED of a Type Ia supernova. This is extinguished by host galaxy dust, redshifted, dimmed by distance, extinguished by Milky Way dust, integrated through photometric passbands at the observation times, and finally perturbed by measurement error, to yield observed SN Ia light curves. The host-dust-extinguished SED $S_s(t,\lambda_r)$ of a supernova $s$, as a function of rest-frame phase $t$ and wavelength $\lambda_r$, is modelled as:
\begin{multline}
    -2.5\log_{10}[S_s(t,\lambda_r)/S_0(t,\lambda_r)] = M_0 + W_0(t,\lambda_r)\\ + \delta M_s +\theta_1^sW_1(t,\lambda_r) + \epsilon_s(t,\lambda_r) +A_V^s\xi(\lambda_r;R_V)
    \label{eq:SEDmodel}
\end{multline}
 (as in \citetalias{mandel20} Eq.\ 12). Here, $S_0(t,\lambda_r)$ is an optical-NIR SN Ia baseline spectral template \citep{hsiao07, hsiao09}; $M_0\equiv-19.5$ is a normalisation constant; $W_0(t,\lambda_r)$ is a broad warping of the baseline template, common to all SNe, to model the average intrinsic SED.
 
 The next three terms model the variations of individual SN Ia SEDs in the absence of dust ($A_V^s = 0)$. $\delta M_s$ is a `grey' (time- and wavelength-independent) brightness offset specific to supernova $s$; $W_1(t,\lambda_r)$ is the first functional principal component, describing the primary SED features that vary between SNe;  $\theta_1^s$ is a scalar coefficient describing the extent to which $W_1(t,\lambda)$ affects the $s$th supernova;  $\epsilon_s(t,\lambda_r)$ is a time- and wavelength-varying residual function, describing SED perturbations particular to SN $s$ that are otherwise unexplained. The statistical properties of these terms are learnt during training and model the covariance structure of the population of intrinsic SN Ia SEDs.  Since colours and luminosities are quantities derived from integrals over the SEDs, intrinsic colour and luminosity correlations across phase and wavelength are captured here.
 
 The last term describes the effect of dust along the line-of-sight in the SN host galaxy on the SN Ia SED. The parameter $A_V^s$ is the $V$-band host dust extinction affecting supernova $s$; and $\xi(\lambda_r;R_V)$ is the \citet{fitzpatrick99} dust law, parametrized by $R_V$. Initially, we assume $R_V$ has a single global value for the sample, but we later explore models where this is split by host galaxy mass, or has a population distribution within a (sub)sample. 
 
 In practice, we represent the $W_0(t,\lambda_r)$, $W_1(t,\lambda_r)$, and $\epsilon_s(t,\lambda_r)$ functions with natural cubic splines, parametrized by matrices of knots, $\bm{W}_0$, $\bm{W}_1$, and $\bm{E}_s$. The main difference from \citetalias{mandel20} (which models a wider wavelength range of 0.35--1.8~$\upmu$m) is the exact configuration of these spline knots. In rest-frame phase, we use the same knot locations as in \citetalias{mandel20}, with knots every 10~d between $-10$ and $40$~d. In wavelength, we place knots at $\bm{l}=(3500, 4900, 6200, 7700, 8700, 9500)$ \AA, corresponding to the centres of the $griz$-bands, plus the outer edges of $g$ and $z$.

After redshifting the host-dust-extinguished SED given by Eq.\ (\ref{eq:SEDmodel}), and accounting for distance, we apply Milky Way dust reddening and extinction using the \citet{fitzpatrick99} law, with $R_{\text{MW}}=3.1$ and $A_{\text{MW}}^s = R_{\text{MW}}\times E(B-V)_{\text{MW}}^s$ computed for each SN using the \citet{schlafly11} reddening map. The SED is then integrated through photometric bandpasses to compute model fluxes. This generative model forms the basis of our likelihood (see \citetalias{mandel20} for exact details, and fig. 1 therein for a graphical representation). We model the population distributions of the latent parameters of individual SNe as:
\begingroup
\allowdisplaybreaks
\begin{align}
    A_V^s &\sim \exponential(\tau_A),\\
    \theta_1^s &\sim N(0,1),\\
    \delta M_s &\sim N(0,\sigma_0^2),\label{eq:delMprior}\\
    \bm{e}_s &\sim N(\bm{0},\bm{\Sigma}_\epsilon),\label{eq:epsilonprior}
\end{align}%
\endgroup
where $\bm{e}_s=\vectorise(\bm{E}_s)$ is the vectorised $\bm{E}_s$ matrix, and $\tau_A$, $\sigma_0$ and $\bm{\Sigma}_\epsilon$ are hyperparameters describing the population mean host galaxy extinction, population standard deviation of the $\delta M_s$ offsets, and population covariance of the residual $\bm{e}_s$ perturbations, respectively. Our weak hyperpriors are as in \citetalias{mandel20}.

\subsubsection{Population Distribution of Dust Law $R_V$}
\label{sec:popRV}
Modelling parameters of individuals as being drawn from a population distribution with unknown hyperparameters to be jointly inferred from the data is known as \emph{partial pooling} \citep{gelman_bda}. Population distribution models for $R_V$ were first introduced by \citet{mandel09,mandel11}. To enable a more direct comparison with the results of \citet{brout20}, we consider an extension of our current model in which the $R_V$ values for individual SNe are drawn from a population distribution, rather than all having a single global value. In this model variant, the $R_V^s$ of an individual SN $s$ is now a latent variable, drawn from a truncated normal (Gaussian)\footnote{Here, $X\sim\text{Trunc-}N(\mu,\sigma^2,a,\infty)$ denotes that the random variable $X$ is drawn from a one-sided truncated normal distribution, with truncation of the lower tail at $a$. The probability density function for $X\geq a$ is \begin{equation*}P(X|\mu,\sigma,a)=\frac{1}{\sigma}\frac{\phi(\xi)}{1-\Phi(\alpha)},\end{equation*} and is zero for $X < a$. Here, $\xi=(X-\mu)/\sigma$, $\alpha=(a-\mu)/\sigma$, and $\phi(z)$ and $\Phi(z)$ are respectively the PDF and CDF of a standard normal random variable $z$.} population distribution,
\begin{equation}
    R_V^s \sim \text{Trunc-}N(\mu_R,\sigma_R^2,0.5,\infty),
    \label{eq:RVsprior}
\end{equation}
with a truncation of the lower tail to force $R_V^s\geq0.5$. A physically-motivated choice of lower bound would be $R_V^s\geq1.2$ \citep[the Rayleigh scattering limit, see][]{draine03}, but we opt for the more liberal choice of $R_V^s\geq0.5$ to match \citet{brout20}. The population mean and variance parameters $\mu_R, \sigma_R^2$ are estimated coherently with all other parameters by augmenting the global posterior density. The simple assumption of a Gaussian population distribution for $R_V^s$ is somewhat motivated by the empirical distribution of $R_V$ values found along different sightlines through the Milky Way \citep{schlafly16}, which is well-described by a narrow Gaussian with mean 3.32 and standard deviation 0.18. While this may not be directly analogous to sightlines to SNe Ia in external galaxies, it is a working assumption that can be elaborated on in future work. Given that our results ultimately suggest a preference for narrow $R_V$ distributions (see \S\ref{sec:pop_RV_results}), we expect our results to be robust to the exact shape of the assumed distribution. One would expect that a Gaussian population distribution would be \textit{more} sensitive to outliers than a heavier-tailed alternative, making an overestimate of the dispersion in the population (which does not appear to have happened here) more of a risk than an underestimate. In future work, we will be able to test alternative possible population distributions (e.g.\ skew-normal, Student's $t$, Gaussian mixtures).

We place a uniform hyperprior on $\mu_R\sim U(1,5)$, and a half-Normal hyperprior on  $\sigma_R\sim\text{Half-}N(0,2^2)$. The latter is selected so as to avoid a hard limit on $\sigma_R$, but to place relatively little prior probability at excessively high values of $\sigma_R>4$. Sensitivity analysis for alternate $\sigma_R$ hyperpriors is demonstrated in Appendix \ref{app:sigmaRprior}.

\subsection{\textit{Partial-Split} Model}
\label{sec:partialsplit}
We also train a version of the model where several key parameters, previously treated as common to all supernovae, are split by host galaxy mass at a critical value, $M_{*,\text{split}},$ to fit for a low- and high-mass version of each. We will refer to this as the \textit{Partial-Split} model. By splitting only selected parameters, the model is able to probe salient differences between the high- and low-mass subsamples, whilst still utilising all of the data for determining the mean intrinsic SED, primary mode of intrinsic SED variation, and intrinsic residual covariance.

The parameters that we split in this way are the $R_V$ (or the population distribution parameters, $\mu_R$ and $\sigma_R$, in the case where supernovae have individual $R_V^s$ values) parametrizing the \citet{fitzpatrick99} host galaxy dust extinction law; the population mean extinction $\tau_A$; and $\sigma_0$, the population standard deviation of the `grey' $\delta M_s$ component of the intrinsic SED. We also introduce a parameter, $\Delta M_0$, which shifts the population mean of the $\delta M_s$ for supernovae in low-mass hosts -- this is functionally equivalent to allowing a constant `mass step'. This means that equation (\ref{eq:delMprior}) becomes
\begin{equation}
    \delta M_s \sim 
    \begin{cases}
        N(0,\sigma_{0, \text{HM}}^2) &\text{ if } M_{*,s} \geq M_{*,\text{split}}\\
        N(\Delta M_0, \sigma_{0,\text{LM}}^2) &\text{ if } M_{*,s} < M_{*,\text{split}}
    \end{cases},
    \label{eq:delMpriorsplit}
\end{equation}
where $M_{*,s}$ is the host galaxy stellar mass of SN $s$, and the subscript LM and HM denote the values of $\sigma_0$ below and above the mass split. The hyperpriors on the low- and high-mass values of $R_V$ (or $\mu_R$, $\sigma_R$), $\tau_A$, and $\sigma_0$ are taken to be the same as in the \textit{No-Split} model. For the low- vs.\ high-mass offset, $\Delta M_0$, we assume a uniform hyperprior within $0.0\pm0.2$~mag, since previous works investigating dependence on host mass \citep[e.g.][]{kelly10, sullivan10, roman18, uddin20, smith20} have found mass step sizes safely within this range. If dust differences fully explain the mass step, we expect $\Delta M_0\approx0$ with the \textit{Partial-Split} model.

\subsection{\textit{Full-Split} Model}
\label{sec:fullsplit}
Finally, we also consider a mode of analysis where entirely separate models (each configured exactly like the \textit{No-Split} model) are trained for the low- and high-mass subsamples, with no shared parameters or information. We will refer to this scheme as the \textit{Full-Split} model. This configuration provides a worthwhile sanity check of the \textit{Partial-Split} model, since it allows all model components to vary between host-mass subsets.  If this is not the case, however, this will lead to weaker inferences, since this effectively doubles the number of parameters and fewer supernovae are available on either side of the mass split.

\section{Data}
\label{sec:data}

\subsection{Foundation DR1}
The first data release of the Foundation Supernova Survey \citep[Foundation DR1;][]{foley18, jones19} presents $griz$ light curves for around 180 cosmologically useful Type Ia supernovae at $z\lesssim0.1$, all observed on the Pan-STARRS-1 system. This sample is far more homogeneous than existing low-redshift SNe Ia samples, with all data having been obtained on a single instrument which is extremely well-calibrated, with accurately determined instrumental properties \citep{stubbs10, schlafly12, magnier13, magnier16}. Unlike previous low-$z$ samples, the SNe followed up by Foundation are mainly discovered by untargeted surveys like ASAS-SN \citep{shappee14} and the PSST \citep{huber15}, meaning the population of host galaxies probed will be less biased (although it is still not totally reflective of the distribution in high-$z$ samples).

\subsection{Pre-Training Cuts}
\citet{foley18} already applied several data cuts to construct the Foundation DR1 cosmology sample of 180 SNe Ia, as detailed in their \S 5.3. In particular, a standard cut on the \textsc{SALT2} apparent colour parameter $c < 0.3$ (equivalent to peak apparent $B-V$), has been applied, ensuring a colour range consistent with the cosmological sample. To define the sample for our present analysis, we additionally restrict the sample to redshifts $0.015<z<0.08$. The lower cut here is to select only SNe Ia in the smooth Hubble flow, limiting the impact of peculiar velocity errors, with the upper limit being the redshift beyond which \citet{foley18} expect observations to be more vulnerable to Malmquist bias\footnote{To ensure that our conclusions are not vulnerable to any redshift-dependent selection effects, we also repeat our analysis with a more conservative redshift cut of $z<0.05$ (which gives a sample of 125 SNe). This does not change any of our conclusions, with the inferred $R_V$ values being only slightly different (higher by about 0.1--0.15, not a statistically significant shift). The Hubble diagram scatter from training and resubstitution on the $z<0.05$ subsample is also not significantly changed compared to the full sample. From a \textit{No-Split} analysis under the $z<0.05$ cut, we estimate a Hubble diagram RMS ($\sigma_{-\text{pv}}$) of 0.131 (0.119)~mag (c.f.\ the first row of Table \ref{tab:hubble_scatter}).}. Requiring $z>0.015$ cuts 8 supernovae, bringing the sample size down to 172, with the $z<0.08$ cut removing a further 5, giving a sample of 167 supernovae. Requiring reliable host masses of $M_{*}>10^{6}\mathrm{M}_\odot$ for all SNe Ia eliminates ten further supernovae, leaving a final sample of 157.

\subsection{Choosing a Mass Split}
\label{sec:choosing}

\begin{table*}
    \centering
    \begin{threeparttable}
        \caption{Summary of the different host galaxy stellar masses at which we split the supernova sample.}
        \label{tab:masscuts}
        \begin{tabular}{l c c c c}\toprule
            Split Point & Host Galaxy Stellar Mass ($\mathrm{M}_\odot$) & Low:High Mass Ratio\tnote{a} & HR Step Size in \textit{No-Split} Run\tnote{b} & $\Delta M_0$ Offset in \textit{Partial-Split} Run\tnote{c}\\ \midrule
            $10^{10}\mathrm{M}_\odot$ & $10^{10}$ & 48:109 & $0.050\pm0.022$ & $0.042\pm0.028$\\
            Median & $10^{10.331}$ & 78:79 & $0.055\pm0.020$ & $0.054\pm0.025$\\
            MLE & $10^{10.286}$ & 75:82 & $0.066\pm0.020$ & $0.072\pm0.024$\\
            \bottomrule
        \end{tabular}
        \begin{tablenotes}
            \item [a] Number of supernovae (out of the 157 in our Foundation cut) above and below each mass split.
            \item [b] Step size derived for each step location based on the Hubble residuals from a training and photometric distance estimation (resubstitution) under the \textit{No-Split} model (see Sections \ref{sec:nosplit} and \ref{sec:choosing}).
            \item [c] The $\text{low} - \text{high}$ host mass magnitude offset, $\Delta M_0$, estimated from the \textit{Partial-Split} posterior analysis (see Sections \ref{sec:partialsplit} and \ref{sec:delM}).
        \end{tablenotes}
    \end{threeparttable}
\end{table*}

\begin{figure}
    \centering
    \includegraphics[width=\linewidth]{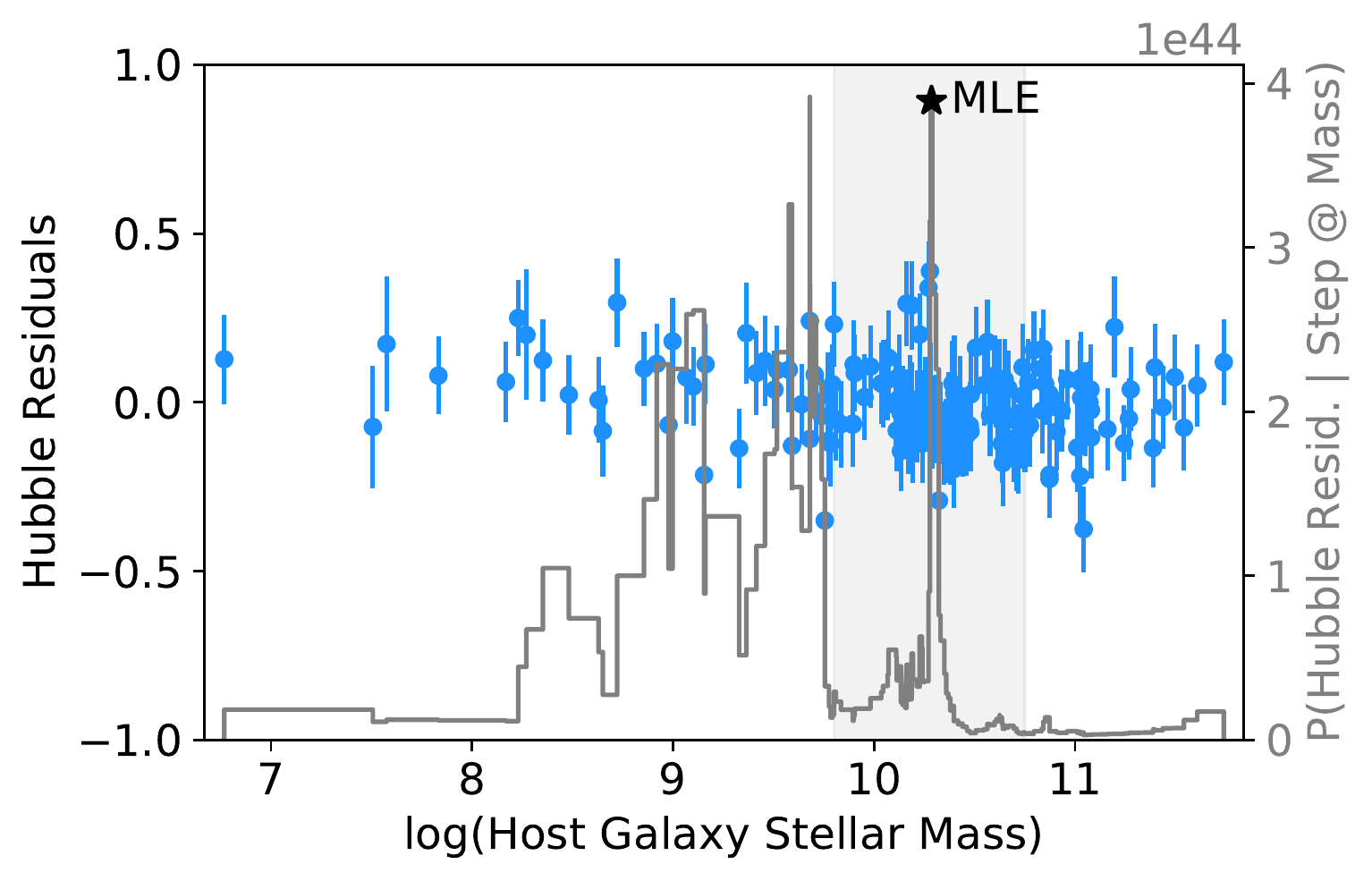}
    \caption{Hubble residuals, $\bm{\Delta\mu}$ plotted along with the marginal likelihood, $P(\bm{\Delta\mu}|\log_{10}M_{*,\text{split}})$, of these data given a step at a particular mass (see Appendix \ref{app:stepmle} for computational details). Indicated with a star is the maximum likelihood step, which we use in some later analyses. The faint grey band covers the interquartile range of the host galaxy masses in the sample.}
    \label{fig:steps}
\end{figure}

We explore several different possibilities for setting the critical host mass at which to split the training set. These are summarised in Table~\ref{tab:masscuts}. Analyses using \textsc{SALT2} \citep[e.g][]{betoule14} typically allow for a step at $M_{*,\text{split}} = 10^{10}\mathrm{M}_\odot$ by convention, so this is one option we use. However it leads to a fairly unbalanced divide, with only 48 of the 157 supernovae in our Foundation sample falling on the low-mass side. As a balanced alternative, we try setting $M_{*,\text{split}}$ to the median host mass of our sample, $10^{10.331}\mathrm{M}_\odot$. Additionally, we choose a split point that is favoured as a mass step location based on Hubble residuals obtained from the \textit{No-Split} \textsc{BayeSN} analysis, where parameters were not separated by host mass. Requiring that the step be located somewhere within the interquartile range (IQR) of host masses, the preferred step location (based on a maximum likelihood estimate -- see Appendix \ref{app:stepmle} for details) is $10^{10.286}\mathrm{M}_\odot$, with a fairly balanced low-:high-mass ratio of 75:82. Figure \ref{fig:steps} illustrates the step location we derive in this way, along with the Hubble residuals it is inferred from. There is a second strong peak in the likelihood at $10^{9.6815}\mathrm{M}_\odot$, which lies outside of the sample IQR (indicated with a grey band on the plot) and gives a highly unbalanced low-:high-mass ratio of 30:127. Although we do not feature it as one of our main choices, repeating our analysis with this split choice does not change our main conclusions.

\section{Results}
\label{sec:results}

\subsection{Primary Intrinsic SED Component}
\label{sec:w1}
\begin{figure}
    \centering
    \includegraphics[width=\linewidth, trim={0 0 2cm 0}, clip=true]{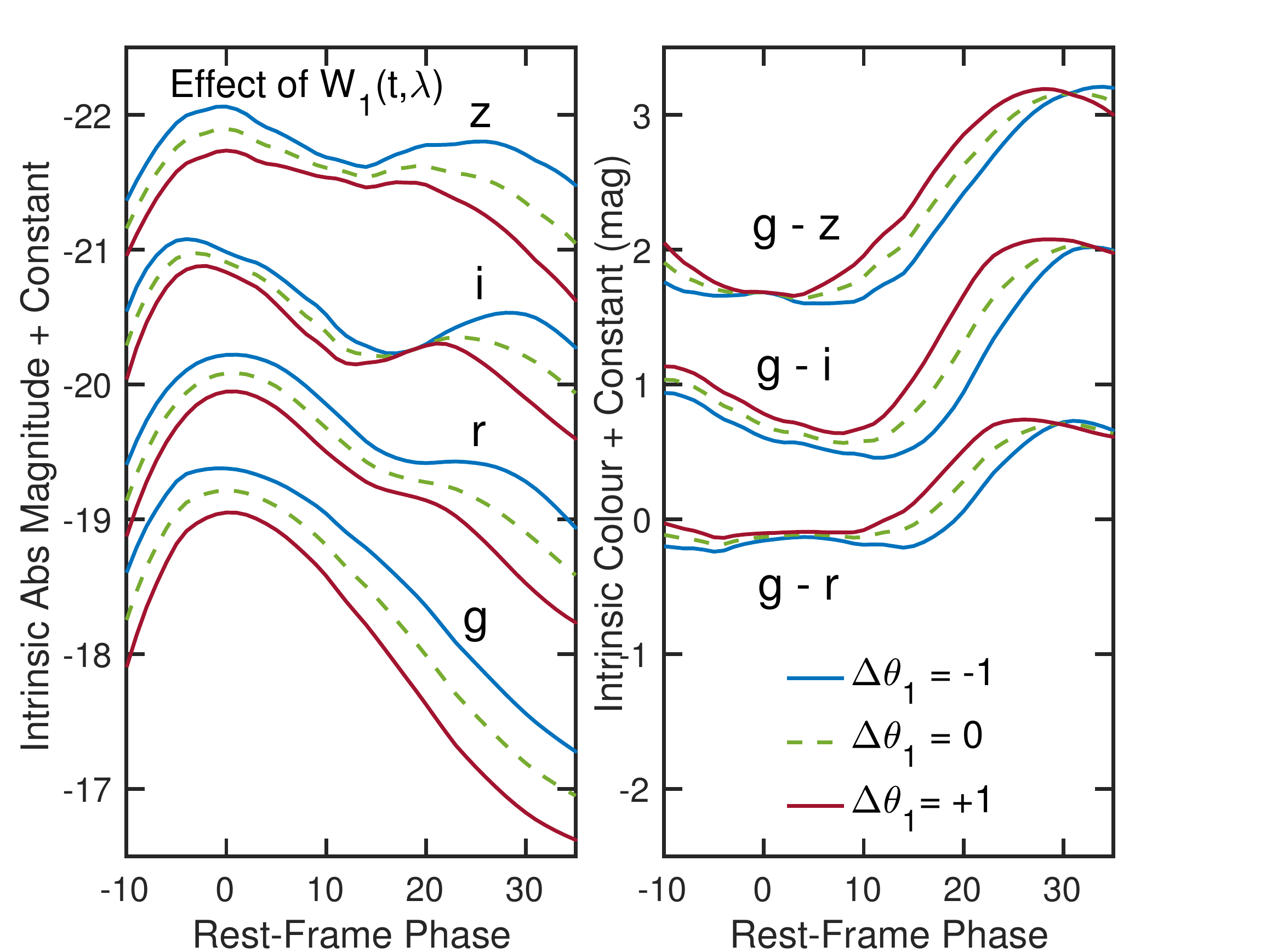}
    \caption{Synthesised rest-frame $griz$ light curves (left panel) and $g-r$, $g-i$, $g-z$, colour curves (right panel) from the \textsc{BayeSN} model trained on the Foundation supernovae, demonstrating the intrinsic effects of the first functional component, $W_1(t,\lambda_r)$. For each passband (or colour), the different coloured lines show the effect of varying $\theta_1$ between $-1$ (blue) and $1$ (red), modifying the extent to which $W_1(t,\lambda_r)$ is present. Vertical offsets are $(0,-1, -2.5,-3.5)$ for $(g,r,i,z)$ and $(0,1.5,2.5)$ for $(g-r,g-i,g-z)$.}
    \label{fig:w1_light_curves}
\end{figure}

The primary mode of intrinsic SED variation within the population is captured by our first functional principal component, $W_1$. The extent to which this is present for a particular supernova in the sample is controlled by that supernova's $\theta_1$ coefficient. To visualise the contribution of $W_1$ to the intrinsic SED, we synthesise rest-frame light curves in the Pan-STARRS-1 $griz$ passbands, varying $\theta_1$, and with all other effects ($A_V$, $\delta M$, $\epsilon$) set to zero. Figure \ref{fig:w1_light_curves} (left panel) shows this for the \textit{No-Split} model.

In the $g$-band, we observe a width-luminosity relation as is typical in optical SN Ia light curves \citep{phillips93}, with increasing $\theta_1$ corresponding to a faster decline. In the $r$ band, we see a weaker version of the same behaviour in the primary maximum. We also see an effect similar to that seen in \citetalias{mandel20}, with the brighter, slower-declining, supernovae having a delayed, more extended, and more prominent secondary bump/maximum -- something more noticeable in the $i$-band. An interesting and surprising difference from the behaviour seen in \citetalias{mandel20} is in the effect of $W_1$ on the $z$-band light curve. The 0.35--1.8~$\upmu$m \citetalias{mandel20} model was trained only on $BVriYJH$ data, meaning its $z$-band behaviour \citepalias[see][fig.\ 6]{mandel20} is an interpolation between the $i$ and $Y$ bands. Indeed, this is apparent, with the \citetalias{mandel20} $z$-band model looking quite similar to the $Y$-band model, with a secondary maximum similar in prominence to (or even brighter than) the first, and a temporary crossing over of the slow and fast declining light curves around the dip between NIR peaks. Here, however, where the model is trained on $z$-band data, we see somewhat different behaviour.  For the dimmest supernovae, the $z$-band light curve almost follows a plateau, declining slowly until around 20~d (approximately in line with the second $i$-band maximum), at which point it begins to fall off more rapidly. For the brighter supernovae, the initial decline appears faster, with a more definite minimum at around 15~d before a small rise to a secondary peak which falls off later than in the dimmer supernovae. Although this is closer in shape to what we would expect in the NIR, the minimum is still very shallow -- far more so than seen for the $i$-, $Y$-, or interpolated $z$-band in \citetalias{mandel20}, or in the $i$-band here.

We also synthesise colour curves, showing the effect of $W_1$ and $\theta_1$ on the intrinsic colour evolution of the supernovae in the sample. The right hand panel of Figure \ref{fig:w1_light_curves} shows rest-frame $g-r$, $g-i$ and $g-z$ intrinsic colour curves for a range of $\theta_1$. An interesting result is that we see very little sensitivity of the $g-r$ and $g-z$ colours around peak to varying $\theta_1$. This suggests that there is little correlation between light curve brightness or decline rate, and these intrinsic colours. For all of the chosen colours, sensitivity to $\theta_1$ increases considerably after peak, particularly around 10--25~days. Additionally, at around 30~days, each colour converges for all values of $\theta_1$, before (most noticeable in $g-z$) flipping over in behaviour (with the supernovae which were previously bluest becoming the reddest, and vice-versa).

\subsection{Residual SED Variation}
\label{sec:delM}

Within the \textsc{BayeSN} model, residual SED variation not explained by the primary intrinsic component, $W_1$, or dust extinction, is captured by two components: the population of `grey' $\delta M_s$ offsets, which are constant in time and wavelength; and the time- and wavelength-varying $\epsilon_s(t,\lambda_r)$ realisations, whose correlation structure is captured by a population covariance matrix, $\bm{\Sigma_\epsilon}$.

\begin{figure}
    \centering
   \includegraphics[width=\linewidth, trim={0 0 2cm 0}, clip=true]{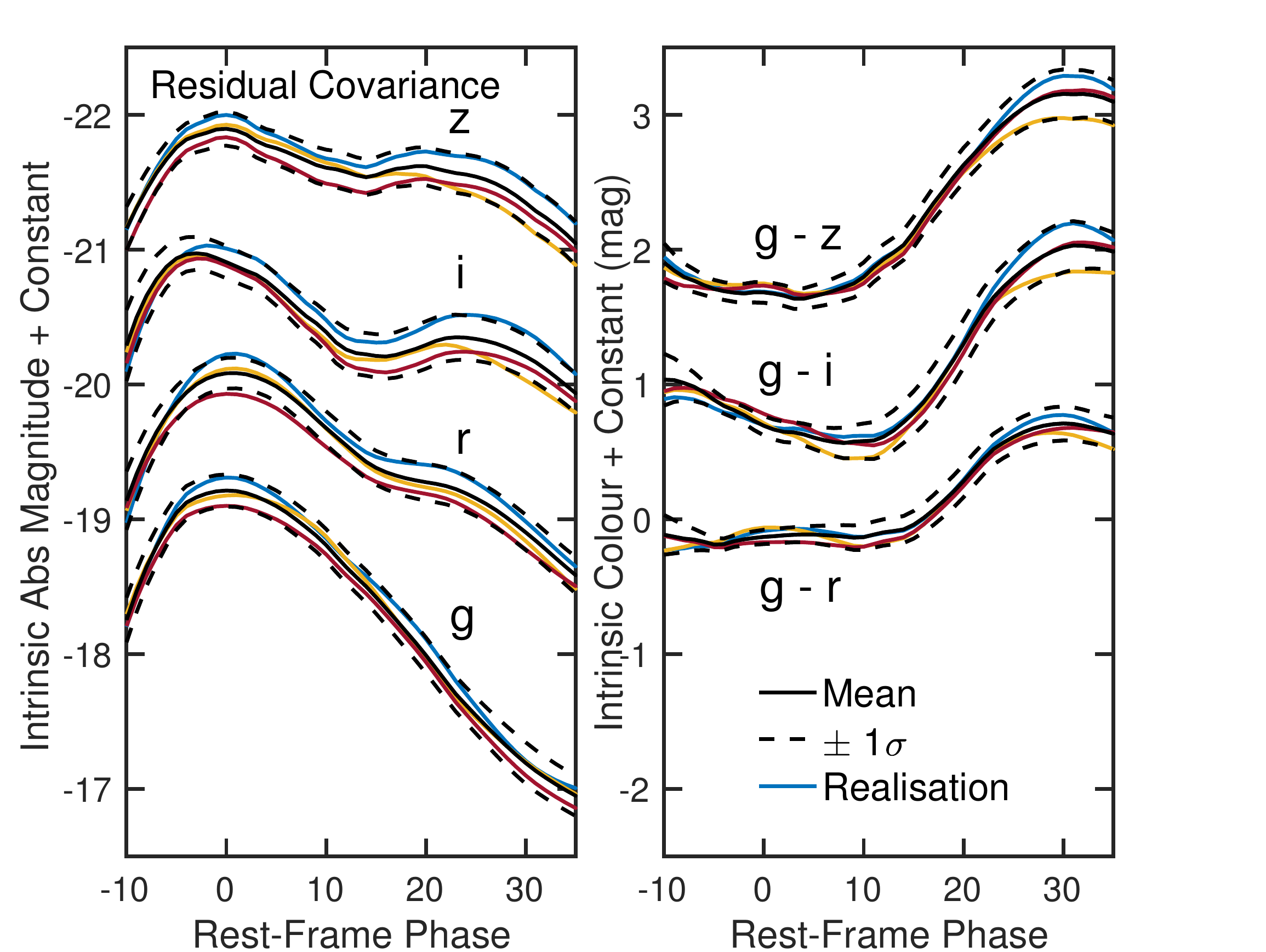}
    \caption{Same as Fig. \ref{fig:w1_light_curves}, but demonstrating the effect of the intrinsic residual components, $\eta_s(t,\lambda_r) = \delta M_s + \epsilon_s(t,\lambda_r)$. The solid black curves show the mean (equivalent to the dashed green lines in Figure \ref{fig:w1_light_curves}), with $\theta_1^s=\delta M_s=\epsilon_s(t,\lambda_r)=0$. The black dashed lines show a $\pm1\sigma$ envelope, and the coloured lines show the effects of the residual functions of three  example SNe Ia.}
    \label{fig:eta_light_curves}
\end{figure}

Taking the two components together, we can study the total level of residual SED scatter within the population. Figure \ref{fig:eta_light_curves} isolates the effect of varying $\eta_s(t,\lambda_r)=\delta M_s + \epsilon_s(t,\lambda_r)$ on rest-frame $griz$ light curves and $g-r$, $g-i$, and $g-z$ colour curves. The light curve residual variance is depicted in Figure \ref{fig:eta_light_curves} (left hand panel). The widths of the $\pm1\sigma$ envelopes, showing the residual variance within the population, are relatively constant near peak, generally showing only  $\sim 0.1$~mag of scatter, and increasing somewhat towards later phases. The colour curve residuals are, by definition, independent of the grey residuals $\delta M_s$ and their variance is depicted in the right hand panel. The residual colour scatter is generally small, but shows especially little scatter near peak (particularly in $g-r$ and $g-z$). Given the relative insensitivity of peak intrinsic $g-r$ and $g-z$ colour to the principal mode of intrinsic SED variation, it seems that these intrinsic colours at peak have small variance across this sample.

When considering the relation to host mass, we will focus on the time- and wavelength-independent residual component. This is summarised by the variance, $\sigma_0^2$, of the Gaussian population distribution of $\delta M_s$, and the shift, $\Delta M_0$, in the mean of this distribution between low- and high-host galaxy masses (see equation \ref{eq:delMpriorsplit}).

\begin{figure}
    \centering
    \includegraphics[width=\linewidth]{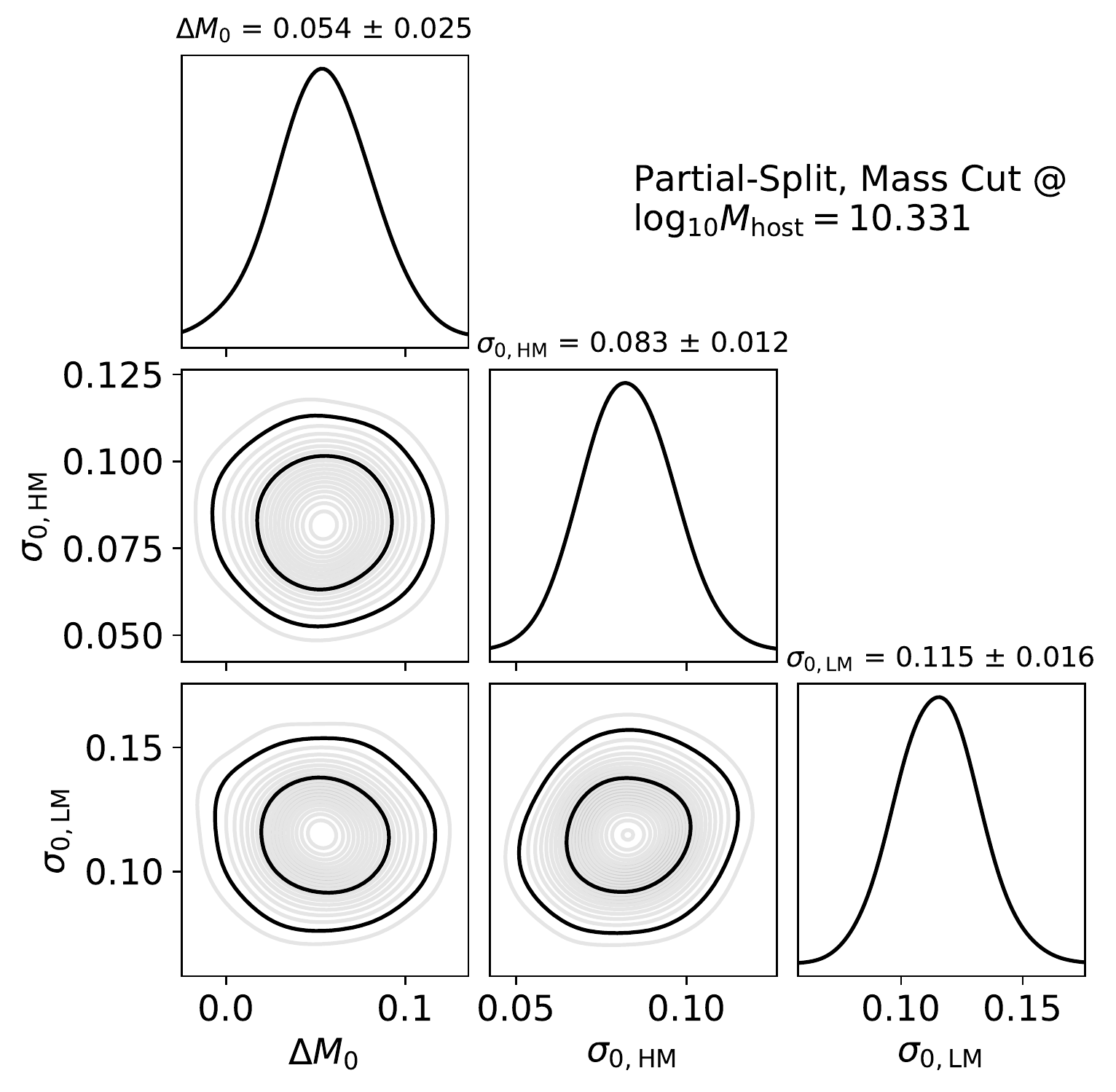}
    \caption{Posterior distribution of the low- vs. high-host-mass magnitude offset, $\Delta M_0$, and the low- and high-host-mass values of $\sigma_0$ -- the population standard deviation of the `grey' component of the intrinsic SED. The median host mass split was used here. Darker contours indicate the regions which contain 68 and 95 per cent of the posterior probability.}
    \label{fig:M0_sigma0_posterior}
\end{figure}

For the \textit{No-Split} analysis where we ignore host galaxy mass, we find that $\sigma_0=0.10\pm0.01$. For the \textit{Partial-Split} analyses where we split $\sigma_0$, $\tau_A$ and $R_V$ by host mass, the outcome depends somewhat on the exact choice of split point, but is qualitatively similar for all of the choices tried here (see Table \ref{tab:masscuts} for a summary of these). For all choices, positive values of $\Delta M_0$ are at least somewhat favoured -- suggesting lower average brightness in low-mass galaxies. For the median host mass split, we estimate $\Delta M_0=0.054\pm0.025$~mag, around $2\sigma$ greater than zero. The difference of $\Delta M_0$ from zero is most significant ($3\sigma$) when the sample is split at the MLE mass split preferred as a step location by the Hubble residuals. For all choices of mass split, the posterior estimate of $\Delta M_0$ is consistent with the step size seen at that mass in Hubble residuals computed from the \textit{No-Split} analysis. For the split choice of $10^{10}\mathrm{M}_\odot$, our inferred $\Delta M_0=0.042\pm0.028$~mag is consistent with, but somewhat less significant than, the mass step of $0.060\pm0.024$~mag found by \citet{jones19} for the Foundation sample. The final two columns of Table \ref{tab:masscuts} summarise the different \textit{Partial-Split} $\Delta M_0$ estimates and \textit{No-Split} step sizes side-by-side.

For all choices of split point (see Table \ref{tab:global_params}, which also repeats the $\Delta M_0$ values), our inferences of $\sigma_0$ are fairly consistent between low- and high-mass hosts. For the median and MLE split points, the posterior mean estimates of $\sigma_0$ are marginally higher in low-mass host galaxies. For the median mass split point, the joint posterior distribution of $\Delta M_0$ and the high- and low-mass values of $\sigma_0$ is shown in Figure \ref{fig:M0_sigma0_posterior}. Our inferences of $\Delta M_0$ and $\sigma_0$ under the \textit{No-Split} and \textit{Partial-Split} models are consistent whether a single $R_V$ or a population distribution is assumed in the full sample or within each subsample.

\begin{table*}
    \centering
    \caption{Posterior means and standard deviations of population parameters for different choices of host mass split.}
    \label{tab:global_params}
    \begin{tabular}{l l c c c c c c c}\toprule
        Model & Split Point & $\Delta M_0$ & \multicolumn{2}{c}{$\sigma_0$} & \multicolumn{2}{c}{$\tau_A$} & \multicolumn{2}{c}{$R_V$}\\ \cmidrule(lr){4-5}\cmidrule(lr){6-7}\cmidrule(l){8-9}
        &&& LM & HM & LM & HM & LM & HM\\\midrule
        \textit{No-Split} & - & - & \multicolumn{2}{c}{$0.10\pm0.01$} & \multicolumn{2}{c}{$0.19\pm0.02$} & \multicolumn{2}{c}{$2.61\pm0.21$}\\
        \textit{Partial-Split} & $10^{10}\mathrm{M}_\odot$ & $0.042\pm0.028$ & $0.10\pm0.02$ & $0.10\pm0.01$ & $0.17\pm0.03$ & $0.22\pm0.03$ & $3.33\pm0.50$ & $2.67\pm0.21$\\
        & Median & $0.054\pm0.025$ & $0.12\pm0.02$ & $0.08\pm0.01$ & $0.19\pm0.03$ & $0.21\pm0.03$ & $2.84\pm0.31$ & $2.58\pm0.23$\\
        & MLE & $0.072\pm0.024$ & $0.11\pm0.02$ & $0.09\pm0.01$ & $0.18\pm0.03$ & $0.22\pm0.03$ & $2.86\pm0.33$ & $2.66\pm0.24$\\
        \textit{Full-Split} & $10^{10}\mathrm{M}_\odot$ & - & $0.15\pm0.02$ & $0.10\pm0.01$ & $0.16\pm0.04$ & $0.22\pm0.03$ & $3.07\pm0.64$ & $2.43\pm0.22$\\
        & Median & - & $0.10\pm0.02$ & $0.09\pm0.01$ & $0.21\pm0.03$ & $0.22\pm0.03$ & $2.83\pm0.35$ & $2.46\pm0.25$\\
        & MLE & - & $0.10\pm0.02$ & $0.09\pm0.01$ & $0.19\pm0.03$ & $0.22\pm0.03$ & $2.86\pm0.36$ & $2.54\pm0.25$\\
        \bottomrule
    \end{tabular}
\end{table*}

For the \textit{Full-Split} configuration, where entirely separate models are trained either side of the split point (see Section \ref{sec:fullsplit}), our $\sigma_0$ estimates (see Table \ref{tab:global_params}) are generally consistent with their counterparts from the \textit{Partial-Split} analyses. For the $10^{10}\mathrm{M}_\odot$ mass split, we estimate $\sigma_0$ in low-mass hosts to be around $2.2\sigma$ higher than in high-mass hosts. For the other splits, the difference is insignificant. There is no explicit $\Delta M_0$ included in the \textit{Full-Split} analysis, since an effective shift in $M_0$ is entirely exchangeable with a shift in the mean intrinsic SED $W_0(t,\lambda_r)$ (see Equation \ref{eq:SEDmodel}). Since, in \textit{Full-Split} mode, the models are trained completely independently on each subsample, we fix each $M_0 = -19.5$ to its default value, and allow each $W_0(t,\lambda_r)$ to absorb any relative magnitude offset.

\subsection{Population Dust Properties}
\label{sec:dustresults}
\subsubsection{Global Dust Law}
In our baseline model, the population host galaxy dust properties are characterised by the global $R_V$ value, and a population mean extinction value, $\tau_A$. When host galaxy stellar mass is ignored, and a single value for each of these parameters is assumed, we find $R_V=2.61\pm0.21$ and $\tau_A=0.19\pm0.02$ -- see the upper left panel of Figure \ref{fig:RV_tauA_posteriors} for a plot of the joint posterior. This $R_V$ value is consistent with the global value of $R_V=2.89\pm0.20$ found by \citetalias{mandel20} for the low-$z$ optical--NIR supernova sample from \citet{avelino19}. The mean extinction value we infer for the Foundation sample is somewhat lower than was found for the sample analysed by \citetalias{mandel20}.

\begin{figure}
    \centering
    \includegraphics[width=\linewidth]{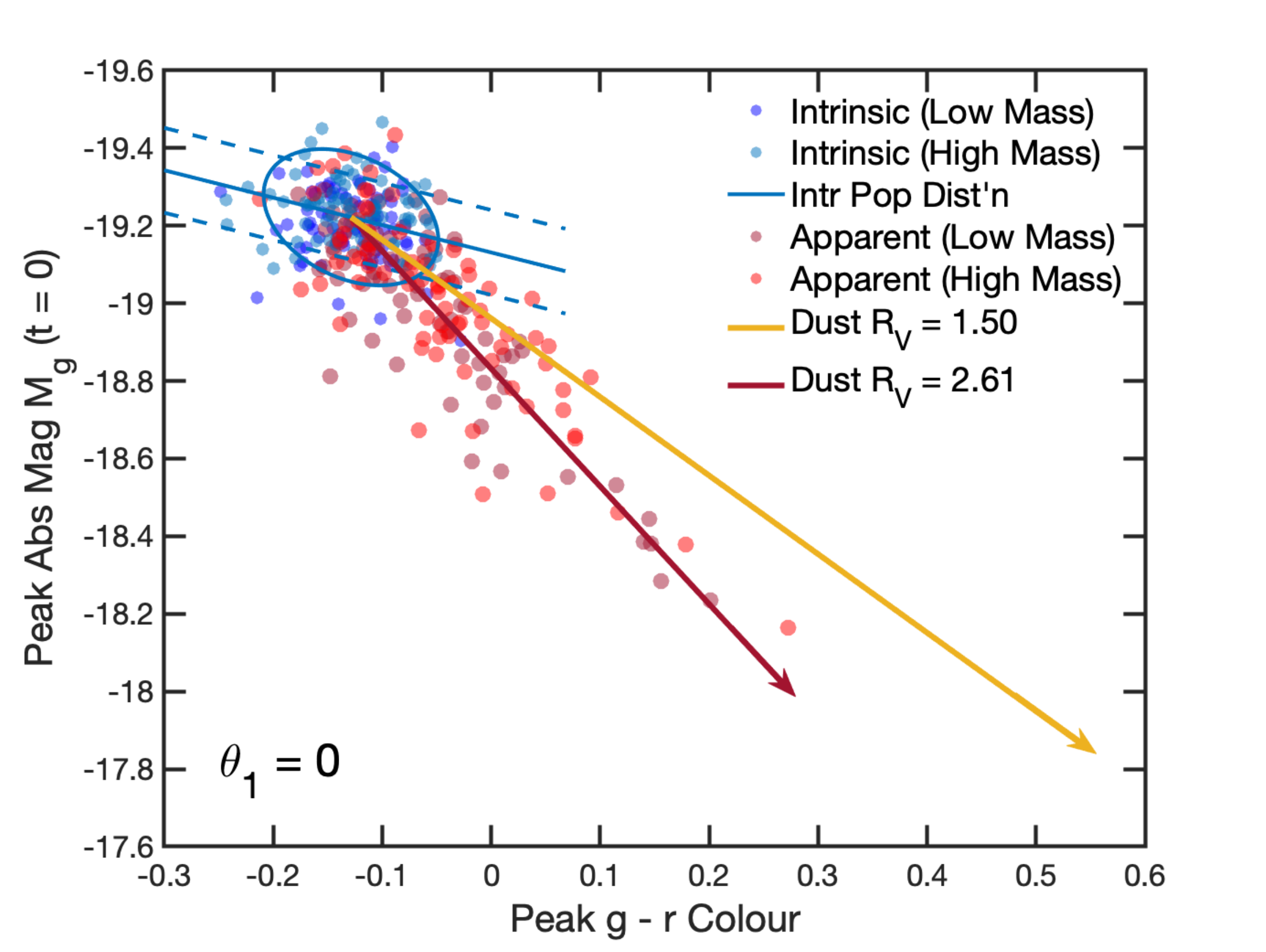}
    \caption{Posterior estimates of peak ($t =0$) $g$-band absolute magnitude vs. $(g-r)$ colour for the unsplit  sample, obtained from the \textit{No-Split} model. Each SN is represented by a red and blue point. Red points indicate extinguished absolute magnitudes and apparent colours. Blue points indicate the inferred intrinsic absolute magnitudes and intrinsic colours. Bright (dark) shaded points indicate SN with high- (low-) mass hosts (where the sample is split at the median host mass of $10^{10.331}\mathrm{M}_\odot$). The blue ellipse depicts the $1\sigma$ locus of the inferred intrinsic population distribution, while the blue solid and dashed lines indicate the inferred mean intrinsic colour--luminosity relation, and intrinsic absolute magnitude scatter around it. The magnitudes and colours have been corrected for the modelled intrinsic SED shape dependence (see Fig. \ref{fig:w1_light_curves}) to $\theta_1 = 0$ using Eq. \ref{eq:SEDmodel}. The reddening--extinction vectors for $A_V = 1$ for two example $R_V$ values ($R_V=1.5$; and our posterior mean $R_V=2.61$ from the full $griz$ light curves) under the \citet{fitzpatrick99} law are shown by the yellow and red arrows emanating from the centre of the intrinsic distribution. Note that we do not expect that all SNe should ``map back'' along the dust vector to the exact centre; rather, the intrinsic properties have a distribution. The average colour--luminosity trend of the reddened SNe Ia agrees with the $R_V = 2.61$ dust law and disfavours $R_V = 1.5$ in both mass bins.}
    \label{fig:colour_dust}
\end{figure}

Our model constrains the dust properties by fitting time-dependent SEDs to the full $griz$ light curves of the SN Ia sample, thereby leveraging the maximal information in the data. However, it is still useful to derive insights from low-dimensional visualisations of the full inference. Figure \ref{fig:colour_dust} shows estimated extinguished rest-frame $g$-band absolute magnitudes and apparent $g-r$ colours at time of maximum for the unsplit sample, obtained from training the \textit{No-Split} model. The SNe Ia in both low- and high-mass hosts exhibit an apparent colour--luminosity trend consistent with the expected reddening vector for our posterior mean $R_V=2.61$ (red arrow). The more reddened SNe Ia are clearly inconsistent on average with a lower $R_V=1.5$ (yellow arrow).  In particular, for a given red apparent colour, the SNe Ia are on average dimmer than would be expected with an $R_V = 1.5$ dust law.

Figure \ref{fig:colour_dust} also shows how our model (similarly to \citealt{mandel17}) decomposes the apparent colour-luminosity distribution into two distinct physical effects: dust reddening-extinction (red arrow, with a slope determined by $R_V$) and an intrinsic colour-luminosity trend (blue line with a shallower slope). This is in contrast to the conventional Tripp formula (Eq. \ref{eqn:tripp}), which models the apparent colour-magnitude relation with a single linear slope $\beta$. The blue contours indicate the finite extent of the inferred intrinsic colour-luminosity distribution. Proper accounting of this intrinsic colour distribution is especially critical for reliable inference of the dust law distribution (c.f.\ Section \ref{sec:pop_RV_results}).

\begin{figure}
    \centering
    \includegraphics[width=\linewidth]{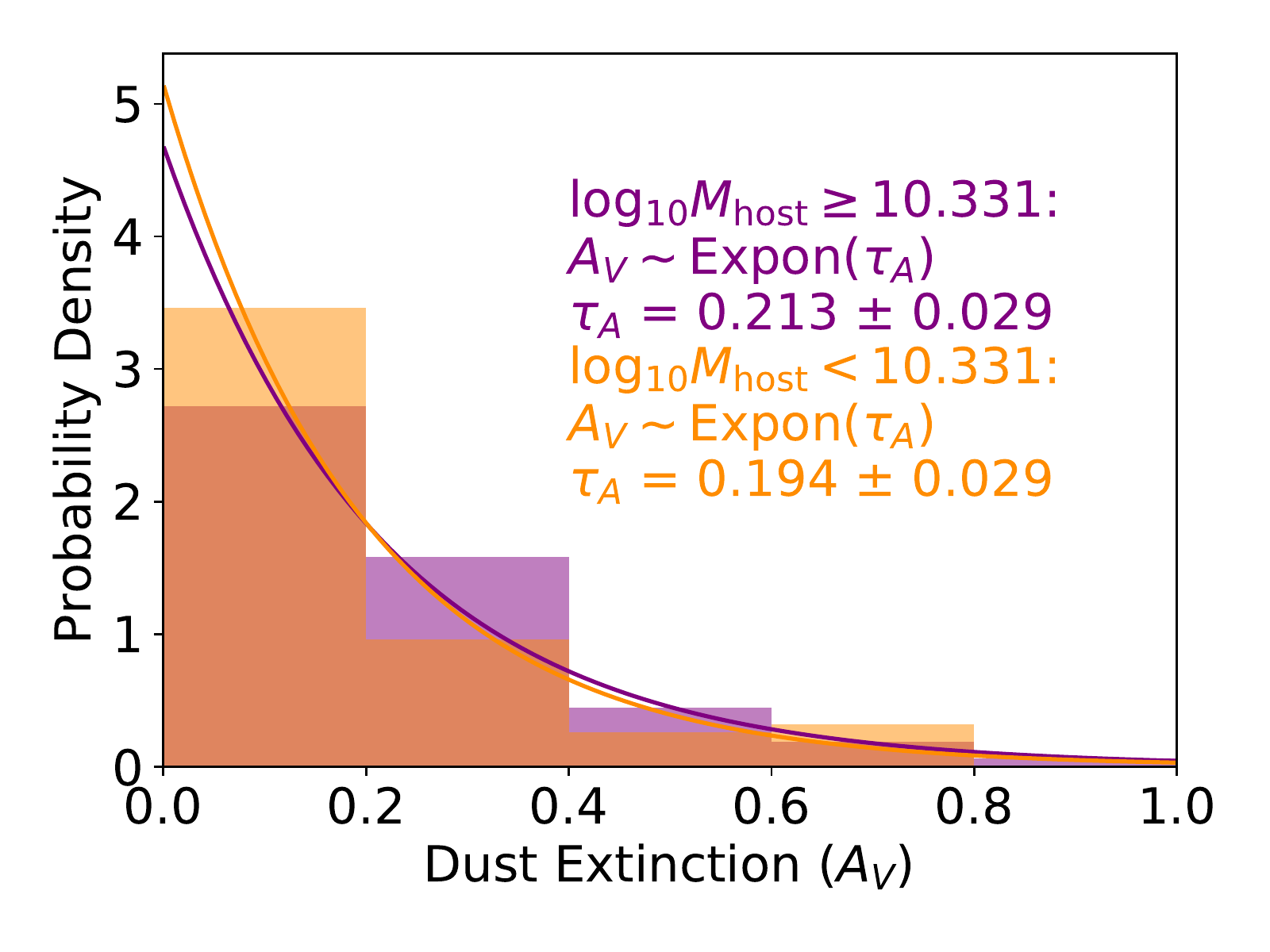}%
    \caption{Population distributions of the inferred dust extinction values of the Foundation sample when split at the median host mass (under the \textit{Partial-Split} model). The solid lines plotted over the histograms show the exponential population distributions parametrized by the inferred mean extinction values for the low- and high- host mass populations.}
    \label{fig:AV_distribution}
\end{figure}

\begin{figure*}
    \centering
    \includegraphics[width=0.5\linewidth]{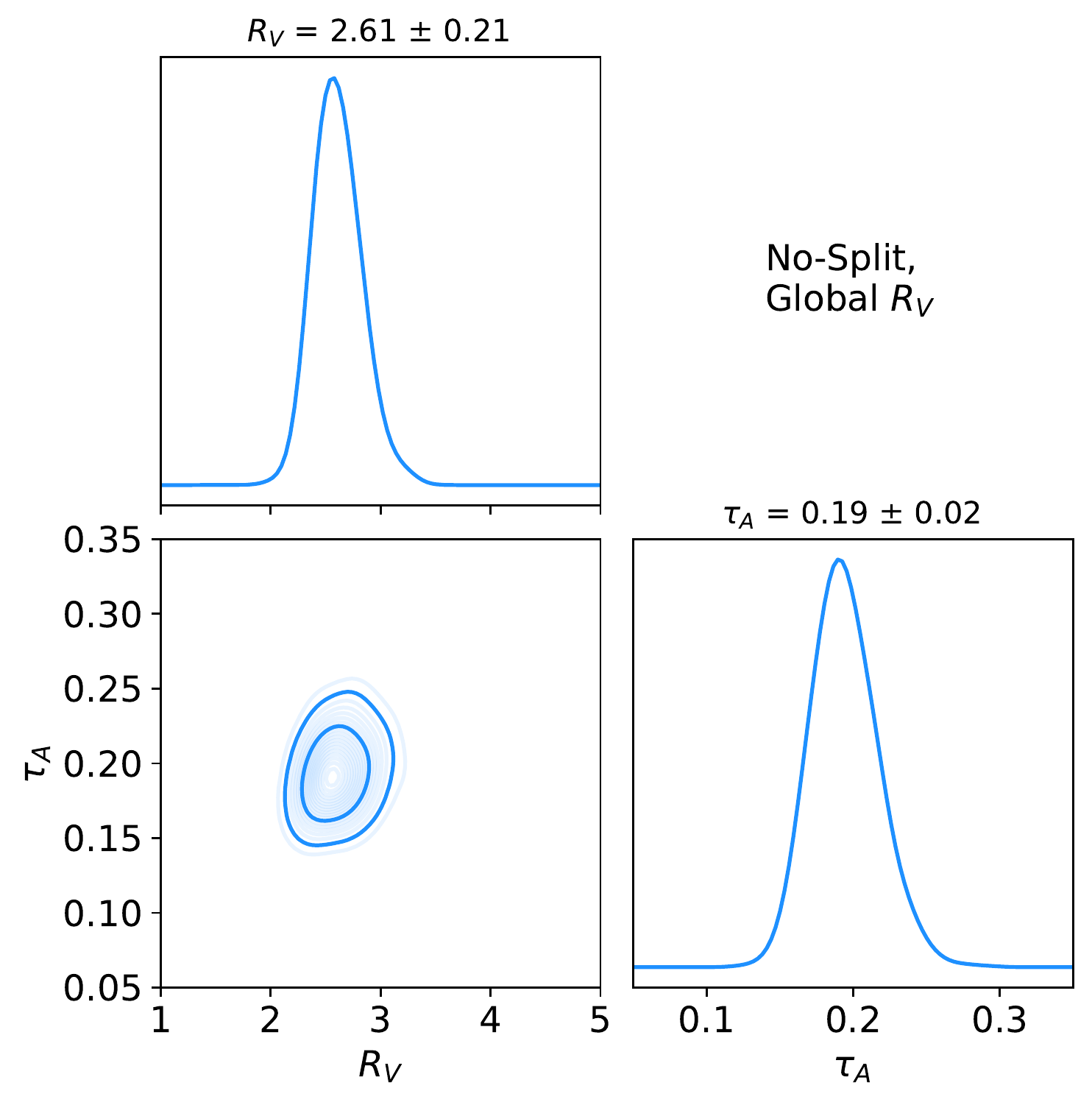}%
    \includegraphics[width=0.5\linewidth]{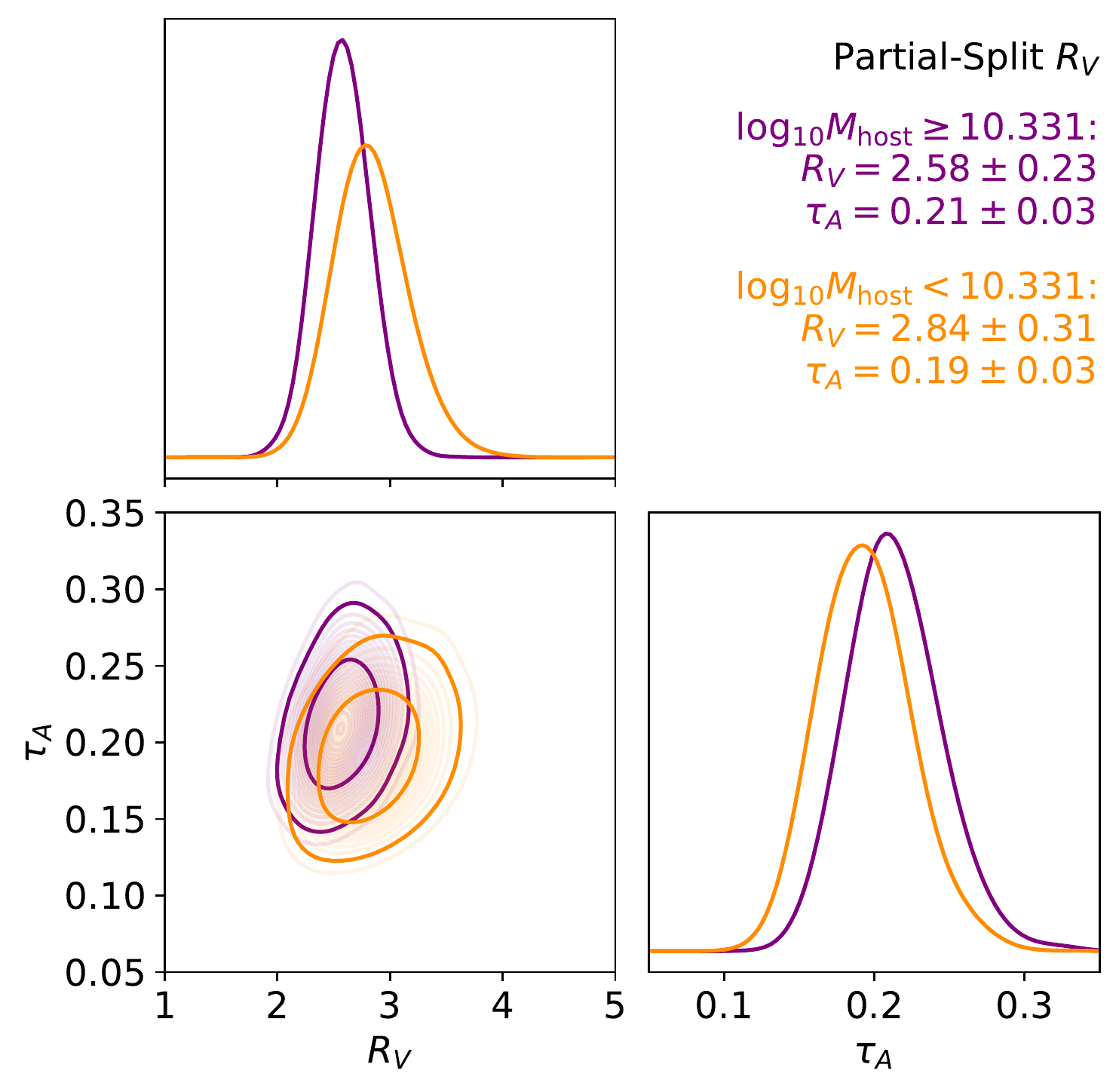}
    \includegraphics[width=0.5\linewidth]{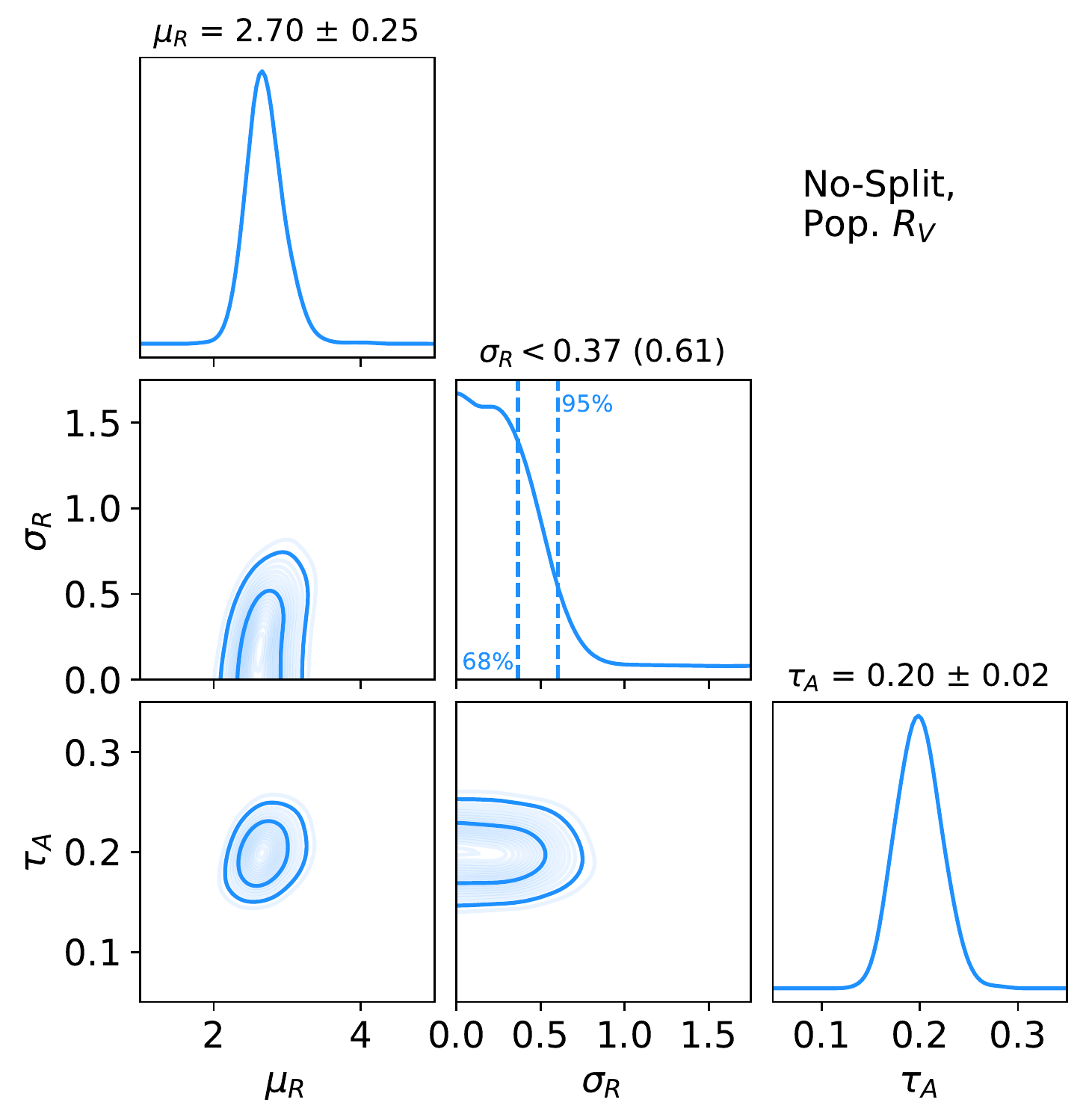}%
    \includegraphics[width=0.5\linewidth]{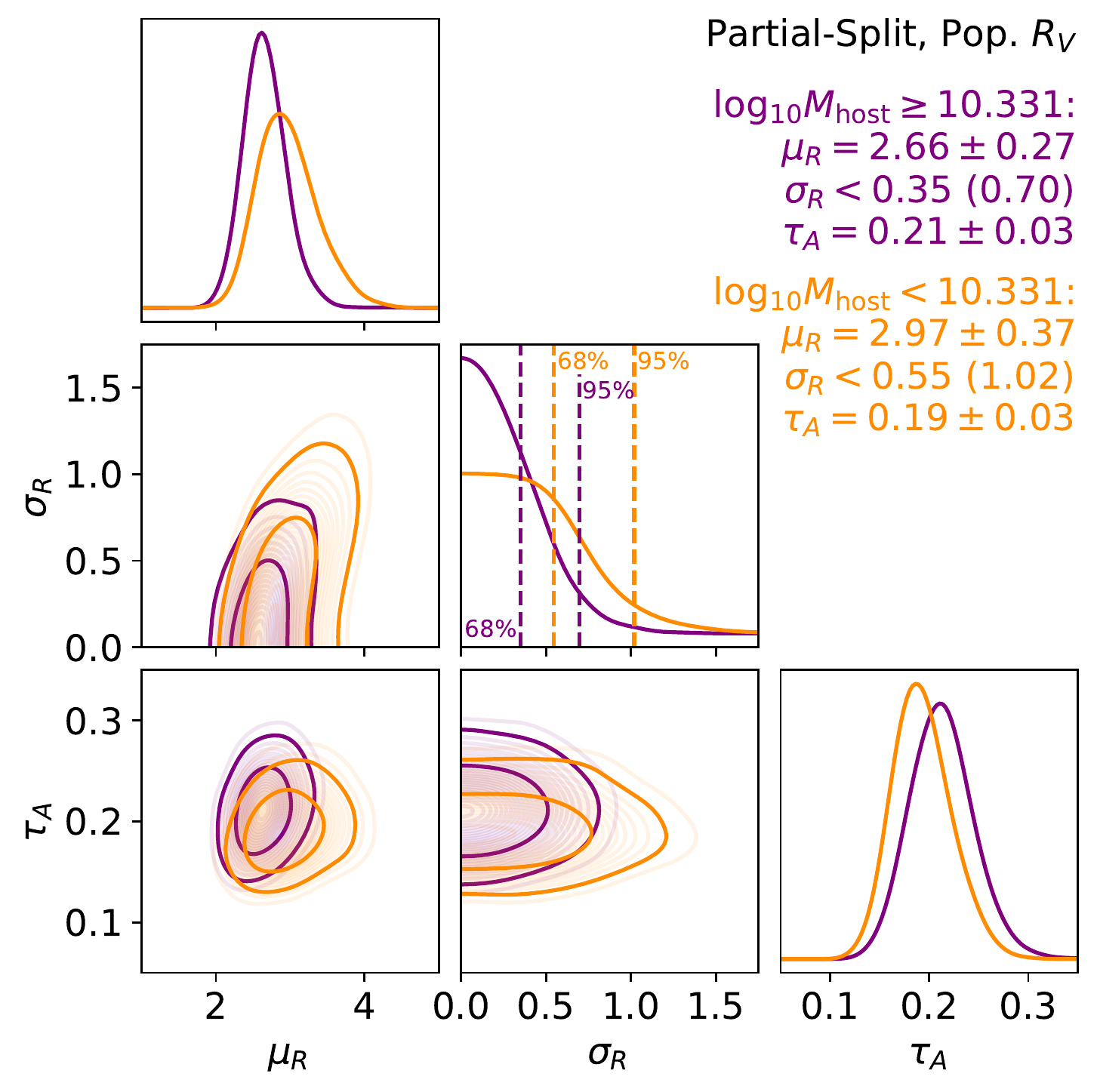}
    \caption{(top row) Posterior distributions of $R_V$ and $\tau_A$ from \textsc{BayeSN} training runs on Foundation DR1 carried out without (left panel) and with (right panel) these parameters split by host galaxy stellar mass at the median. (bottom row) Posterior distributions of $\mu_R$, $\sigma_R$ and $\tau_A$ from the variant of our analysis where we allow a population distribution of $R_V$. Parameter summaries are posterior mean and standard deviation, except for $\sigma_R$, where 68~(95)th percentiles are quoted. These are also indicated on the relevant marginal plots as dashed lines.}
    \label{fig:RV_tauA_posteriors}
\end{figure*}

When we retrain our model with separate $R_V$ and $\tau_A$ values permitted for low- and high-mass hosts, we see little evidence that the extinction distribution is different between the two subsamples. Our estimates of $\tau_A$ are consistent to within $\lesssim1.2\sigma$ between low- and high-mass hosts -- Table \ref{tab:global_params} lists the inferred values for different choices of mass cut. Figure \ref{fig:AV_distribution} shows the exponential extinction population distributions implied by the inferred $\tau_A$ values when the median host mass split is used, along with histograms of the individual posterior mean $A_V$ estimates for the two samples.

We see little evidence for significantly different $R_V$ values either side of the split point. For all choices of split point, the low- and high-mass $R_V$ values are consistent with one another, the global value found when training without a host mass split, and the value inferred from the completely independent \citet{avelino19} sample by \citetalias{mandel20}. Taking the median split point, for example, we find $R_V=2.84\pm0.31$ for the low-mass hosts, and $R_V=2.58\pm0.23$ for the high-mass hosts. This consistency of $R_V$ across the mass split aligns with expectation from the apparent colours and extinguished absolute magnitudes estimated in our \textit{No-Split} analysis (see Fig.\ \ref{fig:colour_dust}). It also aligns with the results of \citet[fig.\ 4]{jones18} who found consistency in the apparent colour--luminosity relation of supernovae in low- and high-mass host galaxies on a sample of 273 SNe Ia (including a subset of Foundation DR1). The joint posterior of $R_V$ and $\tau_A$ for the median mass split is shown in the upper right hand panel of Fig. \ref{fig:RV_tauA_posteriors}. The posterior means and standard deviations for other choices of mass split are listed in Table \ref{tab:global_params}, along with the other global parameters. For the $10^{10}\mathrm{M}_\odot$ split point, which gives less balanced high- and low-mass bins, the ability to constrain $R_V$ well for the low-mass subset is limited by sample size, with a fairly high posterior standard deviation for $R_V$ in this run. Here, we estimate $R_V=3.33\pm0.50$ for hosts with $M_*<10^{10}\mathrm{M}_\odot$, and $R_V=2.67\pm0.21$ for those with $M_*\geq10^{10}\mathrm{M}_\odot$. The estimated difference $\Delta R_V\approx 0.66 \pm 0.54$ is statistically consistent with zero within $\approx1.2\sigma$, but may be suggestive of a larger value. The considerable $R_V$ uncertainty in the low-mass subsample for this more unbalanced split choice likely arises from the dual effect of having fewer SNe generally, and from having fewer of the more extinguished SNe (see Fig. \ref{fig:latent_vs_mass}) that have the most leverage to constrain $R_V$. Nevertheless, for all of the \textit{Partial-Split} runs, $R_V \lesssim 2.2$ tends to be strongly disfavoured on both sides of the mass split, with less than 5 per cent posterior probability in all cases.

From the \textit{Full-Split} analysis, our conclusions about $\tau_A$ and $R_V$ do not change significantly, with all values above and below the various split points being consistent with their \textit{Partial-Split} equivalents. The consistency of $\tau_A$ between high- and low-mass hosts remains within $\lesssim1.2\sigma$. For all three split points, we estimate $R_V$ to be consistent within $1\sigma$ between high- and low-mass hosts, with $R_V\lesssim2$ disfavoured with at least 95 per cent probability in all cases. The consistency of our dust inferences between the \textit{Full-Split} and \textit{Partial-Split} models indicates that our results are robust to how the intrinsic components are estimated.

\subsubsection{Population Distribution of Dust Law $R_V$}
\label{sec:pop_RV_results}
In Figure \ref{fig:colour_dust}, the blue contours indicate the finite extent of the inferred intrinsic  distribution under our global dust law model. This is key to the interpretation of the apparent colour-luminosity distribution and the inference of dust laws. If one naively assumes that the apparent properties of every reddened SN should ``map'' back to the same intrinsic value (e.g.\ the mean of the distribution) under its dust correction, then one might conclude that a wide range of $R_V$ is present. However, this assumption would logically contradict the finding of a intrinsic population distribution with finite width: if the intrinsic covariance is non-zero, then not every reddened SN should simultaneously map back to the same intrinsic point (with zero variance). Instead, our statistical model simultaneously infers and accounts for the nonzero extent of the intrinsic distribution along with the distribution of dust laws, over the full $griz$ light curves. Using this probabilistic approach, we find in this section that the data are consistent with a narrower distribution of dust laws.

Under this mode of analysis, an individual $R_V^s$ is permitted for every supernova, with these drawn from a population distribution described by a Gaussian with mean $\mu_R$, and standard deviation $\sigma_R$, truncated such that $R_V^s\geq0.5$ (see Equation \ref{eq:RVsprior}). This dust law distribution is included alongside our model's usual treatment of the intrinsic colour distribution and dust extinction distribution (see Section \ref{sec:nosplit} for a summary of the technical details), with all aspects being inferred and accounted for simultaneously. For the \textit{No-Split}, and all of the \textit{Partial-Split} configurations where we try this model extension, our inferences of $\mu_R$ are consistent with the global values we found previously. Our $\sigma_R$ posteriors unanimously prefer small values, with the strength at which large values are disfavoured depending on the exact model configurations. Table \ref{tab:pop_RV_params} lists parameter summaries for $\mu_R$ and $\sigma_R$ for the \textit{No-Split} and \textit{Partial-Split} models.

\begin{table}
    \centering
    \begin{threeparttable}
        \caption{Inferences of $R_V$ population parameters.}
        \label{tab:pop_RV_params}
        \begin{tabular}{l c c c c}\toprule
            Split Point\tnote{a} & \multicolumn{2}{c}{$\mu_R$\tnote{b}} & \multicolumn{2}{c}{$\sigma_R$ [68\% (95\%) u.b.]\tnote{c}}\\ \cmidrule(lr){2-3}\cmidrule(l){4-5}
            & LM & HM & LM & HM\\\midrule
            - &\multicolumn{2}{c}{$2.70\pm0.25$} & \multicolumn{2}{c}{0.37 (0.61)}\\
            $10^{10}\mathrm{M}_\odot$ & $3.49\pm0.61$ & $2.72\pm0.23$ & 0.89 (1.87) & 0.32 (0.62)\\
            Median & $2.97\pm0.37$ & $2.66\pm0.27$ & 0.55 (1.02) & 0.35 (0.70)\\
            MLE & $3.03\pm0.43$ & $2.74\pm0.26$ & 0.60 (1.22) & 0.34 (0.64)\\
            \bottomrule
        \end{tabular}
        \begin{tablenotes}
            \item [a] Host mass split point with which the \textit{Partial-Split} model was run. The first line results are from the \textit{No-Split} model.
            \item [b] Posterior mean and std. dev. of the population $R_V$ mean parameter.
            \item [c] 68 (95)\% posterior upper bounds for the population $R_V$ std. dev. parameter.
        \end{tablenotes}
    \end{threeparttable}
\end{table}

The recent work of \citet{brout20} also investigated the possibility that dust in low- and high-mass SN Ia host galaxies has different properties. Similarly to the present work, they modelled host galaxy reddening as following separate exponential population distributions above and below their chosen mass split point of $10^{10}\mathrm{M}_\odot$, with separate Gaussian population distributions of $R_V$ on either side of this split. For their low-redshift sample (including Foundation DR1), they estimate that supernovae in massive hosts have a higher mean level of dust reddening. Our posterior mean $\tau_A$ values for the same choice of split point are suggestive of qualitatively similar behaviour -- something borne out more prominently when looking at the individual reddening values for the supernovae in our sample (see Section \ref{sec:colours} and Figure \ref{fig:latent_vs_mass}) -- although at $\lesssim1.2\sigma$, the difference we find is not statistically significant. For $R_V$, \citet{brout20} reported peak values of $2.75\pm0.35$ and $1.50\pm0.25$ for low- and high-mass hosts, respectively -- a difference of $2.9\sigma$ -- with fairly large population standard deviations of $1.3\pm0.2$ for both subsamples. When not splitting by host galaxy mass, they estimate a mean $R_V$ of $2.0\pm0.2$, and a similarly wide population distribution with $\sigma_R=1.4\pm0.2$.

Under our \textit{No-Split} model, we find an $R_V^s$ population distribution with mean parameter $\mu_R=2.70\pm0.25$. For this configuration, a wide population distribution of $R_V^s$ is strongly disfavoured, with 68 (95) per cent of posterior probability falling at $\sigma_R<0.37~(0.61)$. This disfavours the lower peak value of $\mu_R=2.0\pm0.2$, and wide distribution with $\sigma_R=1.4\pm0.2$ reported by \citet{brout20}. The bottom left panel of Figure \ref{fig:RV_tauA_posteriors} shows the joint posterior of $\mu_R$, $\sigma_R$ and $\tau_A$ (which is insensitive to the adoption of a population distribution of $R_V^s$) for the \textit{No-Split} model.

Under the \textit{Partial-Split} results, when a split point of $10^{10}\mathrm{M}_\odot$ is assumed in line with \citet{brout20}, we find $\mu_R=3.49\pm0.61$ and $\mu_R=2.72\pm0.23$ for supernovae in low- and high-mass hosts, respectively. For the more massive hosts, we find that $\sigma_R<0.32~(0.62)$ with 68 (95) per cent posterior probability. For low-mass hosts, our constraining power is weakened (especially for this choice of split), with the 68 and 95th percentiles of our posterior falling at $\sigma_R=0.89$ and $\sigma_R=1.89$, respectively. The joint $\mu_R$, $\sigma_R$ posterior for low-mass hosts is consistent with its more tightly constrained high-mass counterpart. For host galaxies with $M_*\geq10^{10}\mathrm{M}_\odot$, these results disfavour the wide population distribution of $R_V^s$ with $\sigma_R=1.3\pm0.2$ estimated by \citet{brout20}. For host galaxies with $M_*<10^{10}\mathrm{M}_\odot$, such a diffuse distribution cannot be strongly ruled out, but it is less probable than a distribution with low $\sigma_R$. 

As in the global dust law configuration, our ability to constrain the low-mass host dust behaviour is improved when the sample is split at the median host mass. In this case, the joint posterior of $\mu_R$ and $\sigma_R$ for low-mass hosts is more compressed than for the $10^{10}\mathrm{M}_\odot$ choice of mass split. This brings it much closer to its high-mass companion. Here, we find $\mu_R=2.97\pm0.37$ for low-mass hosts, and $\mu_R=2.66\pm0.27$ for high-mass hosts. For the low-mass subsample, 68 (95) per cent of our posterior probability lies at $\sigma_R<0.55~(1.02)$, whilst for the high-mass subsample we find that $\sigma_R<0.35~(0.70)$ with 68 (95) per cent probability. The bottom right panel of Figure \ref{fig:RV_tauA_posteriors} illustrates the joint posterior of $\mu_R$, $\sigma_R$ and $\tau_A$ for the \textit{Partial-Split} model with the median host mass as split point. For a brief discussion of the prior sensitivity of $\sigma_R$, see Appendix \ref{app:sigmaRprior}.

For all choices of mass split, we obtain $\mu_R$ values consistent with the values of $R_V$ obtained in our global dust law analyses. The $\mu_R$'s across the mass split are consistent within $\lesssim1.2\sigma$. However, the smallest $\Delta M_0=0.042\pm0.028$ $(1.5\sigma)$ occurs for the $10^{10}\mathrm{M}_\odot$ mass split, where we estimate $\Delta \mu_R=0.77\pm0.65$, larger than for the median or MLE mass split. Although this estimated difference in $\mu_R$ is consistent within the statistical uncertainties with zero, it may be suggestive of some possible trade-off between $\Delta \mu_R$ and the size of the mass step, which is qualitatively similar to the trade-off inferred by \citet{brout20}. However, we caution that, for this mass split, our constraint on $\Delta\mu_R$ is more uncertain than for other splits, due to the paucity of more reddened SNe in low-mass hosts. As in the global dust law analysis, $\mu_R\lesssim2.2$ is disfavoured on both sides of the mass split in all cases. This disfavours the low $\mu_R=1.50\pm0.25$ estimated by \citet{brout20} for high-mass host galaxies. Our $\tau_A$ inferences   are insensitive to the inclusion of a population distribution of $R_V^s$.

\subsection{Distribution of SED Shape Parameters}
\label{sec:theta}

It is interesting to investigate how the SED shape parameter, $\theta_1$, is distributed for the low and high mass subsamples. It is only meaningful to do this for the \textit{No-Split} or \textit{Partial-Split} models, where the $W_1$ component modulated by $\theta_1$ is common to both sides of the mass split. We will focus on the \textit{Partial-Split} results here, although those from the \textit{No-Split} model are almost identical. 

\begin{figure}
    \centering
    \includegraphics[width=\linewidth]{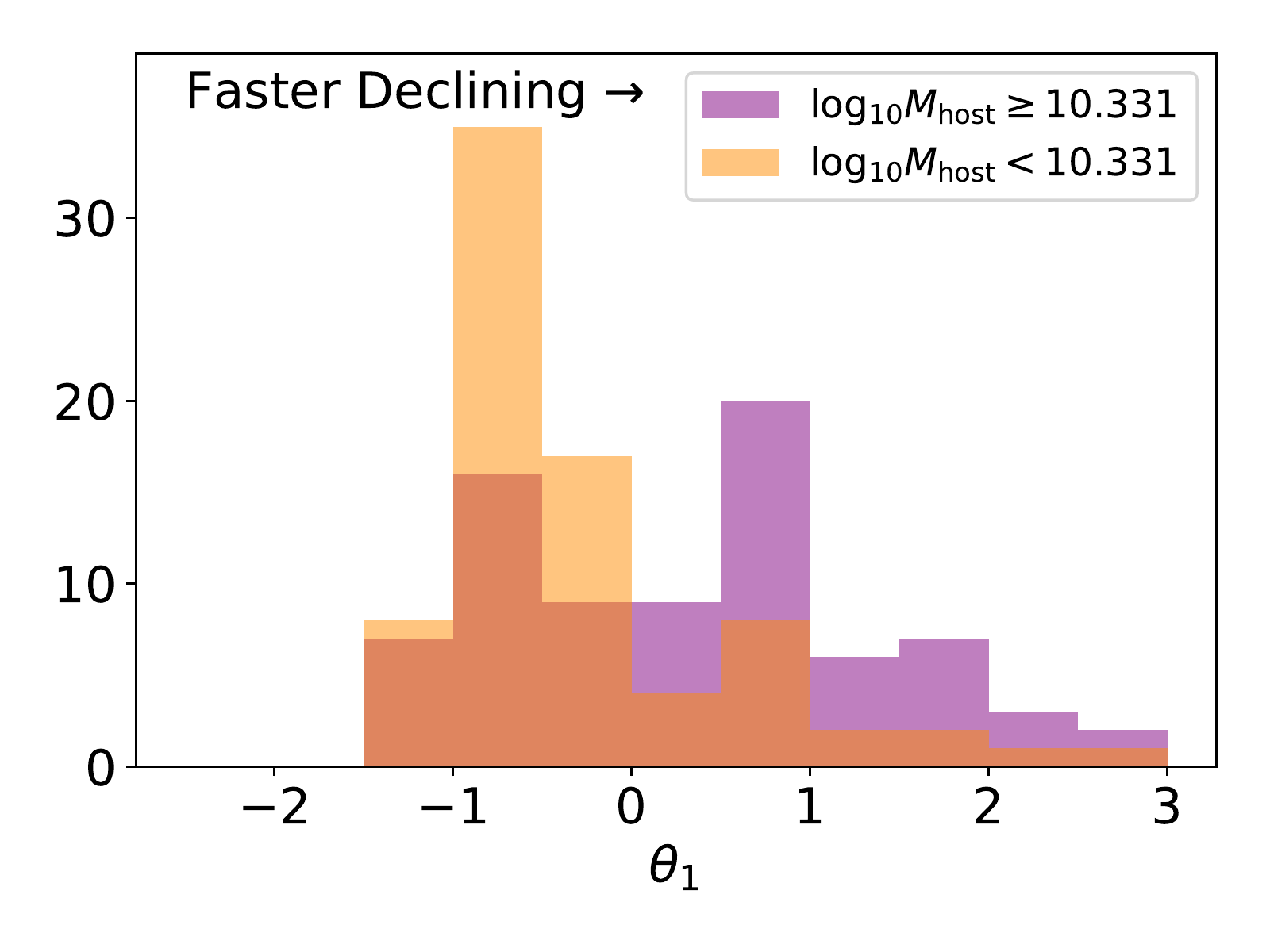}
    \includegraphics[width=\linewidth]{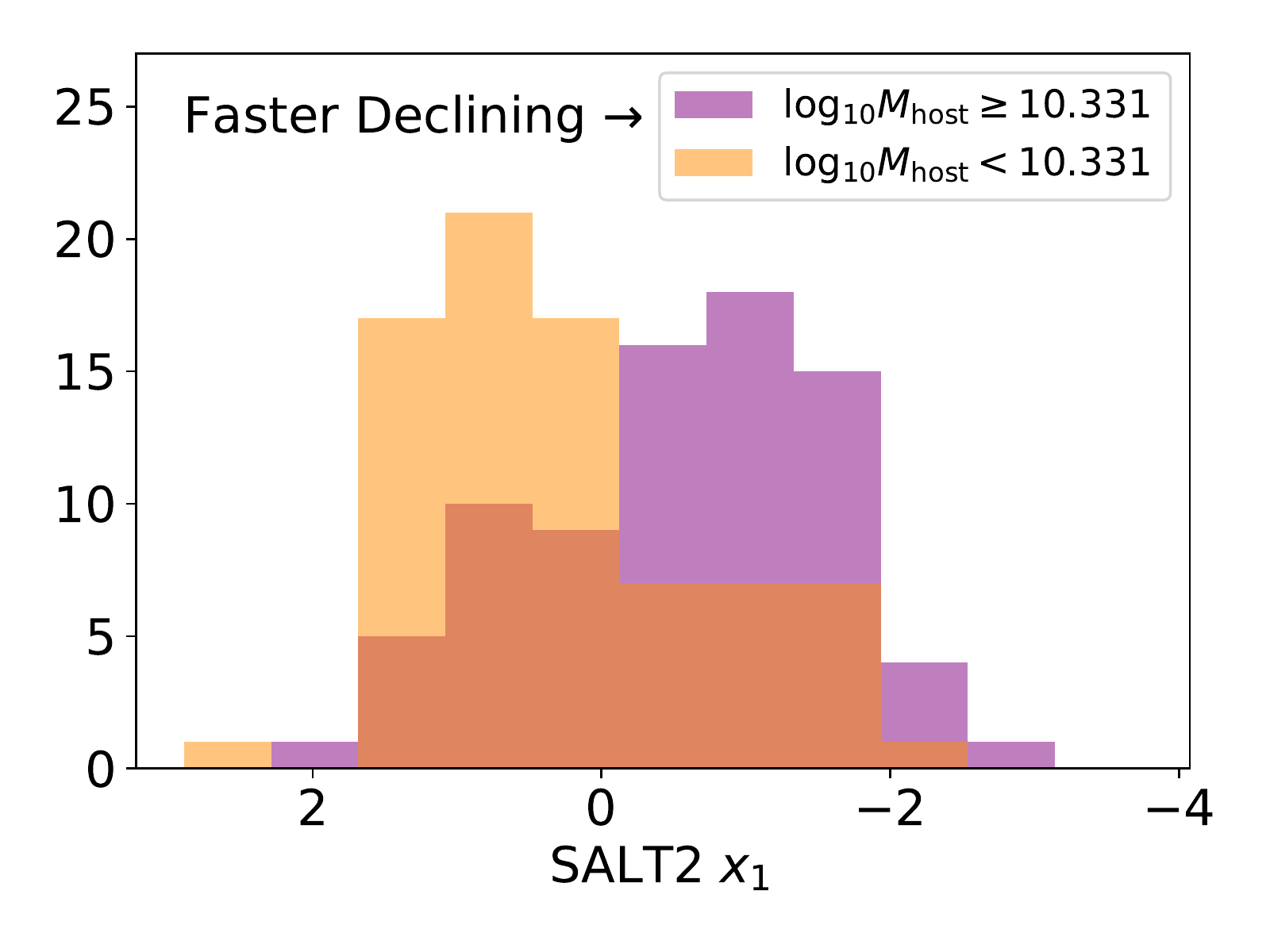}
    \caption{(top panel) Histograms showing how the shape parameter $\theta_1$ -- controlling how strongly the principal SED component, $W_1$, is present for a given supernova -- is distributed for the low- and high- host mass subsamples of Foundation DR1. (bottom panel) Histograms showing the distribution of the \textsc{SALT2} stretch parameter $x_1$ \citep[from][]{foley18,jones19} in the same subsamples. The bins here are those in the top panel, scaled to cover equivalent ranges of parameter space.}
    \label{fig:theta_x1_distributions}
\end{figure}

For all choices of mass split, the SNe in low-mass hosts tend to have $\theta_1$ values which are more concentrated between $-1.5$ and $0$ (i.e. at the bright, slow-declining end of the population), albeit with a tail extending to positive $\theta_1$. High-mass host galaxies seem to have a broader, flatter distribution, with fairly even numbers of SNe falling either side of $\theta_1=0$. The top panel of Figure \ref{fig:theta_x1_distributions} shows histograms of the posterior means of $\theta_1$ from a \textit{Partial-Split} analysis where the sample was divided at the median host galaxy mass.

\begin{figure}
    \centering
    \includegraphics[width=\linewidth]{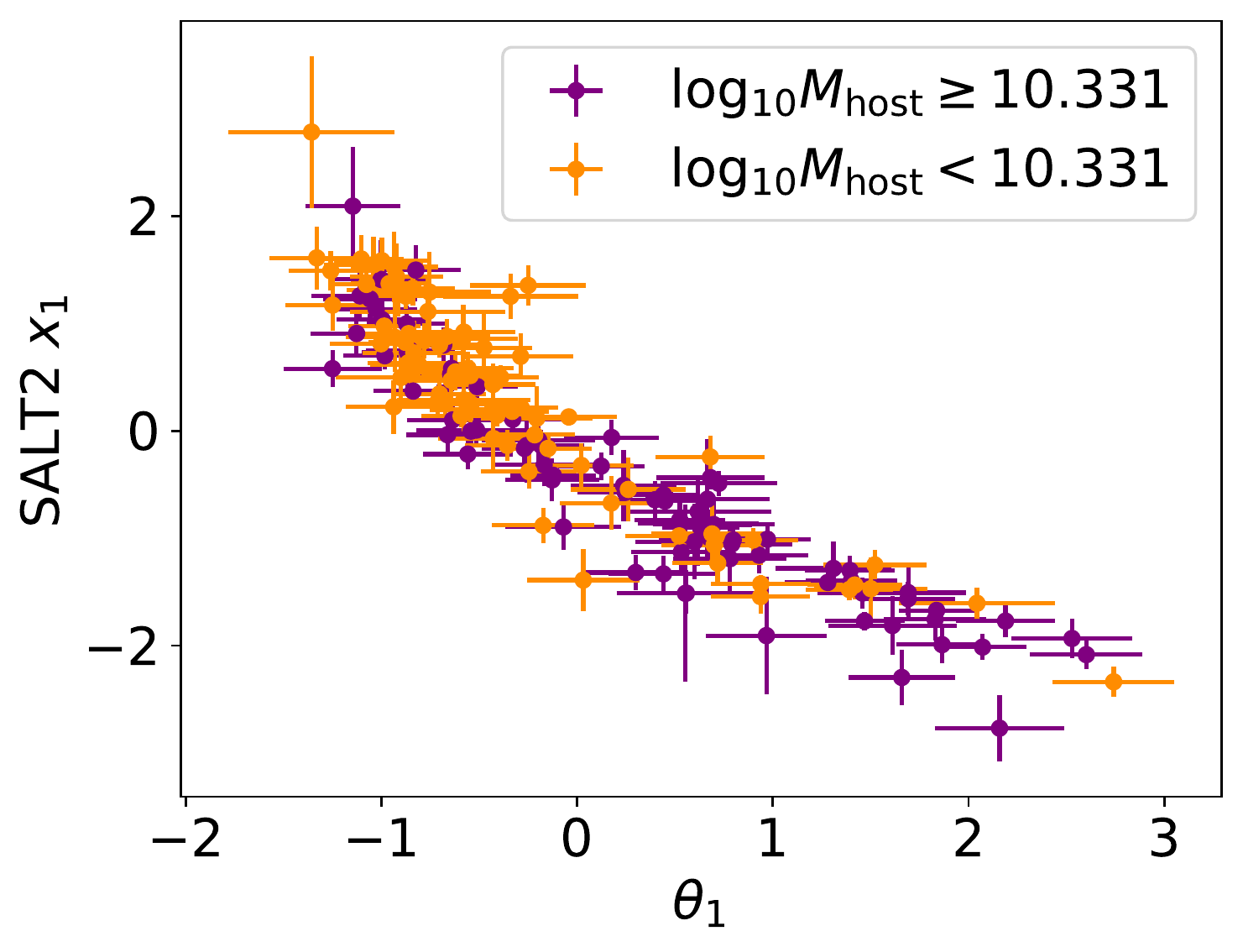}
    \includegraphics[width=\linewidth]{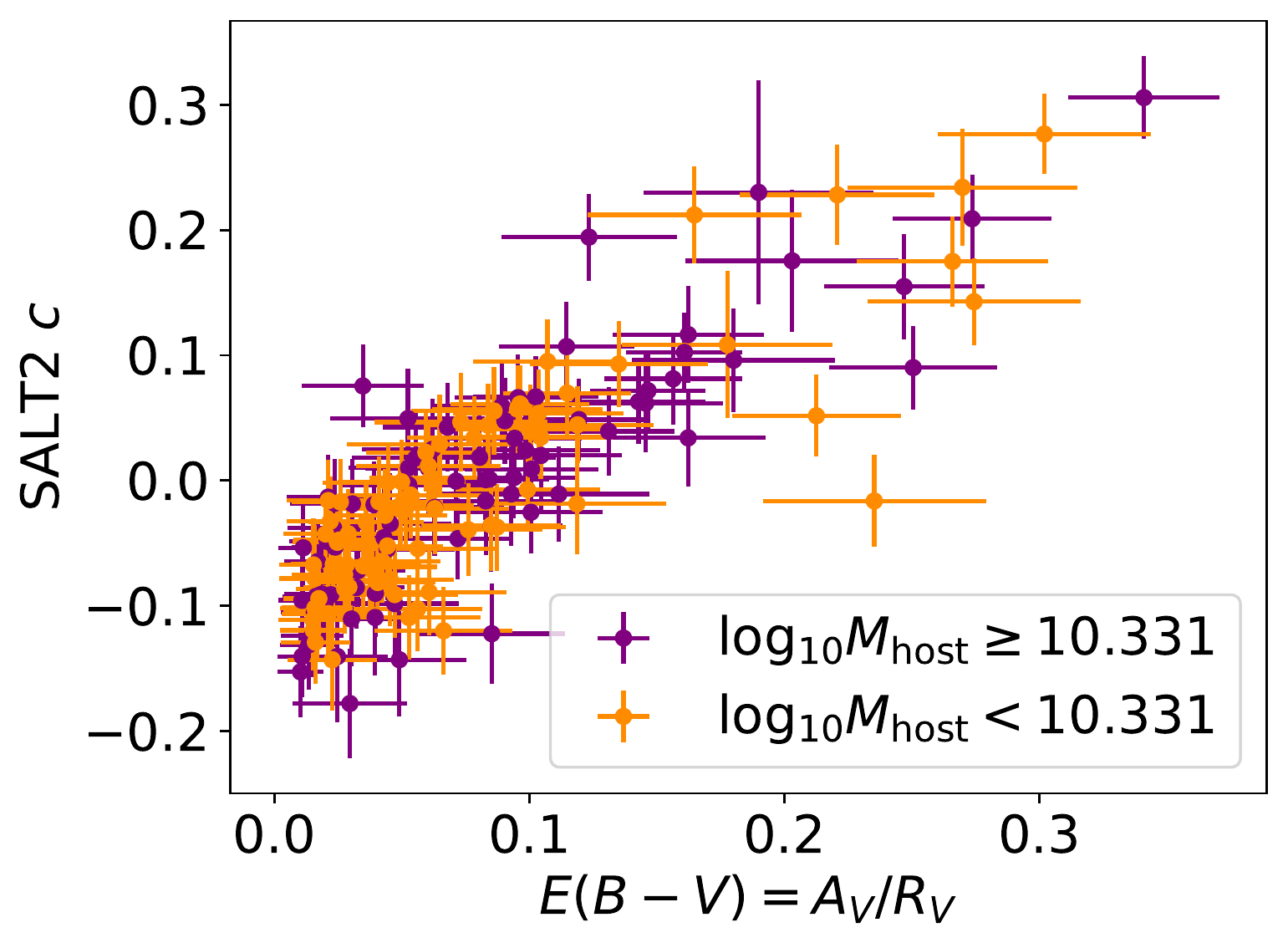}
    \caption{Plots of \textsc{SALT2} parameters against analogous \textsc{BayeSN} quantities, colour coded by host galaxy stellar mass. Values of \textsc{SALT2} parameters from \citet{foley18,jones19}. (top panel) \textsc{SALT2} $x_1$ parameter against the \textsc{BayeSN} $\theta_1$ SED shape parameter. (bottom panel) \textsc{SALT2} colour, $c$, plotted against $E(B-V)=A_V/R_V$ dust reddening derived from \textsc{BayeSN} posterior estimates of $A_V$ and $R_V$.}
    \label{fig:salt2_vs_bayesn}
\end{figure}

We compare this to the behaviour of the \textsc{SALT2} stretch parameter, $x_1$, which captures similar behaviour to \textsc{BayeSN}'s $\theta_1$, with the two being shown to be highly correlated by \citetalias{mandel20}. Indeed, this holds true for the Foundation sample -- the top panel of Figure \ref{fig:salt2_vs_bayesn} plots \textsc{SALT2}'s $x_1$ \citep[from][]{foley18,jones19} against \textsc{BayeSN}'s $\theta_1$ for the 157 supernovae in our Foundation cut. We see a very similar correlation to the one observed in \citetalias{mandel20}. The correlation extends to the way $\theta_1$ and $x_1$ are distributed for the low- and high-mass subsamples -- the bottom panel of Figure \ref{fig:theta_x1_distributions} plots histograms of $x_1$ for the two populations. The bins are plotted from high to low stretch, and scaled so that they are approximately equivalent to the $\theta_1$ bins in the upper panel (i.e. the $n$th bin of the $x_1$ histogram should contain supernovae with similar light curve widths and decline rates to those in the $n$th bin of the $\theta_1$ histogram). Particularly for the low-mass subsample, the $x_1$ and $\theta_1$ distributions are very similar, with the $x_1$ values being concentrated towards the high-stretch, bright and slow-declining supernovae, with a small number of negative stretch cases. This similarity holds true for all choices of mass split point.

These results agree with expectations based on previous works \citep[e.g.][]{neill09, sullivan10, childress13, rigault13, kelsey20}. It is typical to see a tighter stretch distribution in low-mass hosts, which tend to be dominated by the brighter, slow-declining, supernovae, with the dim, fast-decliners only appearing frequently in more massive galaxies. In fact, when we explicitly plot our $\theta_1$ values against host galaxy mass (Figure \ref{fig:latent_vs_mass}, top panel), we see exactly the kind of L-shaped distribution that has been identified in \textsc{SALT2}'s $x_1$ previously \citep{rigault13}, with the quadrant of parameter space corresponding to fast-declining supernovae in low-mass hosts being very sparsely populated.

\subsection{Dust Reddening and Intrinsic Colour}
\label{sec:colours}

\begin{figure}
    \centering
    \includegraphics[width=\linewidth]{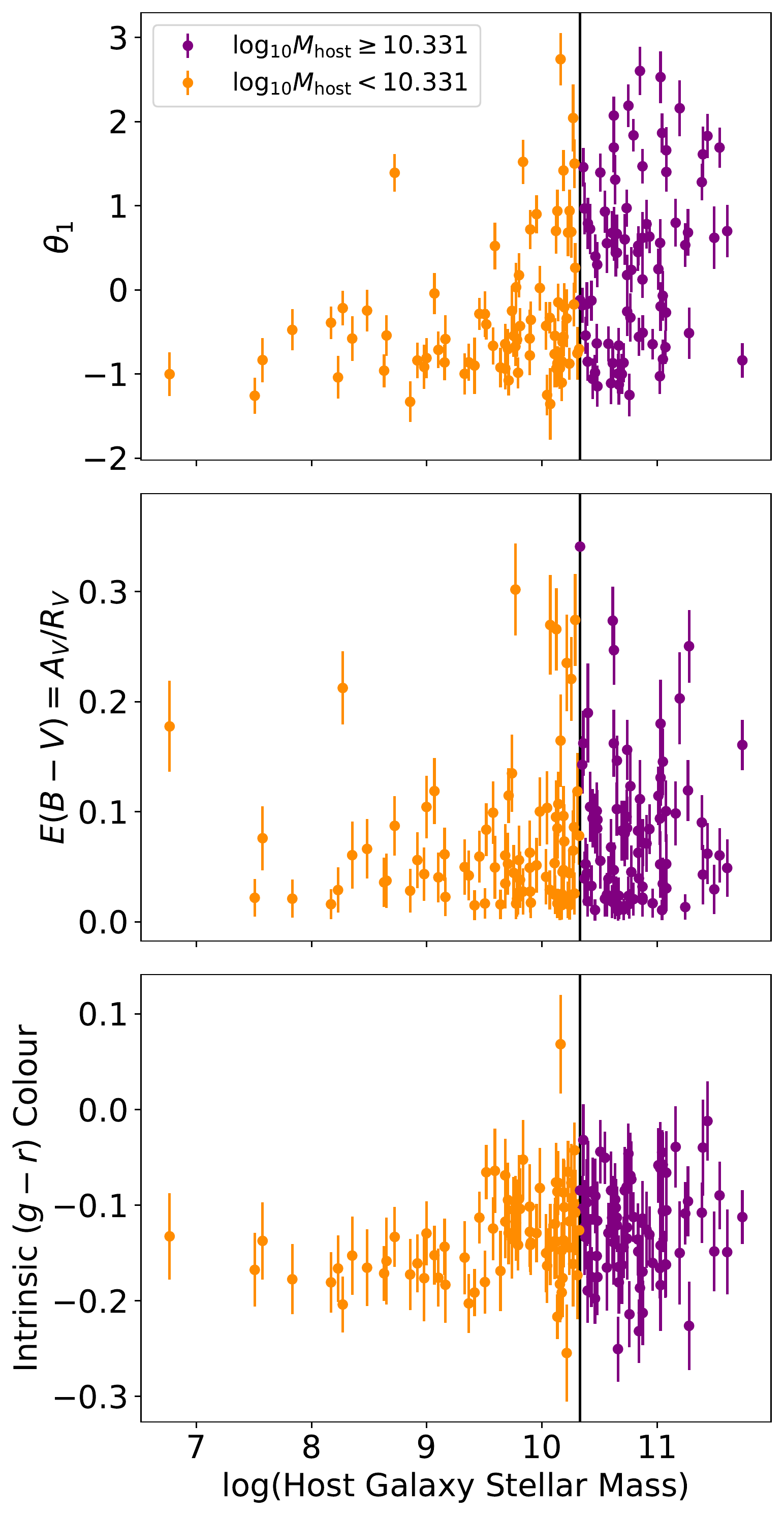}
    \caption{Plots of (top panel) \textsc{BayeSN}'s $\theta_1$ parameter, (middle panel) host galaxy dust reddening, $E(B-V)= A_V/R_V$, and (bottom panel) intrinsic $g-r$ colour vs.\ host galaxy stellar mass. These results are from the \textit{Partial-Split} analysis, with parameters split at the median host mass (depicted as a vertical line).}
    \label{fig:latent_vs_mass}
\end{figure}

Although we have already considered the host extinction population distribution implied by our results, we can also derive explicit values of the inferred reddening due to dust,  $E(B-V) = A_V/R_V$, for all of the supernovae in our sample. This allows a more direct comparison to the \textsc{SALT2} colour parameter, $c$. The bottom panel of Figure \ref{fig:salt2_vs_bayesn} plots \textsc{SALT2} colour, $c$ \citep[from][]{foley18,jones19}, against $E(B-V)$ from the \textit{Partial-Split} \textsc{BayeSN} model. We see that the parameters correlate, albeit not as strongly as $x_1$ and $\theta_1$ (top panel).  For supernovae with low reddening due to dust, we would expect the fitted \textsc{SALT2} colour \citep[an apparent colour that rolls intrinsic and extrinsic effects into a single value -- see][for extensive discussion of this effect]{mandel17} to be dominated by the intrinsic supernova colour, making the deterioration of the $c$ vs.\ $E(B-V)$ relation at $E(B-V) \lesssim 0.1$ unsurprising. For the more extinguished supernovae, the contribution of dust reddening to $c$ becomes dominant, giving rise to the scattered correlation we can see in Figure \ref{fig:salt2_vs_bayesn}.

It is not immediately obvious from Fig. \ref{fig:salt2_vs_bayesn} if $E(B-V)$ correlates with host galaxy mass although we might expect it to from the behaviour of \textsc{SALT2}'s $c$ parameter in previous analyses \citep[e.g.][]{sullivan10,childress13,kelsey20}, which typically see the reddest supernovae only in massive hosts. If we directly plot our computed $E(B-V)$ values against host mass (Figure \ref{fig:latent_vs_mass}, middle panel), some evidence for this behaviour begins to emerge, with very few supernovae with $E(B-V) \gtrsim 0.15$ occurring in hosts less massive than than around $10^{10}\mathrm{M}_\odot$. This mass range is fairly sparsely populated, however, so it is hard to say if this is a genuine difference or a statistical fluctuation in this particular sample.

Since \textsc{BayeSN} separately models intrinsic variation and extrinsic dust effects, we can also investigate if intrinsic supernova colour correlates with host galaxy mass -- something which would be difficult to untangle from dust reddening in an analysis based on \textsc{SALT2} colour. For each supernova in the sample, we compute the posterior mean and standard deviation of its intrinsic $g-r$ colour at peak (i.e. the colour arising purely from the effects of $W_0$, $\theta_1^sW_1$, and $\epsilon_s$). The bottom panel of Figure \ref{fig:latent_vs_mass} plots these derived colours against host galaxy stellar mass. From this, there perhaps appears to be a greater diversity of intrinsic colours found in the more massive galaxies, with these seeming to host both redder and bluer supernovae than their low-mass counterparts. As with $\theta_1$ and $E(B-V)$, however, there are relatively few supernovae below the mass where the behaviour appears to change most strikingly, so such conclusions should be regarded cautiously until analyses are completed on a sample which is even better populated in this mass range.

\subsection{Hubble Diagram Analysis}
\label{sec:hubblediagrams}

\subsubsection{Training and Resubstitution}
\label{sec:resub}
\begin{figure*}
    \centering
    \includegraphics[width=0.5\linewidth]{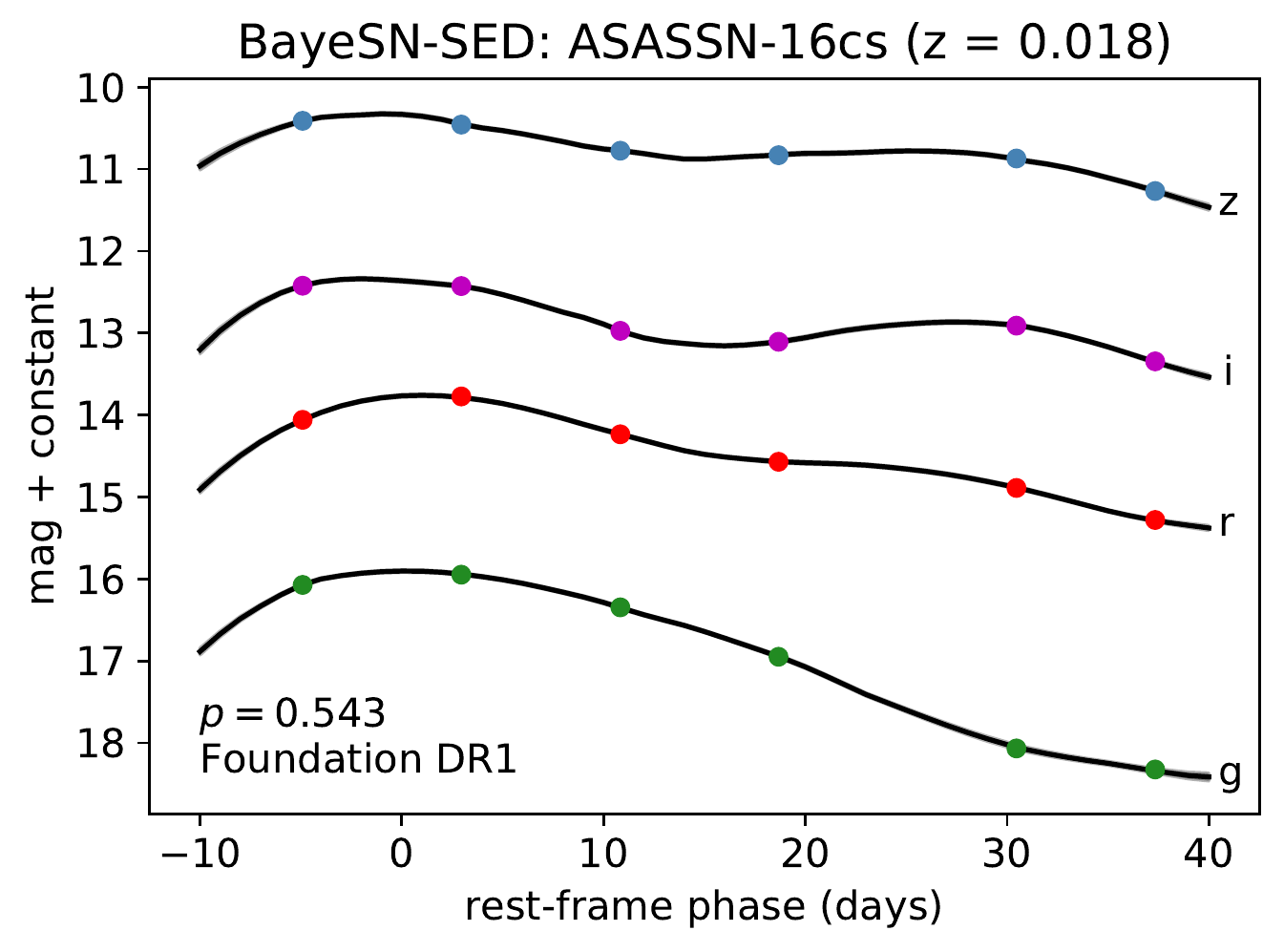}%
    \includegraphics[width=0.5\linewidth]{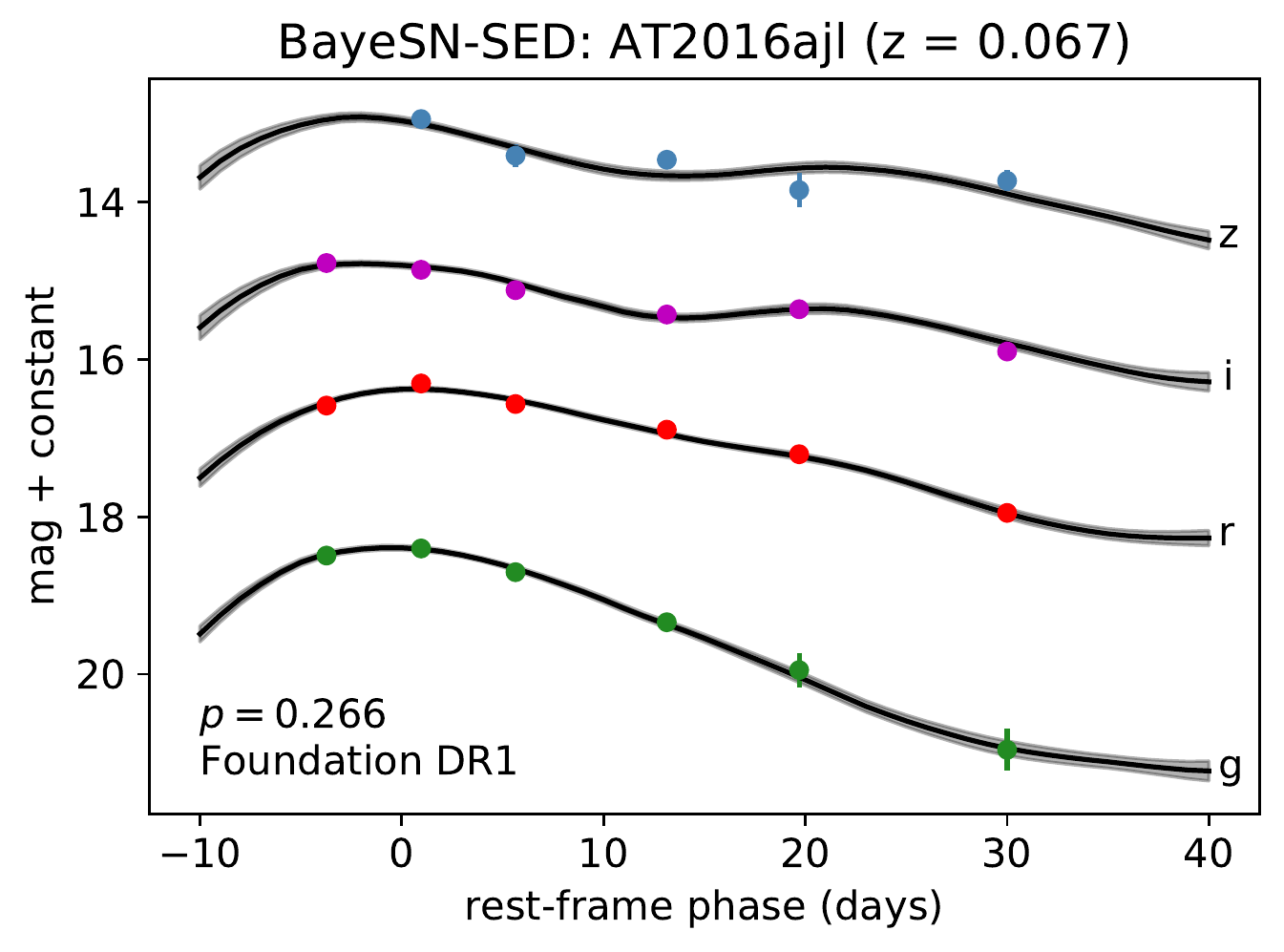}
    \includegraphics[width=0.5\linewidth, trim={0 0 0 1cm}, clip=true]{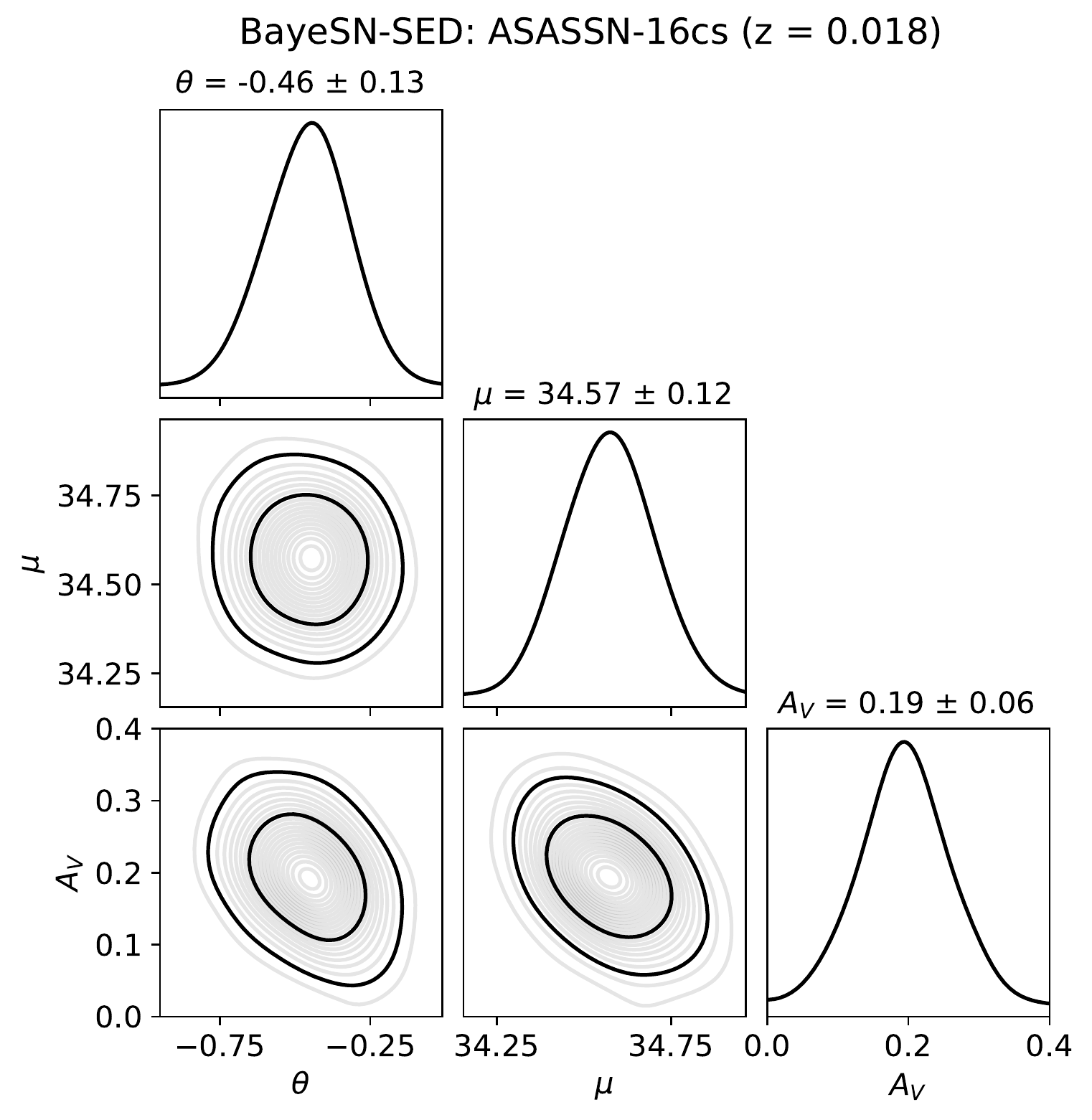}%
    \includegraphics[width=0.5\linewidth, trim={0 0 0 1cm}, clip=true]{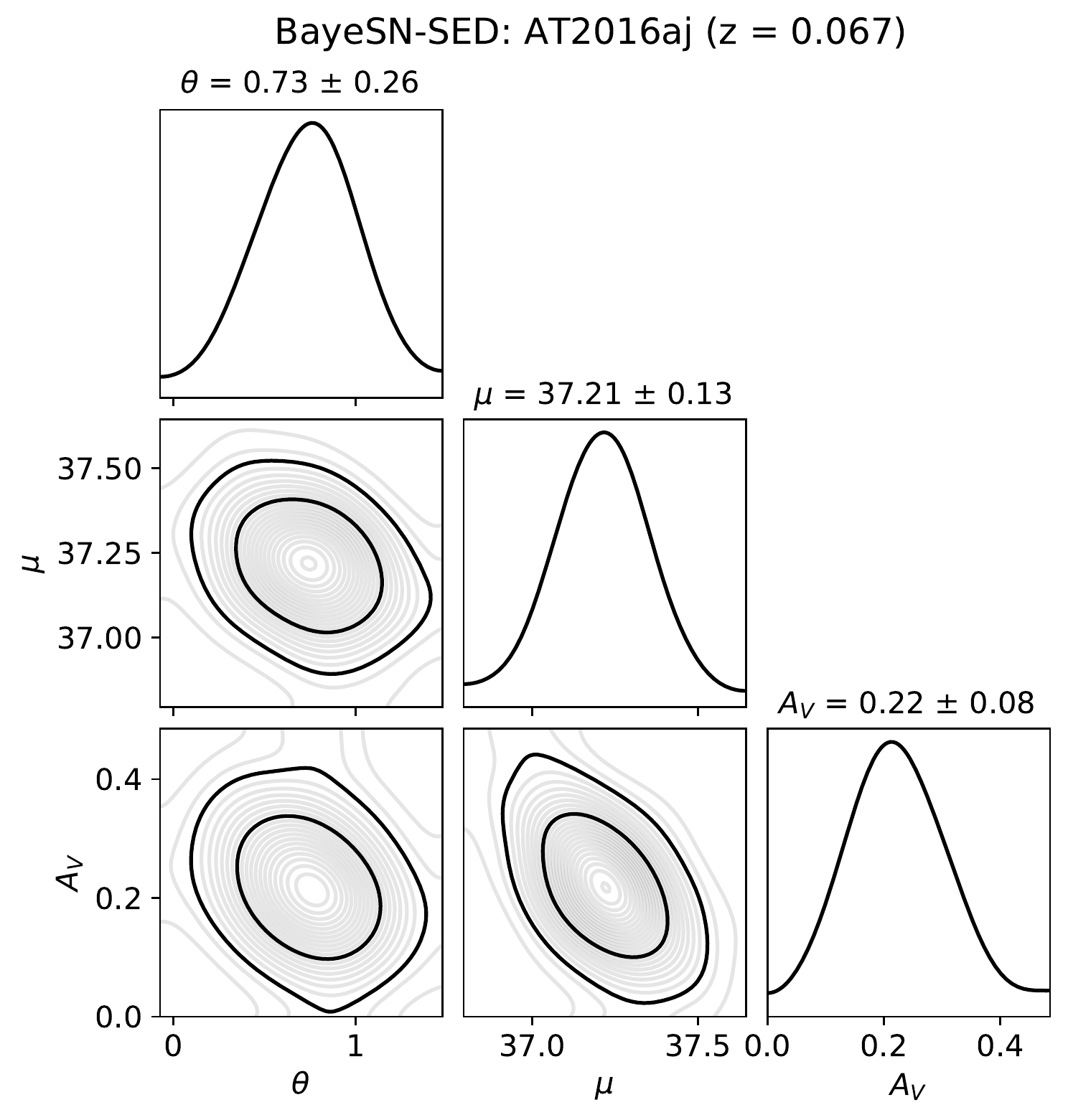}
    \caption{Example photometric distance fit to two of the Foundation DR1 supernovae: ASASSN-16cs (left panels; bright, low-redshift, slower-declining) and AT2016ajl (right panels; dim, higher-redshift, faster-declining). (top panels) Fits to the $griz$ light curves. Vertical offsets are $(0,-2,-4,-6)$ for $(g,r,i,z)$. Indicated in the lower left of the plots are posterior predictive $p$-values (see Appendix \ref{app:pppv} and Figure \ref{fig:pppv_example}) assessing the fitness of the model. An extreme $p$-value close to 0 or 1 would indicate that the model may not be a good fit to the data. (bottom panels) Joint posterior distributions over $\theta_1^s$, $\mu_s^\text{phot}$, and $A_V^S$. The photometric distance estimates used for constructing our Hubble diagram are obtained by marginalising over $\theta_1^s$ and $A_V^s$ (as in the second panel along the diagonal of the corner plot).}
    \label{fig:lc_example}
\end{figure*}

After training each of our models, we fix the global model parameters ($\bm{W}_0$, $\bm{W}_1$, $\bm{\Sigma}_\epsilon$, $\tau_A$, $R_V$, $\sigma_0$, $\Delta M_0$) to their posterior mean values. Then, we `resubstitute' the light curves of the 157 supernovae in our training set, fitting the latent parameters ($\theta_1^s$, $A_V^s$, $\delta M_s$, $\bm{e}_s$) of each supernova independently with no external distance constraints. Marginalising over the latent parameters for each supernova, we thus obtain photometric distance estimates \citepalias[as in][section 2.8]{mandel20}. Figure \ref{fig:lc_example} shows for one SN, ASASSN-16cs, an example light curve fit and the joint posterior over $\theta_1^s$, $A_V^s$, and the photometric distance, $\mu_s^\text{phot}$.

\begin{figure*}
    \centering
    \includegraphics[width=0.5\linewidth]{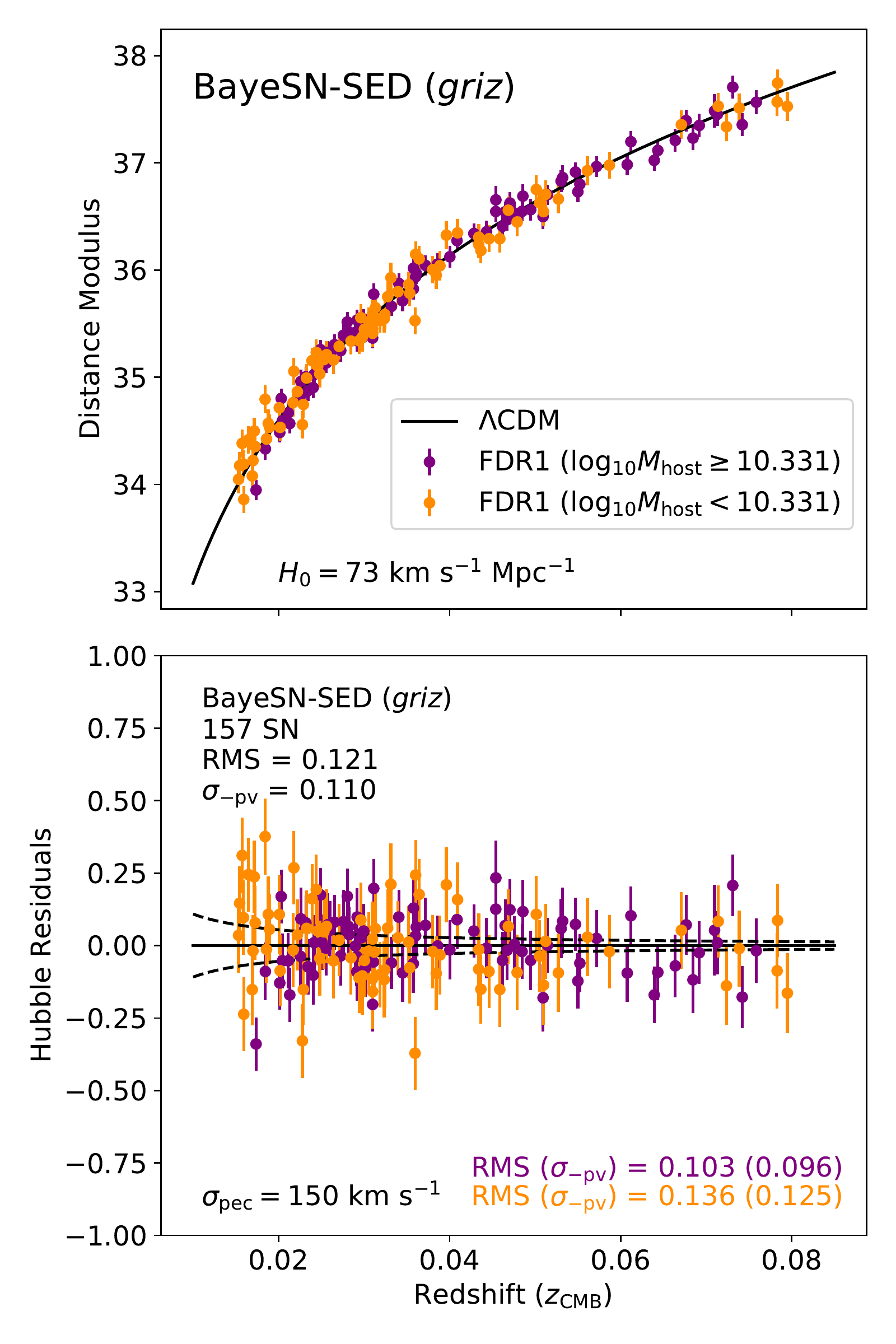}%
    \includegraphics[width=0.5\linewidth]{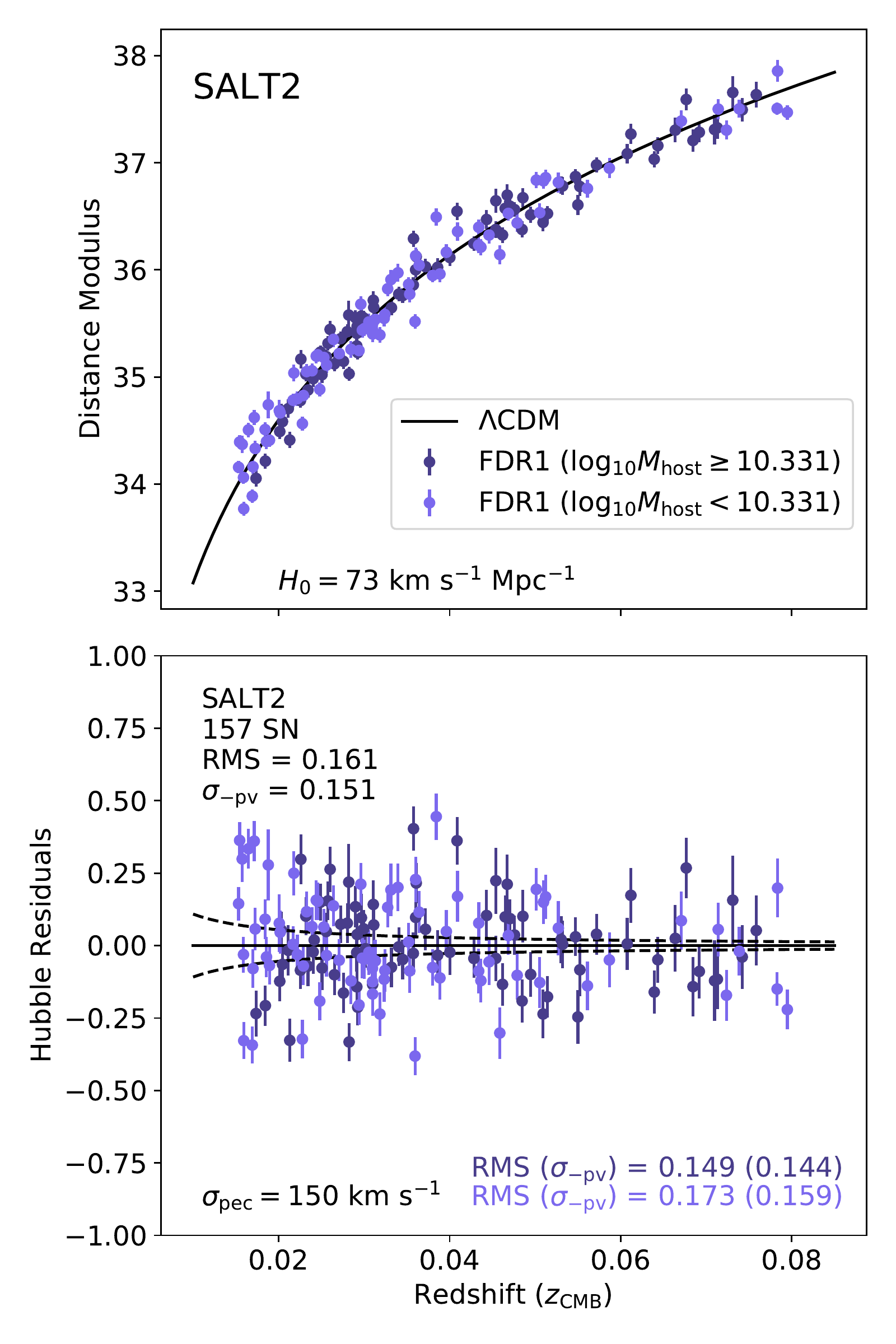}
    \caption{(left panel) Hubble diagram of photometric distances obtained by training on and resubstitution of the Foundation DR1 SNe Ia, using the \textit{Partial-Split} 0.35--0.95~$\upmu$m \textsc{BayeSN} SED model with key parameters split at the median mass cut. (right panel) Hubble diagram based on \textsc{SALT2} fits from \citet{foley18,jones19}, and the Tripp formula with the mass-step applied (Eq. \ref{eq:tripp}). In all cases, dashed lines depict the level of distance uncertainty which can be expected given a peculiar velocity uncertainty of $\sigma_{\text{pec}}=150$~km\,s$^{-1}$.}
    \label{fig:hubble_diagrams}
\end{figure*}

For a given training and resubstitution run, we can construct a Hubble diagram plotting photometric distances, $\mu_s^{\text{phot}}$, against CMB frame redshifts, $z_s^{\text{CMB}}$. During training, we employed external distance constraints \citep[based on a fiducial $\Lambda$CDM cosmology from][]{riess16} centred on $\mu_s^{\text{ext}}=\mu_{\Lambda\text{CDM}}(z_s^{\text{CMB}}; H_0=73.24, \Omega_M=0.28, \Omega_\Lambda=0.72)$, with uncertainties derived from peculiar velocity (assuming $\sigma_{\text{pec}}=150$~km\,s$^{-1}$; \citealp{carrick15}) and spectroscopic redshift uncertainties. To assess the accuracy of our estimates, we compute Hubble residuals for each set of photometric distances from $\mu_s^{\text{phot}} - \mu_s^{\text{ext}}$. Figure \ref{fig:hubble_diagrams} shows Hubble diagrams (upper panels) and residuals (lower panels) constructed in this way from the training and resubstitution of the \textit{Partial-Split} (left column) model, assuming the median host mass as its split point. 

For comparison, in the right column we plot a Hubble diagram computed by applying a linear \citet{tripp98} standardisation,
\begin{equation}
    \mu_s = m_B^s - M_B + \alpha x_1^s - \beta c^s - \gamma I_*^s,
    \label{eq:tripp}
\end{equation}
to the \textsc{SALT2} \citep{guy07,guy10,betoule14} fit results ($m_B^s$, $x_1^s$, $c^s$; $B$-band apparent magnitude, stretch, and colour, respectively) from \citet{foley18, jones19}. Here $I_*^s$ is an indicator variable which obeys $I_*^s=1$ if $M_{*,s}\geq10^{10.331}\mathrm{M}_\odot$ and $I_*^s=0$ otherwise. We fit the values of the Tripp parameters to the full sample and find $M_B=-19.201$, $\alpha=0.125$, $\beta=2.834$, $\gamma=-0.056$, using a simple MCMC sampling of the Tripp distance likelihood using the \textsc{SALT2} fit parameters and external distance constraints\footnote{A Bayesian  method correcting for regression dilution from covariate measurement error (\citealp{kelly07}, see also \citealp{march11}) obtains $\alpha=0.13\pm0.01$, $\beta=3.20\pm0.18$, with a negligible change in RMS ($0.001$).}. When fitting the low- and high-mass host subsamples independently, we find, for all choices of the mass split, the $\alpha$ and $\beta$ coefficients are consistent within the uncertainties with the full-sample values, while the differences in $M_B$ values capture the mass step. Apart from the light curve fit uncertainties and peculiar velocity uncertainties, the SALT2 Hubble diagram has a residual scatter of $\sigma_{\text{res}}=0.134$.

For a given run, we quantify the accuracy of our photometric distance estimates using two metrics: the root mean square (RMS) of the Hubble residuals, and the level of scatter ($\sigma_{-\text{pv}}$) unaccounted for by an assumed peculiar velocity uncertainty of $\sigma_{\text{pec}}=150$~km\,s$^{-1}$ \citepalias[computed as in][eq.\ 32]{mandel20}. Table \ref{tab:hubble_scatter} summarises the values of the two scatter statistics under our different model configurations. In the latter case, the statistics are also computed separately for the portions of the sample above and below the chosen split point. The RMS and $\sigma_{-\text{pv}}$ values for the \textsc{SALT2} Hubble diagram are printed on the right hand panel of Figure \ref{fig:hubble_diagrams}.

Under training and resubstitution, all \textit{No-Split} and \textit{Partial-Split} \textsc{BayeSN} configurations tested yield a total raw RMS training error $<0.125$~mag, with the dispersion after removal of the contribution from peculiar velocity uncertainties being $<0.115$~mag in all cases. For the median mass split (shown in Figure \ref{fig:hubble_diagrams}), we find an RMS ($\sigma_{-\text{pv}}$) of 0.121 (0.110)~mag. These results compare favourably to the larger scatter found in the Hubble diagram derived from the \textsc{SALT2} results, which shows an RMS ($\sigma_{-\text{pv}}$) of 0.161 (0.151)~mag. As in \citet{mandel20}, we sample with replacement 1,000 bootstrapped sets of Hubble residuals from the \textsc{BayeSN} and \textsc{SALT2} photometric distance results, computing the $\Delta\text{RMS}$ each time. For the two Hubble diagrams shown in Figure \ref{fig:hubble_diagrams}, we find that $\Delta\text{RMS}=0.041\pm0.009$~mag. As well as improving on the overall Hubble diagram scatter, by adopting a Bayesian approach, and marginalising over the other latent parameters in obtaining its distance estimates, \textsc{BayeSN} naturally yields well motivated distance uncertainties. Under the \textit{No-Split} model, the distribution of scaled errors, $(\mu_s^{\text{phot}} - \mu_s^{\text{ext}})/\sigma_{\mu,s}^{\text{phot}}$, computed from our photometric distances, $\mu_s^{\text{phot}}$, and their uncertainties, $\sigma_{\mu,s}^{\text{phot}}$, has a standard deviation of 1.07 -- close to the ideal value of 1. The results from the \textit{Partial-Split} analyses are similar. Such agreement is only achieved in the \textsc{SALT2} results if the fitted residual scatter term, $\sigma_{\text{res}}$, is added in quadrature to the distance uncertainties.

One interesting result that emerges from the \textit{Partial-Split} analyses is the somewhat lower level of scatter in the Hubble residuals of supernovae in massive hosts. This could suggest that, despite their greater diversity of inferred properties (see Sections \ref{sec:theta}--\ref{sec:colours} and Figure \ref{fig:latent_vs_mass}), we are more able to effectively standardise supernovae in massive hosts and obtain accurate distances to them. Splitting at the median host mass (which gives the greatest difference of the split choices we consider here), the Hubble diagram RMS for supernovae in high-mass hosts is 0.033~mag lower than in low-mass hosts. Applying bootstrap resampling to estimate the RMS uncertainty in the two subsamples suggests that this is a roughly $2\sigma$ difference, so is not hugely significant. For the difference choices of mass split, the difference in Hubble diagram scatter between mass bins appears to align with the difference in the value of $\sigma_0$ (the level of time- and wavelength-independent residual brightness scatter; see Section \ref{sec:delM}) estimated during training. This is because the mode of SED variation described by $\sigma_0$ contributes to the photometric distance uncertainty. In the SALT2 Hubble diagram, we estimate $\Delta \text{RMS}=0.024\pm0.018$~mag between low- and high-mass hosts, a $1.4\sigma$ difference in the same direction as BayeSN.

The results from our \textit{Full-Split} analyses (see Table \ref{tab:hubble_scatter}) are largely similar to the \textit{Partial-Split} results. The Hubble diagram dispersion for the median and MLE mass splits is fairly consistent between the \textit{Full-Split} and \textit{Partial-Split} runs, for both the low- and high-mass subsamples. The \textit{Full-Split} configuration should be invulnerable to the inference of the intrinsic SED parameters ($\bm{W}_0$, $\bm{W}_1$, $\bm{\Sigma}_\epsilon$) being driven by supernovae in one mass bin at the expense of the characterisation of supernovae in the other. The consistency between our \textit{Full-Split} and \textit{Partial-Split} results thus disfavours this as the cause of the imbalance in Hubble diagram scatter between mass bins in the \textit{Partial-Split} runs.

\begin{table*}
    \centering
    \begin{threeparttable}
        \caption{Hubble diagram scatter for the 157 Foundation supernovae in our sample.}
        \label{tab:hubble_scatter}
        \begin{tabular}{l l l c c c c c c}\toprule
            Scheme\tnote{a} & Model & Split Point & \multicolumn{3}{c}{RMS\tnote{b}} & \multicolumn{3}{c}{$\sigma_{-\text{pv}}$\tnote{c}} \\  \cmidrule(lr){4-6}\cmidrule(lr){7-9}
            & & & All & LM & HM & All & LM & HM\\\midrule
            Resubstitution & \textit{No-Split} & - & 0.124 & - & - & 0.112 & - & -\\
            & \textit{Partial-Split} & $10^{10}\mathrm{M}_\odot$ & 0.121 & 0.125 & 0.119 & 0.109 & 0.115 & 0.106\\
            & & Median & 0.121 & 0.136 & 0.103 & 0.110 & 0.125 & 0.096\\
            & & MLE & 0.118 & 0.129 & 0.106 & 0.107 & 0.117 & 0.098\\
            & \textit{Full-Split} & $10^{10}\mathrm{M}_\odot$ & - & 0.166 & 0.123 & - & 0.158 & 0.111\\
            & & Median & - & 0.127 & 0.104 & - & 0.115 & 0.097\\
            & & MLE & - & 0.128 & 0.109 & - & 0.117 & 0.101\\
            Cross-Validation & \textit{No-Split} & - & 0.134 & - & - & 0.122 & - & -\\
            & \textit{Partial-Split} & $10^{10}\mathrm{M}_\odot$ & 0.134 & 0.142 & 0.130 & 0.123 & 0.133 & 0.118\\
            & & Median & 0.132 & 0.149 & 0.113 & 0.122 & 0.138 & 0.106\\
            & & MLE & 0.129 & 0.142 & 0.115 & 0.119 & 0.131 & 0.107\\
            \bottomrule
        \end{tabular}
        \begin{tablenotes}
            \item [a] Denotes whether photometric distances were obtained through direct training and resubstitution of all supernovae, or via 10-fold cross-validation (see section \ref{sec:hubblediagrams}).
            \item [b] Total root mean square (RMS) of the Hubble residuals.
            \item [c] Dispersion when expected variance due to peculiar velocity uncertainty (assumed to be $\sigma_{\text{pec}}=150~\text{km\,s}^{-1}$; \citealp{carrick15}) is taken out \citep[as in][eq.\ 32]{mandel20}.
        \end{tablenotes}
    \end{threeparttable}
\end{table*}

\subsubsection{Cross-Validation}
Similarly to previous supernova studies \citep{mandel09, mandel11, mandel20, blondin11}, we perform cross-validation (CV) to test the sensitivity of the model to the finite training set and confirm that its performance is not significantly overfit by the re-use of our training set in the resubstitution process. As in \citetalias{mandel20}, we perform 10-fold cross-validation. We retrain our models ten times, each time leaving out 10 per cent of the data, which act as an unseen test set for that model. In this way, we  obtain photometric distance estimates for all 157 supernovae, each one obtained from a model whose training set excluded it. The cross-validated values of our Hubble diagram scatter metrics for the \textit{No-Split} and \textit{Partial-Split} models are given in the bottom rows of Table \ref{tab:hubble_scatter}. Under cross-validation, we find that the RMS ($\sigma_{-\text{pv}}$) for the \textit{No-Split} model rises to 0.134 (0.122), a small increase of 0.01~mag by both scatter metrics-- comparable to what was seen in the \citetalias{mandel20} cross-validation analysis.

For the \textit{Partial-Split} model, the total RMS increase under cross-validation is $\lesssim0.013$~mag. This generally holds on both sides of the mass split, although for the $10^{10}\mathrm{M}_\odot$ mass split the increase on the low-mass side is slightly larger at 0.017~mag. The increase in the total value of $\sigma_{-\text{pv}}$ is consistently $\lesssim0.014$~mag. For the median mass split, we find an RMS ($\sigma_{-\text{pv}}$) of 0.132 (0.122)~mag under cross validation. Since this model under CV is trained on 10\% less data than the same model under resubstitution, we would expect a somewhat less accurate model under CV. Thus, we expect the true RMS out-of-sample error rate of the model trained on the full sample would be somewhere between the two, i.e. $\approx 0.127$ mag.

For the \textsc{BayeSN} cross-validation, we repeatedly retrain a new \textsc{BayeSN} SED model on the training set excluding each held-out fold. Although we can not retrain the full \textsc{SALT2} SED model in the same fashion, we can partially cross-validate the \textsc{SALT2} Hubble diagram. We refit the \citet{tripp98} formula (Eq.\ \ref{eq:tripp}) to the same 9 of 10 training folds as in our \textsc{BayeSN} analysis, using the ($\alpha$, $\beta$, $M_B$, $\gamma$) each time to compute the photometric distances in the held-out fold. The \textsc{SALT2} RMS rises from 0.161 to 0.166~mag under this partial cross-validation. Applying the same bootstrapping procedure as in section \ref{sec:resub}, we find  $\Delta\text{RMS}=0.034\pm0.009$ between the cross-validated \textsc{SALT2} and \textsc{BayeSN} Hubble diagrams, a $3.8 \sigma$ improvement. This improvement in the accuracy of photometric distance estimates is a good indication that the \textsc{BayeSN} SED model is an effective statistical description of the $griz$ SN Ia light curves of the Foundation sample.

\subsection{Simulation-Based Validation}
As a sanity check of our code, we conduct simulation-based tests. We simulate Foundation DR1-like $griz$ light curve datasets from the \textsc{BayeSN} forward model. We then train new \textsc{BayeSN} models on these simulations to check that the original parameters are recovered.

For example, to check our $R_V$ inferences, we generate simulations with all population hyperparameters ($\bm{W}_0$, $\bm{W}_1$, $\bm{\Sigma}_\epsilon$, $\sigma_0$, $\tau_A$), except for $R_V$, fixed to their posterior mean values from the \textit{No-Split} analysis. We generated 3 simulations, each of which has a different input true global host dust law $R_V = (2.0, 3.0, 4.0)$.  We then trained new \textsc{BayeSN} models on these simulated data. We successfully recovered posterior estimates $\hat{R}_V=(1.88\pm0.11,2.87\pm0.19,4.27\pm0.34)$ after training on the simulated datasets. 

Additionally, we generated further simulations with input $R_V$ distributions with population parameters $(\mu_R, \sigma_R) = (2.0, 1.0)$ and $(3.0, 1.0)$, to verify that we would be able to detect a wide population distribution \citep[\`a la][]{brout20} in a Foundation-like sample. We repeated our population inference on these new simulated datasets. For the input population with $(\mu_R,\sigma_R)=(2.0,1.0)$, our posterior estimates of the population parameters were $(\hat{\mu}_R,\hat{\sigma}_R)=(2.30\pm0.42,1.59\pm0.47)$. For the input population with $(\mu_R,\sigma_R)=(3.0,1.0)$, our posterior estimates were $(\hat{\mu}_R,\hat{\sigma}_R)=(3.06\pm0.30,0.86\pm0.26)$. From the results of the two runs with $\mu_R= 2.0$ or $3.0$, we would conclude with 95\% posterior probability that $\sigma_R>0.93$ or $0.50$, respectively. Thus, we would detect an $R_V$ distribution of non-zero width with high confidence.

\section{Conclusions}
\label{sec:conclusions}
We have used the $griz$ light curves of 157 Type Ia supernovae from the first Foundation Supernova Survey data release \citep{foley18,jones19} to train a \textsc{BayeSN} SN Ia SED model, continuous over 0.35--0.95~$\upmu$m. By training on a homogeneous SN sample -- with all data in Foundation DR1 having been obtained on the Pan-STARRS-1 system -- we ensure that the resulting model is affected by minimal cross-calibration systematics. This is the first statistical model for SN Ia in the rest-frame $z$-band to have been trained on $z$-band light curves. Foundation DR1 is the first large homogeneous sample of rest-frame $z$-band observations of SNe Ia, and was not available to previous continuous SED models \citep[e.g. \textsc{SALT2};][]{guy07,guy10,betoule14} or models for light curves in discrete passbands (e.g.\ \citealp{mandel09,mandel11}; and \textsc{SNooPy}; \citealp{burns11}). By training on rest frame $z$-band data, we improve on the \citetalias{mandel20} SED model at these wavelengths. Using our model, we compute photometric distance estimates for the Foundation sample, presenting the first Hubble diagram based on full rest-frame $griz$ light curves. The value added by the $iz$-band data, and the advantages of our modelling approach, combine to yield an excellent total RMS of 0.121~mag. This result is robust under 10-fold cross-validation, where the total RMS only rises to 0.132~mag.

The \textsc{BayeSN} framework incorporates the host galaxy dust extinction law (assuming \citealp{fitzpatrick99}) at the SED level, properly modelling its impact at all phases and wavelengths. In this way, it leverages the colour--luminosity information in the full $griz$ light curves of the sample, enabling robust inferences about the $R_V$ value(s) parametrizing the dust law. When fitting for a single global dust law for the full Foundation sample, we estimate $R_V=2.61\pm0.21$. This result, based on an independent sample, aligns well with many previous analyses of other low-$z$ SN Ia samples with apparent colours consistent with the cosmological cut (peak apparent $B-V\lesssim0.3$) \citep[e.g.][]{riess96, phillips99, folatelli10, chotard11, foley11,mandel11, mandel17, mandel20, burns14, sasdelli16, leget20, arima21}, some of which utilised spectroscopic or NIR data to help break the degeneracy between intrinsic variation and dust effects. The recent SN Ia ``sibling'' analysis of \citet{biswas21} estimated a similar $R_V \approx2.5$, albeit for a slightly more reddened supernova (AT2019lcj; $c_\text{SALT2}=0.57\pm0.04$, $E(B-V)_\text{SNooPy}=0.63\pm0.07$, $\log_{10}(M_*/\mathrm{M}_\odot)=10.48^{+0.11}_{-0.46}$) than we have considered here. Allowing a population distribution of $R_V$ values, parametrized by a mean, $\mu_R$, and standard deviation, $\sigma_R$, we find $\mu_R=2.70\pm0.25$ and a strong preference for small dispersion, i.e. $\sigma_R < 0.61$ at 95\% posterior probability, consistent with our global dust law results.

Beyond this, we also used the Foundation data to train a modified version of the \textsc{BayeSN} model which separates several key parameters by host galaxy stellar mass, including the average dust extinction, and the dust law $R_V$ (or parameters describing its population distribution). When assuming a single $R_V$ for each mass group, we estimate $R_V=2.84\pm0.31$ for host galaxies less massive than the median, and $R_V=2.58\pm0.23$ for those more massive. Both are consistent with our single global value of $R_V=2.61\pm0.21$. For choices of mass split other than the median, we still infer values of $R_V$ between low- and high-mass hosts that are consistent within $\lesssim1.2\sigma$ (see Table \ref{tab:global_params}). In all cases, we find that $R_V\lesssim2.2$ is excluded from the posterior at the 95 per cent level, and that the results tend to be consistent with the global-$R_V$ result. When a population distribution of $R_V^s$ values is allowed either side of the median mass split, we find a mean of $\mu_R=2.97\pm0.37$ for less massive hosts, and $\mu_R=2.66\pm0.27$ for more massive hosts, both consistent with our mass-agnostic analysis. For all choices of mass split, our $\mu_R$ estimates are consistent to within $\sim1.2\sigma$ across the split. In all cases, the mean values are consistent with their counterparts from the global-$R_V$ analyses, and values of $\mu_R\lesssim2.2$ tend to be strongly disfavoured on both sides of the mass split. We find that the width $\sigma_R$ of the $R_V^s$ population distribution is consistent with  small values (see lower panels of Fig.\ \ref{fig:RV_tauA_posteriors}). For high-mass hosts, or for the whole population when host mass is ignored, wide distributions of $R_V^s$ are disfavoured, with $\sigma_R\lesssim0.7$ with at least 95\% posterior probability in all cases. For low-mass hosts, constraining power is weakened by having smaller sample sizes, and fewer SNe with $E(B-V)\gtrsim0.15$, so a wide $R_V$ population distribution cannot be completely ruled out. While it cannot be ruled out that some individual SNe with low-to-moderate reddening may be affected by dust laws with unusually low $R_V$ (e.g. SN 2012et with $R_V\approx 1.7^{+0.6}_{-0.5}$; c.f.\ caveats in \citealp{amanullah15}), we find no strong evidence in this sample that this is true for the  aggregate, which can be adequately explained with $R_V \approx \text{2.5--3}$. Our population mean estimates, $\mu_R$, are slightly below the average Milky Way value \citep[e.g.][]{schlafly16} by at most $2.4\sigma$.

Looking at the extinction population distribution either side of the median host galaxy mass, we estimate an average host galaxy dust extinction of $\tau_A=0.19\pm0.03$ for low masses, and $\tau_A=0.21\pm0.03$ for high masses. Our estimates of $\tau_A$ are consistent to within $\lesssim1.2\sigma$ between low- and high-mass host galaxies for all choices of mass split, and are largely insensitive to whether a single $R_V$ or a population distribution is assumed within each subsample.

Even allowing for dust to differ between low- and high-mass host galaxies, we still find evidence for some level of mass step-like behaviour. We see a magnitude offset between supernovae in low- and high-mass host galaxies which suggests that the latter group are brighter on average by a $\Delta M_0$ between 0.04 and 0.07~mag. This range is consistent with the size of mass steps found by previous analyses at a range of mass splits from $\sim10^{10}$--$10^{10.8}\mathrm{M}_\odot$ \citep[e.g.][]{kelly10,sullivan10,betoule14,roman18,uddin20,kelsey20,smith20}, including the \citet{jones19} study of the Foundation data. For all choices of mass split point, the offset we estimate is consistent with the corresponding Hubble residual step seen in our joint, mass-agnostic, analysis.

\citet{brout20} invoked differing $R_V$ distributions between low- and high-mass host galaxies as a possible cause of the mass step. They reported that wide $R_V$ distributions ($\sigma_R = 1.3 \pm 0.2$) centred on $\mu_R=2.75\pm0.35$ in low-mass hosts and $\mu_R=1.50\pm0.25$ in high-mass hosts could explain away the $\sim0.06$~mag mass step they otherwise saw in their data. In our analysis of the Foundation data, we find that consistency of $R_V$ between low- and high-mass hosts is favoured, with a preference for narrow population distributions. While the estimated difference in mean $R_V$ is more uncertain when the sample is split at $10^{10}\mathrm{M}_\odot$ (due to the paucity of more reddened SNe in low-mass hosts), we infer that a non-zero mass step and small difference in mean $R_V$ are preferred for all split choices. A plausible explanation for the mass step should account for its empirical observation over a range of mass splits. Consequently, our results disfavour significantly different $R_V$ distributions between low- and high-mass galaxies as an explanation of the mass step. A lack of correlation between the mass step and host galaxy dust properties has also been suggested by recent work \citep{ponder20, uddin20} finding mass steps in the near-infrared (where sensitivity to dust properties should be minimised) comparable to those in the optical \citep[but see also][]{johansson21}. This is further pointed to by the results of \citet{gonzalezgaitan20}, who find that a mass step persists even when separate apparent colour--luminosity relations are allowed in low- and high-mass galaxies. Complementary evidence for a non-dust explanation of SN Ia host galaxy dependence is also provided by the Hubble residual v.s.\ specific star formation rate (sSFR) results of \citet[\S 5.6]{hand21}, and by recent work favouring  two progenitor populations \citep[``prompt'' and ``delayed''; see e.g.][]{mannucci05, mannucci06, scannapieco05} as the cause of the mass/local sSFR step \citep{briday21}.

In this paper, we have presented analysis and evidence demonstrating that allowing for differences in dust properties in low- and high-mass host galaxies does not explain the host-mass step of SN Ia magnitudes in the Foundation DR1 sample.  While this is a relatively large, homogeneous, and excellent nearby sample, owing to the well-known problem of induction, we cannot rule out the possibility that other supernova samples from other surveys may exhibit different characteristics. Future investigations into these scientific questions and the refinement of our statistical modelling and inference techniques will both be greatly enhanced by the growth of well-calibrated SN Ia datasets. In particular, future data releases from the Foundation Supernova Survey and the Young Supernova Experiment \citep[YSE;][]{jones20} will increase the already large sample of low-$z$ SNe Ia with Pan-STARRS light curves. Complementing this, the Carnegie Supernova Project-II \citep[CSP-II][]{phillips19} will augment the previous CSP-I data releases \citep{contreras10, stritzinger11, krisciunas17} to provide further high quality observations at $z\lesssim0.1$. This will include vital near-infrared photometry, which can be particularly valuable in untangling the complex puzzle of supernova-host correlation \citep{ponder20, uddin20, johansson21}. With larger samples, we will be able to determine more surely whether any small estimated differences between the dust properties of low- and high-mass hosts become statistically significant.

In future work, more sophisticated modelling of SN Ia correlations with host mass -- e.g.\ functional regression to examine the time- and wavelength-dependence of SED components against host mass -- is a natural next step, and will be possible within the \textsc{BayeSN} framework. As well as this, we will be able to apply our analysis scheme to study correlations with other global or local environmental properties (e.g. star formation rate, \citealp{sullivan06, lampeitl10, dandrea11, childress13, campbell16, uddin17, rigault13, rigault15, rigault18}; colour, \citealp{roman18}; metallicity, \citealp{gallagher08, dandrea11, hayden13, pan14, campbell16, moreno16b, moreno16a}; age, \citealp{gallagher08, gupta11, pan14, campbell16, rose19, rose20}; or morphology \citealp{hamuy96, hamuy00, sullivan03, hicken09b, pruzhinskaya20}). Incorporation of spectroscopic indicators of intrinsic colour \citep[e.g. ejecta velocity,][]{foley11, mandel14, dettman21} may provide additional insight.  We are working to carefully merge the present SED model with that of \citetalias{mandel20} extending through NIR $H$-band, and also to extend it to the ultraviolet.

The parallel development of both the data and modelling techniques will be critical to a future understanding of SNe Ia and their correlations with their host galaxies. The proper understanding and modelling of these effects will in turn be critical to fully exploit future data from the \textit{Nancy Grace Roman Space Telescope}'s supernova surveys, and Vera C.\ Rubin Observatory's Legacy Survey of Space and Time.

\section*{Acknowledgements}
We thank the anonymous referee for their helpful comments. We thank Dan Scolnic and Dillon Brout for useful discussions. We also thank the Foundation Supernova Survey team for their work in producing and making public the DR1 dataset.

ST was supported by the Cambridge Centre for Doctoral Training in Data-Intensive Science funded by the UK Science and Technology Facilities Council (STFC). KSM acknowledges funding from the European Research Council under the European Union’s Horizon 2020 research and innovation programme (ERC Grant Agreement No. 101002652). This project has been made possible through the ASTROSTAT-II collaboration, enabled by the Horizon 2020, EU Grant Agreement No. 873089. Support for this work was provided by NASA through the NASA Hubble Fellowship grant \#HF2-51462.001 awarded by the Space Telescope Science Institute, which is operated by the Association of Universities for Research in Astronomy, Inc., for NASA, under contract NAS5-26555. SMW was supported by the STFC. GN was supported by the University of Illinois at Urbana-Champaign and the Center for Astrophysical Surveys at the National Center for Supercomputing Applications.  

This work made use of the Illinois Campus Cluster, a computing resource that is operated by the Illinois Campus Cluster Program (ICCP) in conjunction with the National Center for Supercomputing Applications (NCSA) and which is supported by funds from the University of Illinois at Urbana-Champaign.

\section*{Data Availability}
 

The data for the 180 supernovae in the Foundation DR1 cosmology sample \citep{foley18, jones19} are publicly available at \url{https://github.com/djones1040/Foundation_DR1}. Code and model files will be made available at \url{https://github.com/bayesn}.



\bibliographystyle{mnras}
\bibliography{bib} 



\appendix

\section{Searching for a Mass Step}
\label{app:stepmle}
As described in Section \ref{sec:choosing}, our \textit{Partial-Split} and \textit{Full-Split} analyses require us to define the host galaxy stellar mass which delineates low and high masses. As well as the median, and the conventional choice of $10^{10}\mathrm{M}_\odot$, we also try a split point motivated by the data. Specifically, this is the maximum likelihood location of a `mass step' (computed subject to the constraint that the step lies within the interquartile range of host masses) for the Hubble residuals from our \textit{No-Split} analysis. 

For a given set of Hubble residuals, $\bm{\Delta\mu} = (\Delta\mu_1,\dots,\Delta\mu_N)$, we assume that the residuals are consistent with a step function, with mean $m_<$ for host masses of $M_* < M_{*,\text{step}}$, and mean $m_>$ for host masses of $M_* \geq M_{*,\text{step}}$. We allow for the presence of some residual scatter, $\sigma_{\text{res}}$, about the step function which covers any variance in the Hubble residuals not explained by their individual errors. The likelihood for the Hubble residual of a supernova, $s$, in a host galaxy with stellar mass $M_{*,s}$ will be
\begin{multline}
    P(\Delta\mu_s|m_<,m_>,\sigma_{\text{res}},\log_{10} M_{*,\text{step}}) \\=
    \begin{cases}
        N(\Delta\mu_s|m_<,\sigma_s^2+\sigma_{\text{res}}^2) \qquad & \text{if } M_{*,s} < M_{*,\text{step}}\\
        N(\Delta\mu_s|m_>,\sigma_s^2+\sigma_{\text{res}}^2) \qquad & \text{if } M_{*,s} \geq M_{*,\text{step}}.
    \end{cases}
\end{multline}
To identify the optimal step location, $M_{*,\text{step}}$, for a sample of Hubble residuals, we compute the marginal likelihood, $P(\bm{\Delta\mu}|\log_{10} M_{*,\text{step}})$, of the Hubble residuals given the step location. The joint likelihood of all the data can be reduced to a product of low and high mass likelihoods,
\begin{multline}
    P(\bm{\Delta\mu}|m_<,m_>,\sigma_{\text{res}},\log_{10} M_{*,\text{step}}) \\= S_<N(m_<|\nu_<,\sigma_<^2) \times S_>N(m_>|\nu_>,\sigma_>^2),
\end{multline}
where
\begin{equation}
    \nu_< = \sigma_<^2\left[\sum_{\substack{i\\M_{*,i}<M_{*,\text{step}}}}\frac{\Delta\mu_i}{\sigma_i^2+\sigma_{\text{res}}^2}\right],
\end{equation}
\begin{equation}
    \sigma_<^2 = \left[\sum_{\substack{i\\M_{*,i}<M_{*,\text{step}}}}\frac{1}{\sigma_i^2+\sigma_{\text{res}}^2}\right]^{-1},
\end{equation}
and
\begin{equation}
    S_< = \sqrt{\frac{2\pi\sigma_<^2}{\prod_{i}2\pi(\sigma_i^2+\sigma_{\text{res}}^2)}}\exp\left[-\frac{1}{2}\left(\sum_i\frac{\Delta\mu_i^2}{\sigma_i^2+\sigma_{\text{res}}^2} - \frac{\nu_<^2}{\sigma_<^2}\right)\right],
    \label{eq:lkhdcoeff}
\end{equation}
and similarly for $\nu_>$, $\sigma_>$, $S_>$. Here, $\nu_<$ and $\nu_>$ are just the variance weighted means of the Hubble residuals either side of the step, with $\sigma_<$ and $\sigma_>$ being the standard errors on these. 

The marginal likelihood we are interested in can be found by imposing priors on $m_<$, $m_>$ and $\sigma_{\text{res}}$, and integrating over these parameters. We assume conditional priors of $P(m_<|\sigma_{\text{res}}) = N(m_<|0,1)$ and $P(m_>|\sigma_{\text{res}}) = N(m_>|0,1)$, which are weakly informative over the likely range of heights either side of the step. These conjugate priors mean that the marginalisation over $m_<$ and $m_>$ is analytic, giving
\begin{equation}
    P(\bm{\Delta\mu}|\sigma_{\text{res}},\log_{10} M_{*,\text{step}}) = S_<\tilde{S}_< \times S_>\tilde{S}_>,
\end{equation}
with $S_<$ (and equivalently, $S_>$) as given by Equation (\ref{eq:lkhdcoeff}),
and 
\begin{equation}
    \tilde{S}_< = \sqrt{\frac{1}{2\pi(\sigma_<^2 +1)}}\exp\left[-\frac{\nu_<^2}{2(\sigma_<^2 + 1)}\right],
\end{equation}
with $\tilde{S}_>$ being computed similarly. Introducing a uniform prior, $P(\sigma_{\text{res}})=U(\sigma_{\text{res}}|0,0.2)$, on the residual scatter parameter, we can numerically marginalise over this to compute the marginal likelihood of interest,
\begin{multline}
    P(\bm{\Delta\mu}|\log_{10} M_{*,\text{step}}) = \int^\infty_{-\infty}P(\bm{\Delta\mu}|\sigma_{\text{res}},\log_{10} M_{*,\text{step}})P(\sigma_{\text{res}})\,\mathrm{d}\sigma_{\text{res}}
    \\=\frac{1}{0.2}\int^{0.2}_0 S_<\tilde{S}_< \times S_>\tilde{S}_> \,\mathrm{d}\sigma_{\text{res}},
    \label{eq:marglkhdb}
\end{multline}
for a given step location, $\log_{10} M_{*,\text{step}}$. 

For the \textit{No-Split} Hubble residuals, we evaluate Equation (\ref{eq:marglkhdb}) for a range of host masses to find the step position(s) for which those Hubble residuals are most probable.

\section{Hyperprior Choice in Dust Law Population Analysis}
\label{app:sigmaRprior}
\begin{figure}
    \centering
    \includegraphics[width=\linewidth]{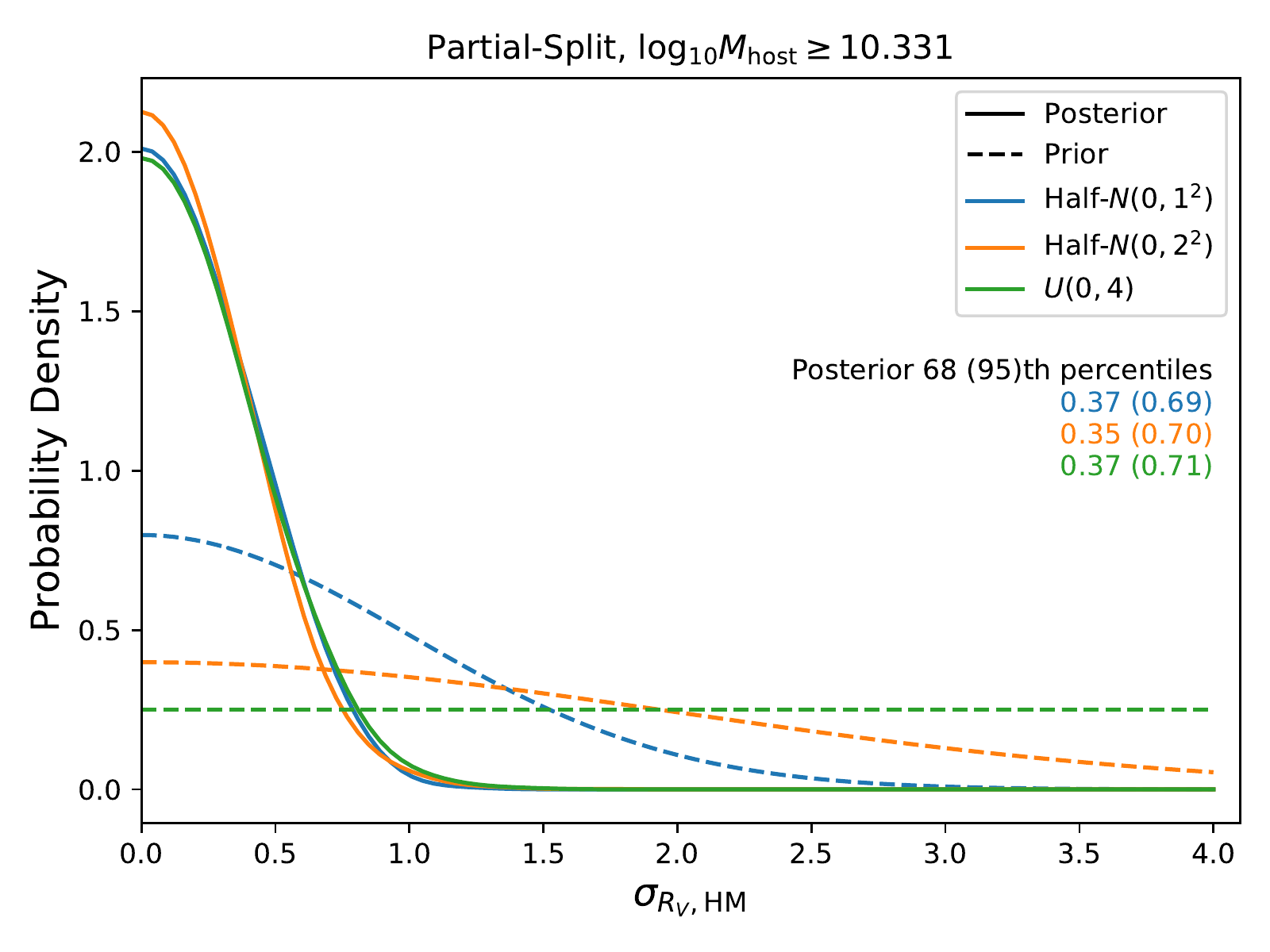}
    \includegraphics[width=\linewidth]{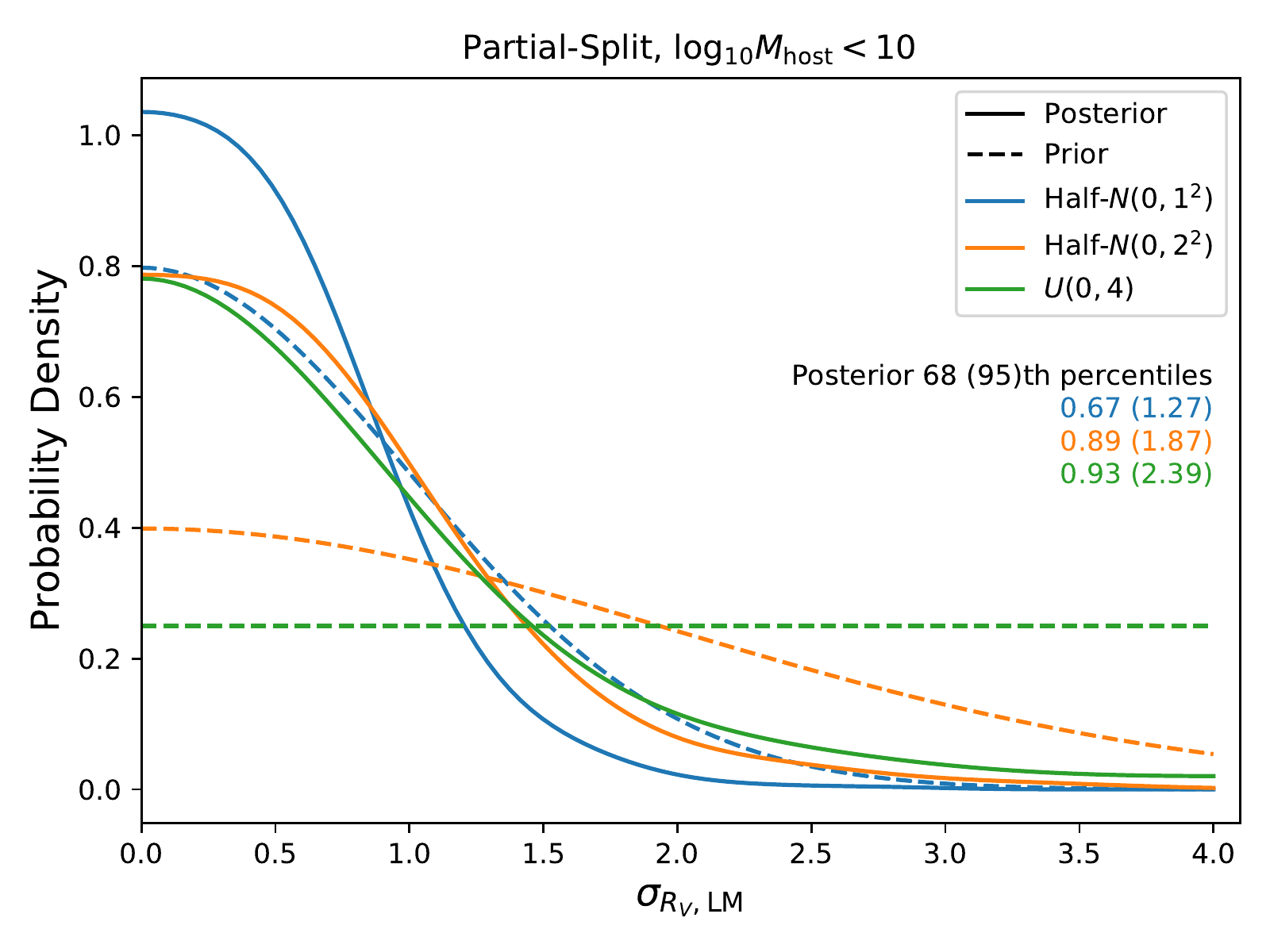}
    \caption{Plot illustrating the effect of hyperprior choice on the inference of $\sigma_R$ under the \textit{Partial-Split} model. Solid lines show the marginal posterior distributions of $\sigma_R$ for three choices of prior, with dashed lines plotting the corresponding priors. (top panel) Posterior distribution of $\sigma_R$ for galaxies more massive than the population median. Prior sensitivity is negligible in this case. The orange curve corresponds to our fiducial prior choice (corresponding to the purple $\sigma_R$ marginal distribution in Figure \ref{fig:RV_tauA_posteriors}). (bottom panel) Constraints on $\sigma_R$ for host galaxies less massive than $10^{10}\mathrm{M}_\odot$, a more unbalanced mass split. This illustrates the `worst-case scenario' in prior sensitivity.}
    \label{fig:sigmaR_prior}
\end{figure}

In the versions of our analysis allowing for a population (or populations) of dust law parameters, an important choice is the prior on the $R_V^s$ population distribution width, $\sigma_R$. As described in Section \ref{sec:popRV}, our default choice is a half-Normal prior, with a standard deviation of 2, as this penalises unreasonably large $\sigma_R$ without introducing an arbitrary hard cutoff. To test the sensitivity of our results to this choice, we repeat our analyses with two other hyperprior choices, $\sigma_R\sim\text{Half-}N(0,1^2)$, and $\sigma_R\sim U(0,4)$. The results of the \textit{No-Split} analysis, and the high mass side of the \textit{Partial-Split} analyses, are largely insensitive to the hyperprior choice. For the low mass side of the \textit{Partial-Split} analysis -- particularly for the more unbalanced choices of split point -- where constraints on the $R_V^s$ population properties are weaker, prior sensitivity is increased. The lower panel of figure \ref{fig:sigmaR_prior} shows the marginal posterior distribution of the low-mass value of $\sigma_R$ for the three hyperprior choices we consider (which are also plotted for comparison). These results are from the \textit{Partial-Split} model, with an assumed host galaxy mass split of $10^{10}\mathrm{M}_\odot$. This split point gives a rather unbalanced low-:high-mass ratio of 48:109 (see Table \ref{tab:masscuts}), meaning this plot reflects one of the `worst-case scenarios' for $\sigma_R$ hyperprior sensitivity. However, even in this case, we see that the posterior density is fairly insensitive between the weaker choices of a half-normal $\text{Half-}N(0,2^2)$ or uniform $\sigma_R\sim U(0,4)$ hyperprior. For comparison, the upper panel of Figure \ref{fig:sigmaR_prior} shows the marginal posteriors of the high-mass $\sigma_R$ when splitting at the median host mass. This case is much less prior-sensitive.

\section{Posterior Predictive Checks for Photometric Distance Fits}
\label{app:pppv}
When fitting a trained model to (potentially unseen) light curves for the purpose of photometric distance estimation (as in Section \ref{sec:hubblediagrams}), a means of assessing model fitness is valuable. As well as visually comparing the posterior distribution of model light curves to the data (Figure \ref{fig:lc_example}), we also compute a posterior predictive $p$-value \citep{gelman96, rubin84,meng94} as a numerical estimate of model fitness. This can be readily computed from a set of posterior samples, and provides a natural way of assessing fitness whilst averaging over the model parameters. The $p$-value is defined as the probability that a test quantity is more extreme when computed based on replicated data (simulated from the posterior predictive distribution) than it is when computed based on the data. This is well defined even when the test quantity is a function of the model parameters (as would be the case with $\chi^2$).

\begin{figure}
    \centering
    \includegraphics[width=\linewidth]{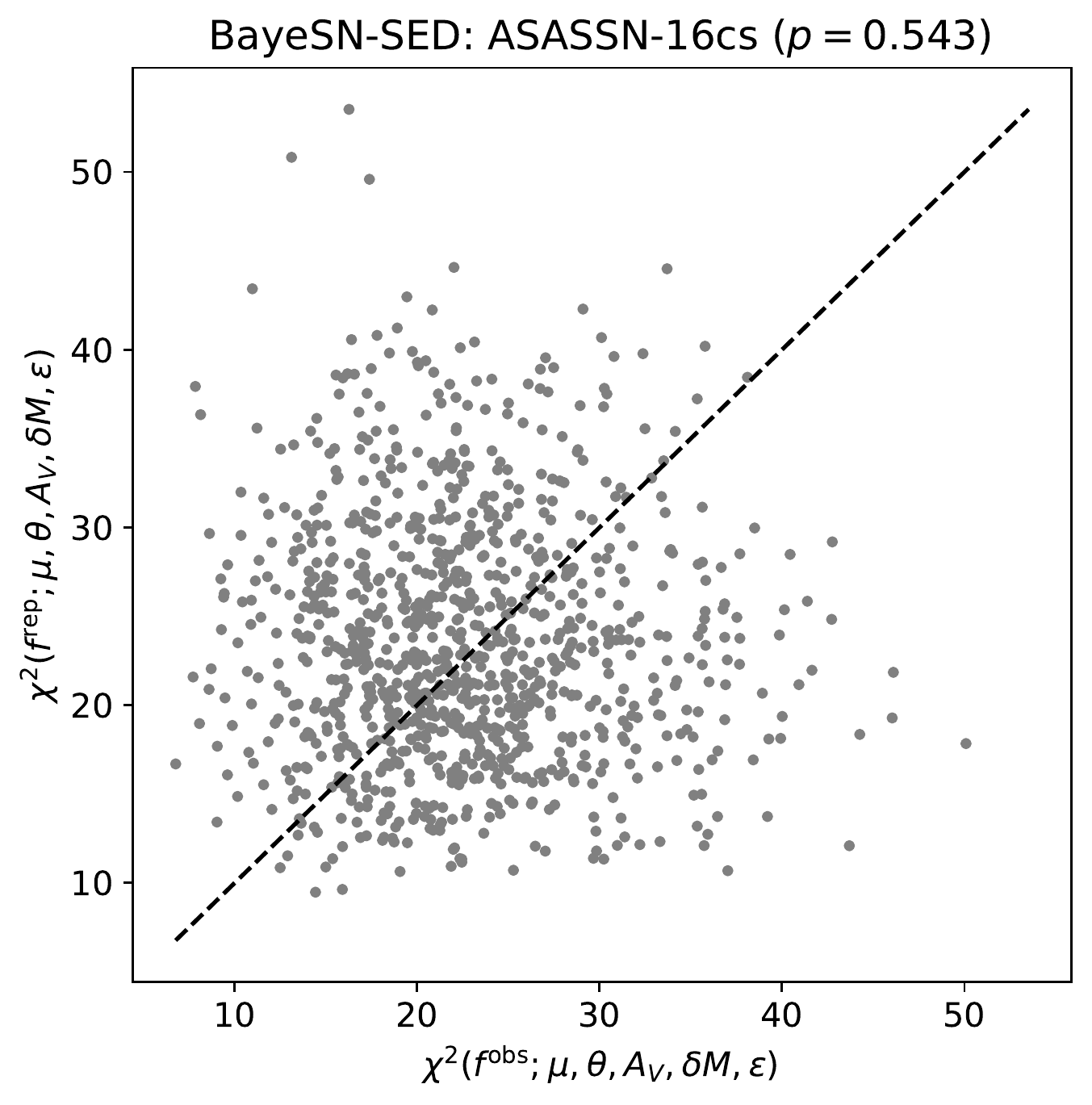}
    \includegraphics[width=\linewidth]{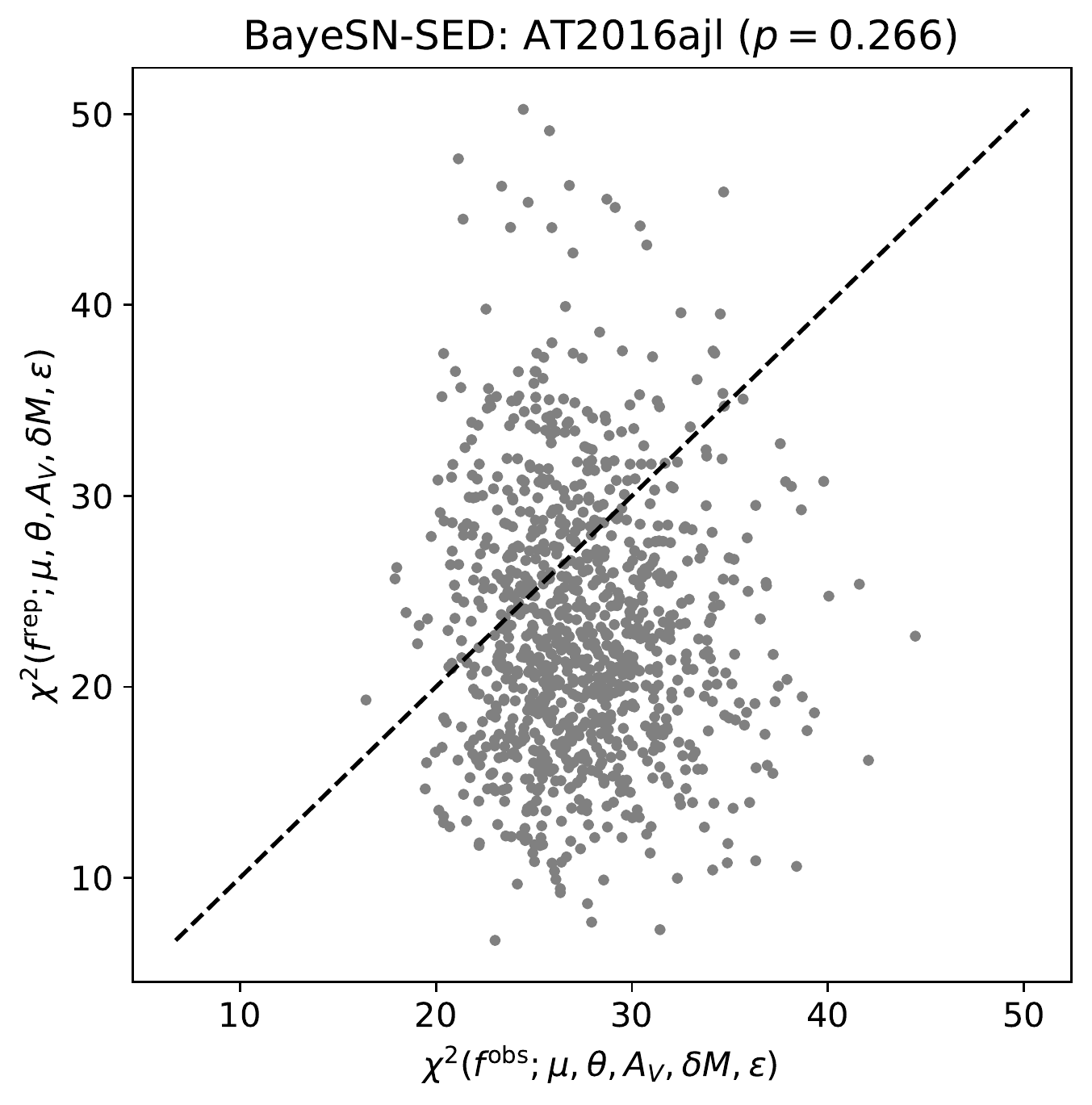}
    \caption{Illustration of the posterior predictive $p$-value estimation for the two light curve fits in Figure \ref{fig:lc_example}. Each point corresponds to a single posterior sample, with the $y$-axes plotting posterior predictive values, $\chi^2(\bm{f}^\text{rep}_m(\bm{\phi}_m);\bm{\phi}_m)$, of our test quantity, and the $x$-axes showing realised values, $\chi^2(\bm{f}^\text{obs};\bm{\phi}_m)$, computed based on the data. The fraction of samples lying above the dashed $1:1$ line gives our $p$-value estimate.}
    \label{fig:pppv_example}
\end{figure}

In our case, we choose to use
\begin{equation}
    \chi^2(\bm{f}; \bm{\phi}) = \sum_{n=1}^N\frac{[f_n - f_n^\text{model}(\bm{\phi})]^2}{\sigma_n^2}
\end{equation}
as our test statistic\footnote{We have also tried the mean squared error (MSE), and found this to be a reasonable alternative.}, where $\bm{\phi}=(\mu_s^\text{phot}, \theta_1^s, A_V^s, \delta M_s, \bm{e}_s)$ represents our model parameters; $\bm{f}=(f_1,\dots,f_N)$ is the vector of fluxes being tested; $f_n^\text{model}$ is the model flux computed at the time, and in the passband, of $f_n$, given the parameters $\bm{\phi}$; and $\sigma_n$ is the flux error associated with $f_n$. For a given supernova, we have a set of $N$ flux observations, $\bm{f}^\text{obs}=(f_1^\text{obs},\dots,f_N^\text{obs})$, with associated errors, $\bm{\sigma}=(\sigma_1,\dots,\sigma_N)$. Conditional on $\bm{f}^\text{obs}$, we will also have a set of $M$ samples, $(\bm{\phi}_1,\dots,\bm{\phi}_M)$, from the posterior distribution $P(\bm{\phi}|\bm{f}^\text{obs})$. For the $m$th posterior sample, $\bm{\phi}_m$, we can compute a vector of model fluxes, $\bm{f}^\text{model}_m(\bm{\phi}_m)$, and a vector of replicated observations, $\bm{f}^\text{rep}_m(\bm{\phi}_m)$, where
\begin{equation}
    f^\text{rep}_{m,n}(\bm{\phi}_m) \sim N(f^\text{model}_{m,n}(\bm{\phi}_m), \sigma_n^2)
\end{equation}
for $n = 1,\dots,N$. Then, we can numerically estimate our posterior predictive $p$-value,
\begin{equation}
    p=P[\chi^2(\bm{f}^\text{rep}(\bm{\phi});\bm{\phi})\geq\chi^2(\bm{f}^\text{obs};\bm{\phi})|\bm{f}^\text{obs}],
\end{equation}
by computing the fraction of posterior samples for which $\chi^2(\bm{f}^\text{rep}_m(\bm{\phi}_m);\bm{\phi}_m)\geq\chi^2(\bm{f}^\text{obs};\bm{\phi}_m)$. An extreme $p$-value (close to 0 or 1) would suggest that the model may not describe the data well.

Figure \ref{fig:pppv_example} illustrates this process visually for the two light curve fits featured in Figure \ref{fig:lc_example}. For each supernova, we have plotted the distribution of $\chi^2(\bm{f}^\text{rep}(\bm{\phi});\bm{\phi})$ vs. $\chi^2(\bm{f}^\text{obs};\bm{\phi})$, based on 1000 posterior samples of $[\bm{\phi}, \bm{f}^\text{rep}(\bm{\phi})]$. The estimated $p$-value is computed from the fraction of samples lying above the dashed $1:1$ line.

\bsp	
\label{lastpage}
\end{document}